\documentclass[3p,review]{elsarticle}
\usepackage{geometry}                
\usepackage{graphicx}
\usepackage{subfig}
\usepackage{amssymb}
\usepackage{amsmath}
\usepackage{epstopdf}
\usepackage{color}
\usepackage{bm}
\usepackage{mathrsfs} 
\usepackage{hyperref}
\usepackage{wrapfig}
\usepackage{textcomp}
\usepackage{tabularx}
\usepackage{comment}
\usepackage{subeqnarray}
\usepackage{mathtools}
\usepackage{multirow}
\usepackage{placeins}
\newcolumntype{Y}{>{\centering\arraybackslash}X}

\usepackage[nameinlink,noabbrev]{cleveref}
\makeatletter
\def\@seccntformat#1{\@ifundefined{#1@cntformat}%
   {\csname the#1\endcsname\quad}
   {\csname #1@cntformat\endcsname}
}
\makeatother

\makeatletter
\def\ps@pprintTitle{%
   \let\@oddhead\@empty
   \let\@evenhead\@empty
   \let\@oddfoot\@empty
   \let\@evenfoot\@oddfoot
}
\makeatother

\usepackage{scalerel}[2014/03/10]
\usepackage{stackengine}

\begin{document}

\begin{frontmatter}
 \title{  Application of the ASBM-SA closure in a turbulent flow over a hump in the presence 
of separation control.
}
\author[1,2]{C.~F.~Panagiotou}
\author[3]{F.~S.~Stylianou}
\author[5,6]{E.~Gravanis}
\author[5,6]{E.~Akylas}
\author[3,4]{S.~C.~Kassinos}
\address[1]{ Laboratory Of Environmental Engineering (GAIA), Nireas-International Water Research Centre, 
             University of Cyprus, Nicosia, Cyprus }
\address[2]{ Department of Civil \& Environmental Engineering, University of Cyprus, 
             Nicosia, Cyprus }
\address[3]{ Computational Sciences Laboratory (UCY-CompSci), Nireas-International 
             Water Research Centre, University of Cyprus, Nicosia, Cyprus}
\address[4]{ Department of Mechanical \& Manufacturing Engineering, University of 
             Cyprus, Nicosia, Cyprus}                           
\address[5]{ Department of Civil Engineering \& Geomatics, Cyprus University of Technology, Limassol, Cyprus }
\address[6]{ Eratosthenes Centre of Excellence, Cyprus University of Technology, Limassol, Cyprus }
\begin{abstract}
We demonstrate the coupling between the Algebraic Structure-Based Model (ASBM) and the 
one-equation Spalart–Allmaras (SA) model, which provides an easy route to bringing structure 
information in engineering turbulence closures. The estimation ability of the hybrid model
was tested for a flow over a hump model with no-flow control and steady suction. 
ASBM-SA model produced satisfactory predictions for the streamwise Reynolds stress component, 
while a qualitative agreement with the experiments was achieved for the transverse component.
Regarding the shear stress component, ASBM-SA closure provides improved predictions compared 
to SA in the entire domain.   
\end{abstract}

\end{frontmatter}

\pagestyle{plain}

\newcount\ndots
\def\drawline#1#2{\raise 2.5pt\vbox{\hrule width #1pt height #2pt}}
\def\spacce#1{\hskip #1pt}
\def\solid{\drawline{24}{.5}\nobreak\ }
\def\bdash{\hbox{\drawline{4}{.5}\spacce{2}}}
\def\dashed{\bdash\bdash\bdash\bdash\nobreak\ }
\def\bdot{\hbox{\drawline{1}{.5}\spacce{2}}}
\def\dotted{\hbox{\leaders\bdot\hskip 24pt}\nobreak\ }
\def\chndash{\hbox {\drawline{8.5}{.5}\spacce{2}\drawline{3}{.5}\spacce{2}\drawline{8.5}{.5}}\nobreak\ }
\def\chndot{\hbox {\drawline{9.5}{.5}\spacce{2}\drawline{1}{.5}\spacce{2}\drawline{9.5}{.5}}\nobreak\ }
\def\chndotdot{\hbox {\drawline{8}{.5}\spacce{2}\drawline{1}{.5}\spacce{2}\drawline{1}{.5}\spacce{2}\drawline{8}{.5}}\nobreak\ }
\def\chndotdotdot{\hbox {\drawline{8}{.5}\spacce{2}\drawline{1}{.5}\spacce{2}\drawline{1}{.5}\spacce{2}\drawline{1}{.5}\spacce{2}\drawline{8}{.5}}\nobreak\ }
\def\trian{\raise 1.25pt\hbox{$\scriptscriptstyle\triangle$}\nobreak\ }
\def\circle{$\circ$\nobreak\ }
\def\diam{$\diamond$\nobreak\ }
\def\solidcircle{$\bullet$\nobreak\ }

\def\smalltriangle{$\scriptstyle\triangle\textstyle$\nobreak\ }
\def\smallplus{$\scriptstyle + \textstyle$\nobreak\ }
\def\smalltimes{$\scriptstyle\times\textstyle$\nobreak\ }
\def\smallnabla{$\scriptstyle\nabla\textstyle$\nobreak\ }
\def\square{${\vcenter{\hrule height .4pt
        \hbox{\vrule width .4pt height 3pt \kern 3pt
        \vrule width .4pt}
        \hrule height .4pt}}$\nobreak\ }
\def\plus{\raise 1.25pt \hbox{$\scriptscriptstyle +$}\nobreak\ }
\def\x{\raise 1.25pt \hbox{$\scriptscriptstyle \times$}\nobreak\ }
\def\ldash{\hbox {\drawline{7}{.5}\spacce{2}\drawline{7}{.5}\spacce{2}\drawline{7}{.5}}\nobreak\ }
\def\lchndash{\hbox {\drawline{15}{.5}\spacce{3}\drawline{7}{.5}}\nobreak\ }
\def\tsolid{\drawline{24}{1.2}\nobreak\ }

\def\graytrian{\raise 1.25pt
   \hbox to 3bp{
\hfill}\nobreak\ }
\def\trian{\raise 1.25pt
   \hbox to 3bp{
\hfill}\nobreak\ }
\def\solidtrian{\raise 1.25pt
   \hbox to 3bp{
\hfill}\nobreak\ }

\def\graytriand{\raise 1.25pt
   \hbox to 3bp{
\hfill}\nobreak\ }
\def\triand{\raise 1.25pt
   \hbox to 3bp{
\hfill}\nobreak\ }
\def\solidtriand{\raise 1.25pt
   \hbox to 3bp{
\hfill}\nobreak\ }

\def\square{\raise 1.pt
   \hbox to 3bp{
\hfill}\nobreak\ }
\def\graysquare{\raise 1.pt
   \hbox to 3bp{
\hfill}\nobreak\ }
\def\solidsquare{\raise 1.pt
   \hbox to 3bp{
\hfill}\nobreak\ }

\def\circle{\raise 1.pt
   \hbox to 3bp{
\hfill}\nobreak\ }
\def\solidcircle{\raise 1.pt
   \hbox to 3bp{
\hfill}\nobreak\ }
\def\graycircle{\raise 1.pt
   \hbox to 3bp{
\hfill}\nobreak\ }

\def\dotcirc{$\cdots\ $\circle$\cdots$\ }

\def\dashx {\bdash\bdash\smalltimes\bdash\bdash}

\def\chndashx {\drawline{8.5}{.5}\spacce{2}
\drawline{3}{.5}$\scriptstyle\times\textstyle$\drawline{8.5}{.5}\spacce{2}
\drawline{3}{.5}\nobreak\ }

\def\solidcclose{\drawline{10}{.5}\nobreak\raise
  0.5pt\hbox{$\bullet$}\drawline{10}{.5}\nobreak\ }

\def\solidsclose{\drawline{10}{.5}\nobreak\raise
  0.5pt\hbox{\solidsquare}\drawline{10}{.5}\nobreak\ }

\def\solidtclose{\drawline{10}{.5}\nobreak\raise
  0.5pt\hbox{\solidtrian}\drawline{10}{.5}\nobreak\ }

\def\solidcopen{\drawline{10}{.5}\nobreak\raise
  0.5pt\hbox{\circle}\drawline{10}{.5}\nobreak\ }

\def\solidsopen{\drawline{10}{.5}\nobreak\raise
  0.5pt\hbox{\square}\drawline{10}{.5}\nobreak\ }

\def\solidtopen{\drawline{10}{.5}\nobreak\raise
  0.5pt\hbox{\trian}\drawline{10}{.5}\nobreak\ }

\def\solidx{\drawline{10}{.5}\nobreak\raise
  0.5pt\hbox{\x}\drawline{10}{.5}\nobreak\ }

  
\font\msakkk=msam10
\def\diamsol{{\msakkk \char7}}
\def\diamop{{\msakkk \char6}}
\def\starsol{{\msakkk \char70}}
\def\triansolu{{\msakkk \char78}}
\def\triansold{{\msakkk \char72}}
\def\triansolr{{\msakkk \char73}}
\def\triansoll{{\msakkk \char74}}
\def\trianopu{{\msakkk \char77}}
\def\trianopd{{\msakkk \char79}}
\def\trianopr{{\msakkk \char66}}
\def\trianopl{{\msakkk \char67}}
\def\squarsol{{\msakkk \char4}}
\def\squarop{{\msakkk \char3}}

\def\mydash{\hbox{\drawline{2}{.5}\spacce{2}}}
\def\shdashed{\mydash\mydash\mydash\mydash\mydash\mydash\nobreak\ }

\def\bdot{\hbox{\drawline{.5}{.5}\spacce{1}}}
\def\dotted{\hbox{\leaders\bdot\hskip 24pt}\nobreak\ }

\newcommand{\AC}[1]{\vspace{.5cm} [A COMPLETER: #1]}
\newcommand{\CTRClass}{{\it{\bf ctr\_summer.cls }}}
\newcommand{\noi}{\par}

\section{Introduction}\label{introduction}
The class of RANS models most often used in engineering applications is that of Eddy 
Viscosity Models (EVM). One of the most popular EVM is the Spalart–Allmaras (SA) one-equation
model \cite{SA94}. The SA model is often favored by practicing engineers 
because it exhibits superior robustness, low CPU time requirements and substantially lower 
sensitivity to grid resolution compared to two-equation models. On the other hand, one has 
to recognize that, despite its computational and implementational attractiveness, the eddy 
viscosity assumption is also the source of some of the most important performance limitations.
For example, like other EVM, the SA model fails to capture important flow features, such as 
turbulence anisotropy or the effects of mean or system rotation.
A common feature of the classical closure approaches described so far is the assumption that 
all key information about the turbulence is contained in the scales of the turbulence and in 
the turbulence stress tensor. However, one should consider that the turbulent stresses 
contain information only about the componentality of the turbulence, i.e. about the directions 
in which the turbulence fluctuations associated with large-scale eddies are most energetic. 
Thus, traditional closures do not take into account the morphology of the energy-containing 
eddies. Yet, eddies tend to organize spatially the fluctuating motion in their vicinity. In 
doing so, they eliminate gradients of fluctuation fields in some directions (those in which 
the spatial extent of the structure is significant) and enhance gradients in other directions
(those in which the extent of the structure is small). Thus, associated with each eddy are 
local axes of dependence and independence that determine the dimensionality of the local 
turbulence structure. This structure dimensionality information is complementary to the 
componentality information contained in the Reynolds stresses, and as Kassinos \& Reynolds 
\cite{TF61} and Kassinos et al. \cite{kas2001} have shown, it is dynamically important.
A detailed description of the complete set of the turbulence structure tensors is given in 
several works \cite{kas2001, Stylianou2015, Panagiotou2016}.

The significant effect that the morphology of the large-scale structures has on the evolution of  
turbulent statistics \cite{Shraiman2000} has motivated the development of structure-based 
models. These models can be classified into two categories: differential and algebraic. The 
first category involves solving a set of transport equations to evaluate structure tensors. 
Simplified models have been proposed that are applicable at the homogeneous limit 
\cite{Kassinos96, Kassinos2012, Panagiotou2014}, while a more sophisticated differential model was proposed by 
Poroseva et al. \cite{Poroseva00} that was tested in a turbulent flow passing through a 
cylindrical pipe that rotates around its longitudinal axis.
Furthermore, structure-based models were recently constructed to evaluate scalar transport,
ranging from simple passive scalars \cite{Panagiotou2016, Panagiotou2020} to strongly stably 
stratified flows \cite{Panagiotou2017}.

The second category refers to algebraic approaches, which are based on assumptions that lead 
to constitutive equations.
The Algebraic Structure-Based Model (ASBM) \cite{kas2006, TF85} is an engineering 
structure-based turbulence model that follows this second approach. It is a fully realizable 
two-equation structure-aware model that provides the full Reynolds stress tensor.
Panagiotou \& Kassinos \cite{Panagiotou2015} presented a successful coupling between the 
ASBM with the one-equation SA model. Their intention was to combine the numerical robustness 
and stability of the SA model along with the deeper physical content of the ASBM.
The performance of the hybrid model, called ASBM-SA, was evaluated in several standard benchmark 
cases, ranging from simple fully-developed channel flows to a flow over a steep hill, achieving
an overall good agreement between model and experimental predictions. 
Hence, the aim of this study is to evaluate the performance of the ASBM-SA closure to a more
complex case, in particular the case of turbulent flow over a two-dimensional hill with and 
without separation control.

\section{Coupling between ASBM and the SA model}\label{coupling}

The main limitation of SA (and any other one-equation model) is that it does not
provide of a complete set of turbulence scales. On the other hand, the ASBM closure relies
on the availability of suitable turbulence scales and this was the key stumbling block in trying
to couple the ASBM and SA closures.
The Bradshaw hypothesis \cite{Bradshaw67} has been used as the starting point for transforming 
turbulence scales $\kappa-\epsilon$ closures, where $\kappa$ and $\epsilon$ are the turbulent kinetic
energy and energy dissipation rate respectively, to one equation models. Here, our objective 
is to use the same phenomenology in order to extract these scales from the SA one-equation 
closure. A complete description of this formulation and on how the coupling between the two 
closures has been achieved can be found in Panagiotou \& Kassinos \cite{Panagiotou2015}.

\section{Results and discussion}\label{introduction}

\subsection{Outline}

\begin{figure}[h!]
 \centering
\includegraphics[width=1.0\textwidth]{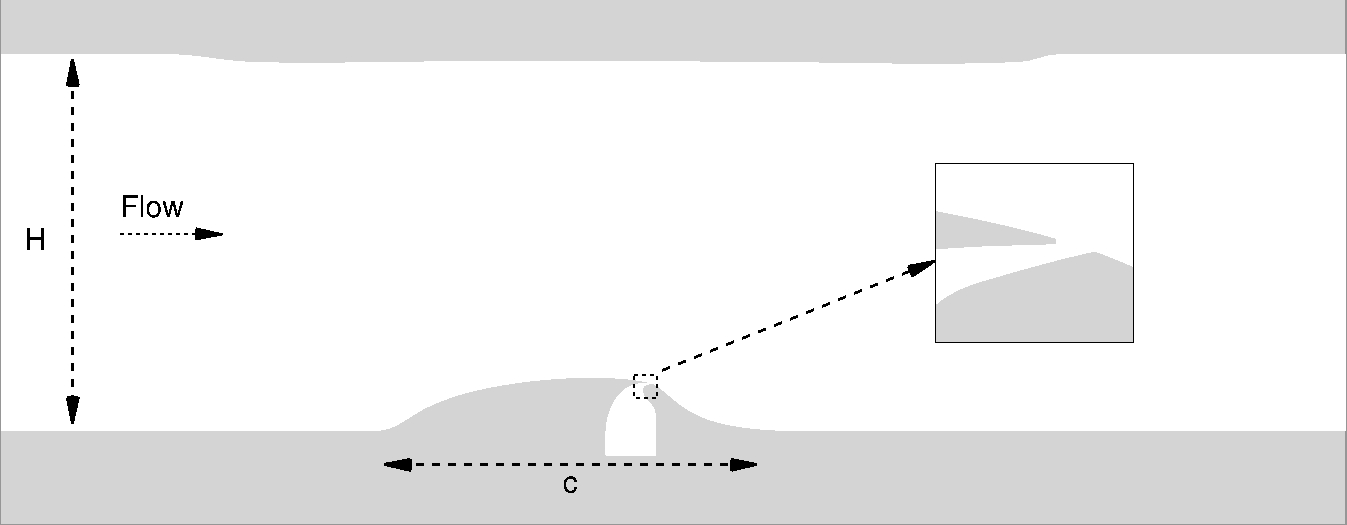}
\caption{Sketch of the geometry, with a  modification along the top-surface such as to 
account for the side-plate effects, as described in \cite{Rumsey2014}.}
\label{fig:sketch_domain}
\end{figure}

In this study we have considered  the case of no-flow control, as well as the case where active-control is applied via 
steady-suction. During the CFDVAL2004  Workshop \cite{Rumsey2004}, these two cases were 
investigated in depth to assess  the performance of popular engineering turbulence models 
in strongly separated flows subjected to favorable/adverse pressure gradients.  
Experiments have been conducted in the NASA Langley Transonic Cryogenic Tunnel by Greenblatt 
et al. \cite{Greenblatt2004}. The shape of the hump is that of a ``Modified Glauert-Goldschmied" 
hill, similar to the one used by Seifert and Pack \cite{Seifert2002}. The experiments are 
nominally two-dimensional (2D), despite the presence of three dimensional (3D) effects near 
the side end-plates. The scenarios involved both uncontrolled and controlled flow (steady 
suction) for Reynolds  numbers ($Re$) ranging from 0.37 up to 1.1 million,  corresponding 
to Mach numbers ($M$) ranging from 0.04 up to 0.12. One no-flow control case and one 
active-control case were selected for the extraction of detailed experimental measurements. 
Figure~\ref{fig:sketch_domain}  shows the geometry of the whole domain, including  a 
detailed view of the flow control slot. The chord length of the hump is denoted as $c$, the 
height of the domain $H$ is $90 \%$ the chord length, while the maximum height of the hump 
is approximately $0.13 \ c$.  The slot is located near $x/c  \approx 0.65 $,  where the slot 
width  $h$ is $0.00187 \ c$. Detailed information  regarding the geometry, computational 
grids and the relevant experiments  can be found in \cite{Rumsey2014}. 
One of the conclusions reached during the CFDVAL2004 Workshop is that blockage effects 
stemming from the presence of side plates need to be accounted for in simulations, otherwise 
the computed pressure coefficients levels exhibit significant discrepancy relative to the 
experiments, especially over the hump. Thus, the top tunnel surface around hump location is 
modified so as to reflect the change in the tunnel cross-sectional area due to the presence 
of the side-plates, as described in \cite{Rumsey2014}. 
\FloatBarrier

\subsection{Validation cases}

\subsubsection{Turbulent boundary layer}

\begin{figure*}[h!]
\centering
\subfloat[]
{\label{fig:bl_ux_grid_analysis}
\includegraphics[scale=0.13]{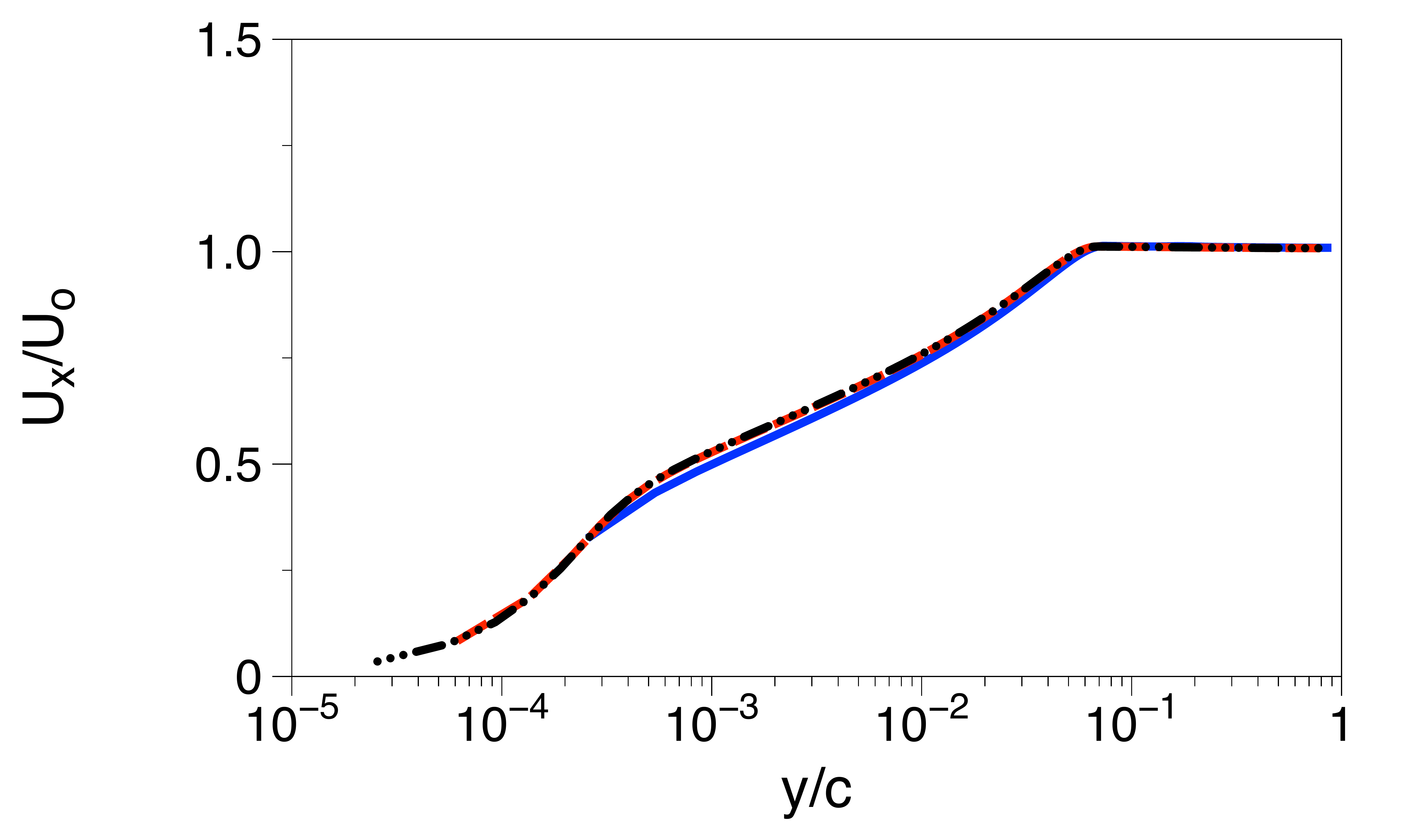}} 
\subfloat[]
{\label{fig::bl_nut_grid_analysis}
\includegraphics[scale=0.13]{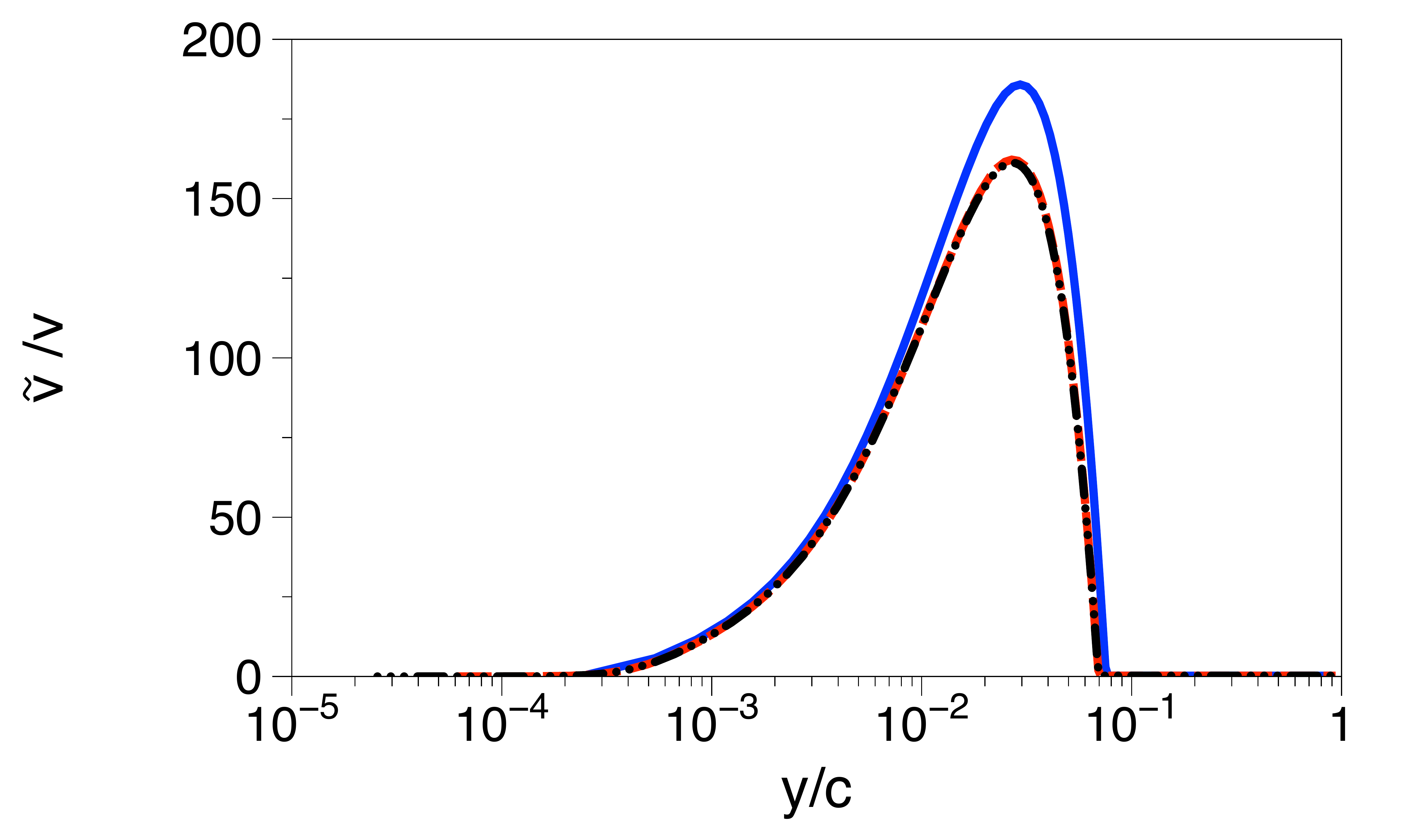}} 
 \label{fig::bl_grid_analysis}
\caption{Grid-convergence analysis for a spatially developing turbulent boundary layer at 
$Re_{\delta} \approx 68,200$.  SA model predictions for (a) the streamwise mean velocity 
and (b) the pseudo-viscosity. Comparison is made  among three different grids: Grid 1 
($\solid$) ; Grid 2 ($\dashed$) ; Grid 3 ($\chndotdotdot$).  }
\end{figure*}

\begin{figure}[h!]
 \centering
\includegraphics[scale=0.2]{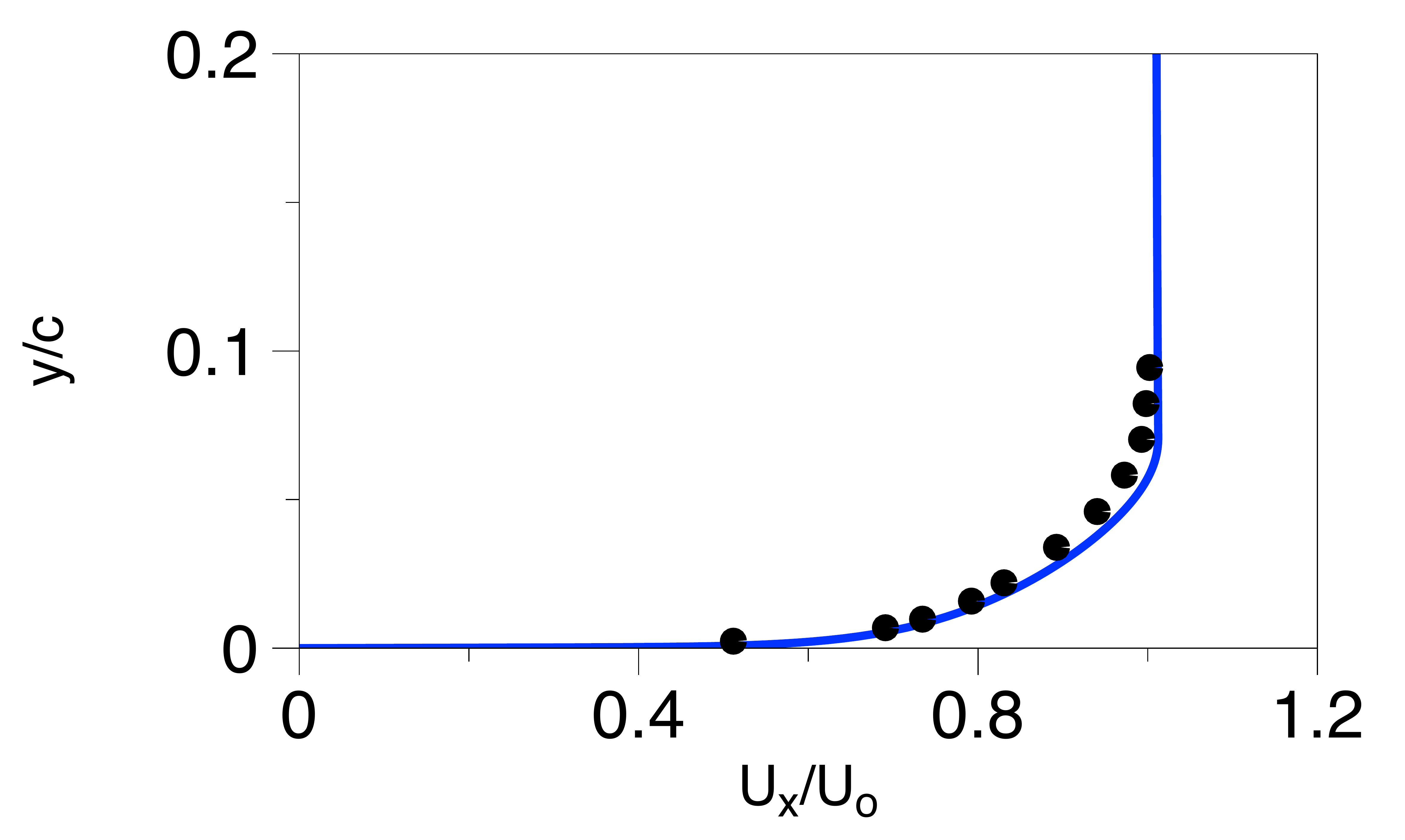}
\caption{SA model predictions (lines) for the streamwise mean velocity. Comparison is made 
to the experiments (symbols) of Greenblatt et al. \cite{Greenblatt2004}. }
 \label{fig:BL_UX_PROFILE}
\end{figure}

As a first step, we performed steady computations of a spatially developing boundary layer 
flow over a flat plate for flow conditions that correspond to the experiments of Greenblatt 
et al. \cite{Greenblatt2004}. 
Profiles of the converged solution at a specific streamwise location were extracted and then used as 
inlet boundary conditions for the  cases involving the 2D hump. The desired Reynolds number 
is $Re_{\delta} \approx 68200$ based on the freestream velocity $U_{\infty}$, where subscript 
$\infty$ denotes freestream values, and the boundary 
layer thickness $\delta \approx 0.074 \,c $. At the inlet, Dirichlet boundary conditions are 
imposed for the mean streamwise velocity $U_{x}$  and the pseudo-viscosity $\tilde{\nu}$ variables, 
such as ${ \tilde{\nu}_{\infty} }/{\nu} \approx 3$  and $U_{\infty}=0.1 \ M$, yielding a freestream 
Reynolds number $Re_{\infty}=929,000$ based on the chord length $c$ of the hump, air viscosity $\nu$ 
and the freestream velocity $U_{\infty}$. 
At the outlet, a penalty condition is imposed to prevent the occurrence of reflectional effects while ensuring
mass conservation. A slip condition was imposed at the top surface, a no-slip condition at 
the bottom wall surface and periodic conditions along the spanwise direction. In order to 
obtain grid-independent solutions, three different meshes of increasing resolution were 
considered. For each mesh, geometric functions were used to define the normal distribution 
of the nodes, while uniform spacing has been  adopted along the streamwise direction. 
Grid 1 contains a non-uniform mesh of size 120 x 90 x 1 along the streamwise, wall-normal 
and spanwise directions respectively.  
The corresponding size for Grid 2 is  130 x 120 x 1 and for Grid 3 is 140 x 150 x 1.  The 
finest grid yields a value of $y^{+}$ around 0.5 for the wall-adjacent cell at the location
of the extracted data. Figures~\ref{fig:bl_ux_grid_analysis}-b show predictions using the 
SA closure for the streamwise mean velocity and pseudo-viscosity respectively.  
In Figure~\ref{fig:BL_UX_PROFILE} we show a comparison between the predictions of the SA 
model using Grid 3  as the baseline grid, and the experimental data for the streamwise mean 
velocity $U_{x}$, yielding a good agreement. 
\FloatBarrier

\FloatBarrier

\subsubsection{ No-flow control }

\begin{figure*}[h!]
\centering
\subfloat[]
{\label{fig:panoramic_view}
\includegraphics[scale=0.25]{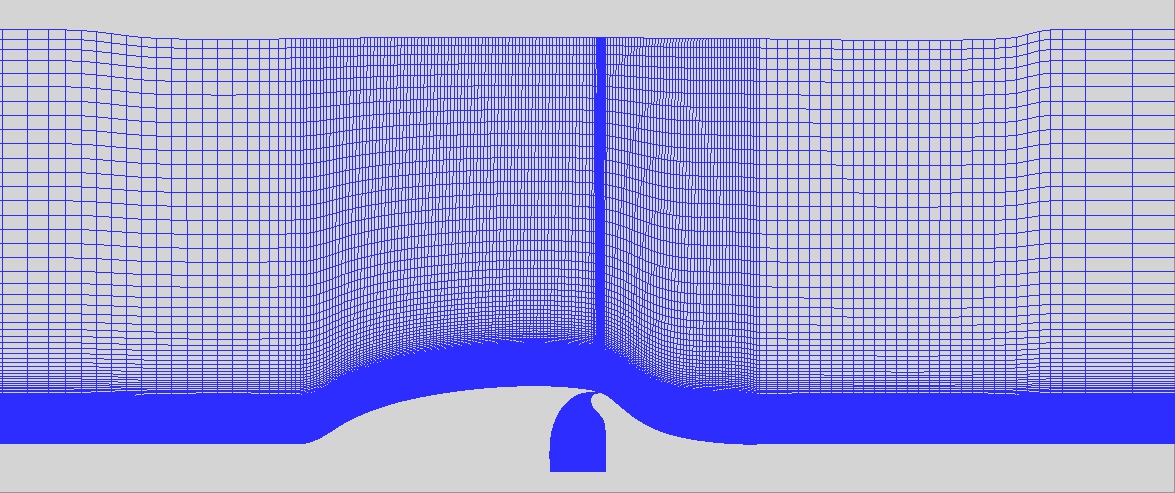}} \\
\centering
\subfloat[]
{\label{fig:zoom_cavity}
\includegraphics[scale=0.25]{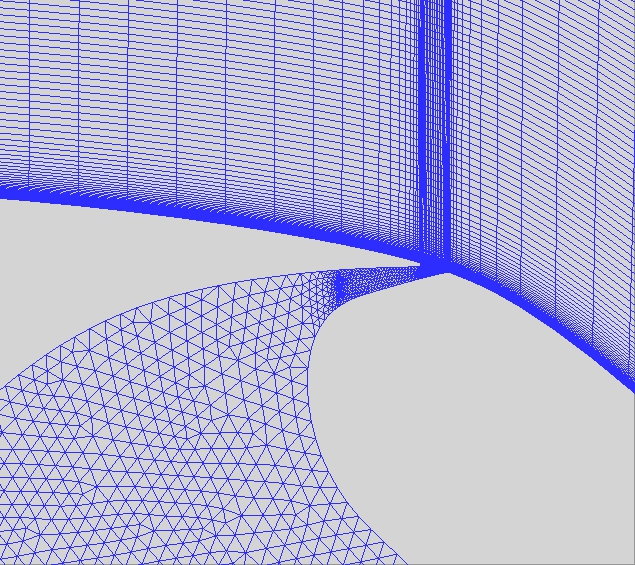}} 
\subfloat[]
{\label{fig:zoom_slot}
\includegraphics[scale=0.25]{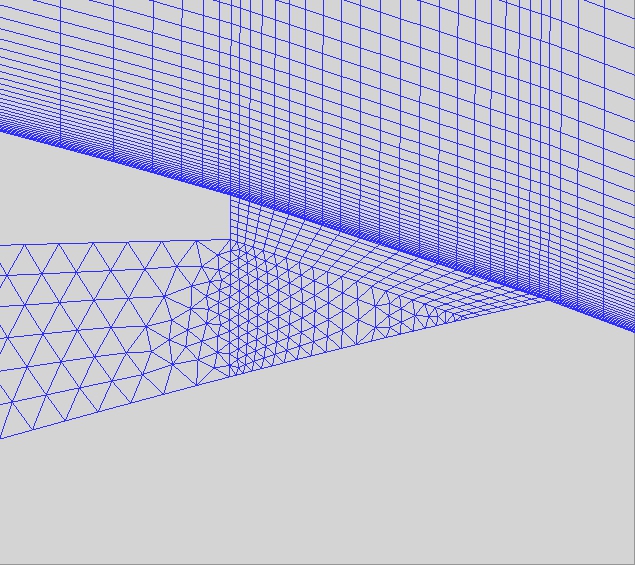}} 
\caption{ Unstructured computational grid with details of  the slot region.} 
 \label{fig:hill_meshing}
\end{figure*}

\begin{figure*}[h!]
\centering
\subfloat[]
{\label{fig:xc_cp_mesh_comp}
\includegraphics[scale=0.13]{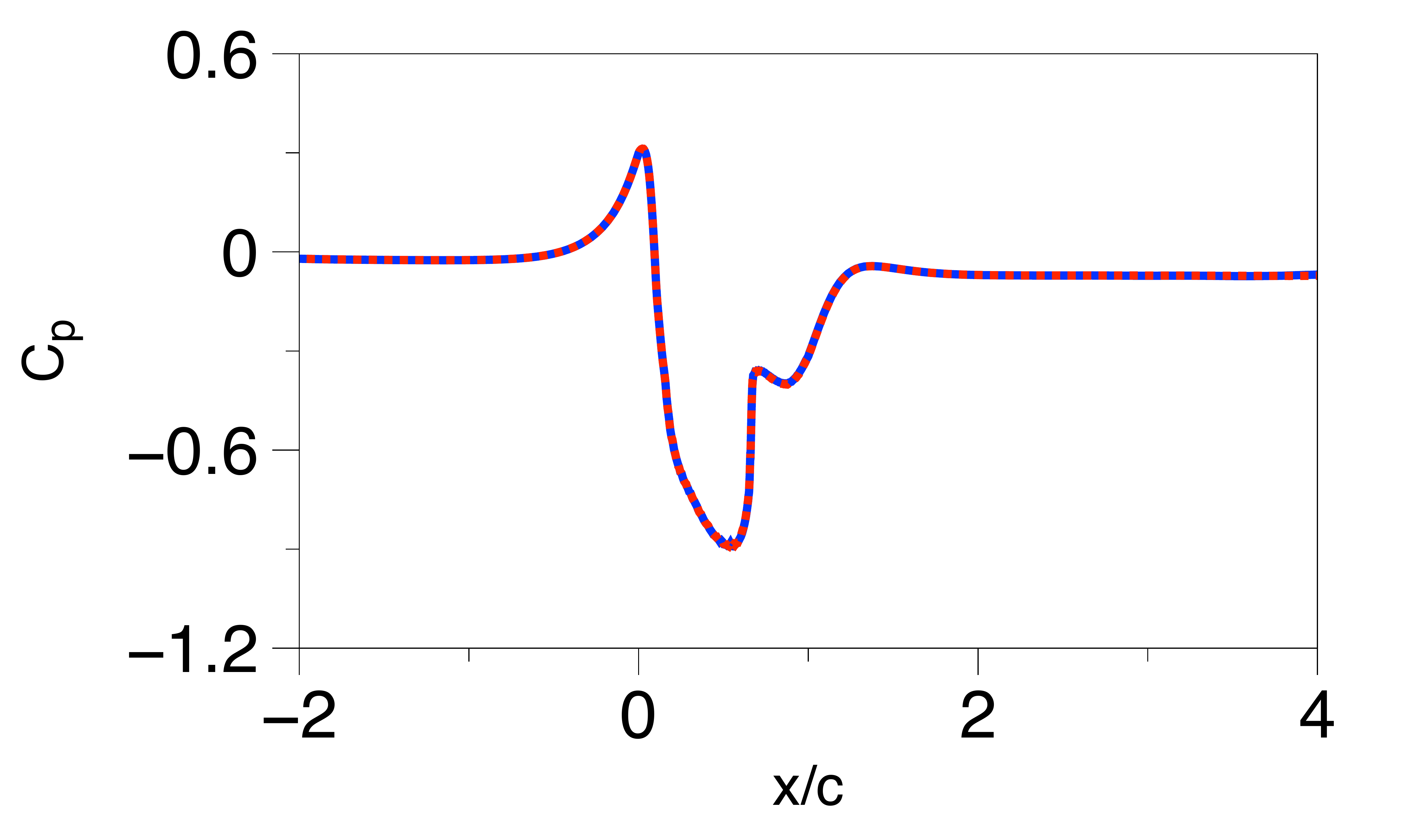}} 
\subfloat[]
{\label{fig:xc_cf_mesh_comp}
\includegraphics[scale=0.13]{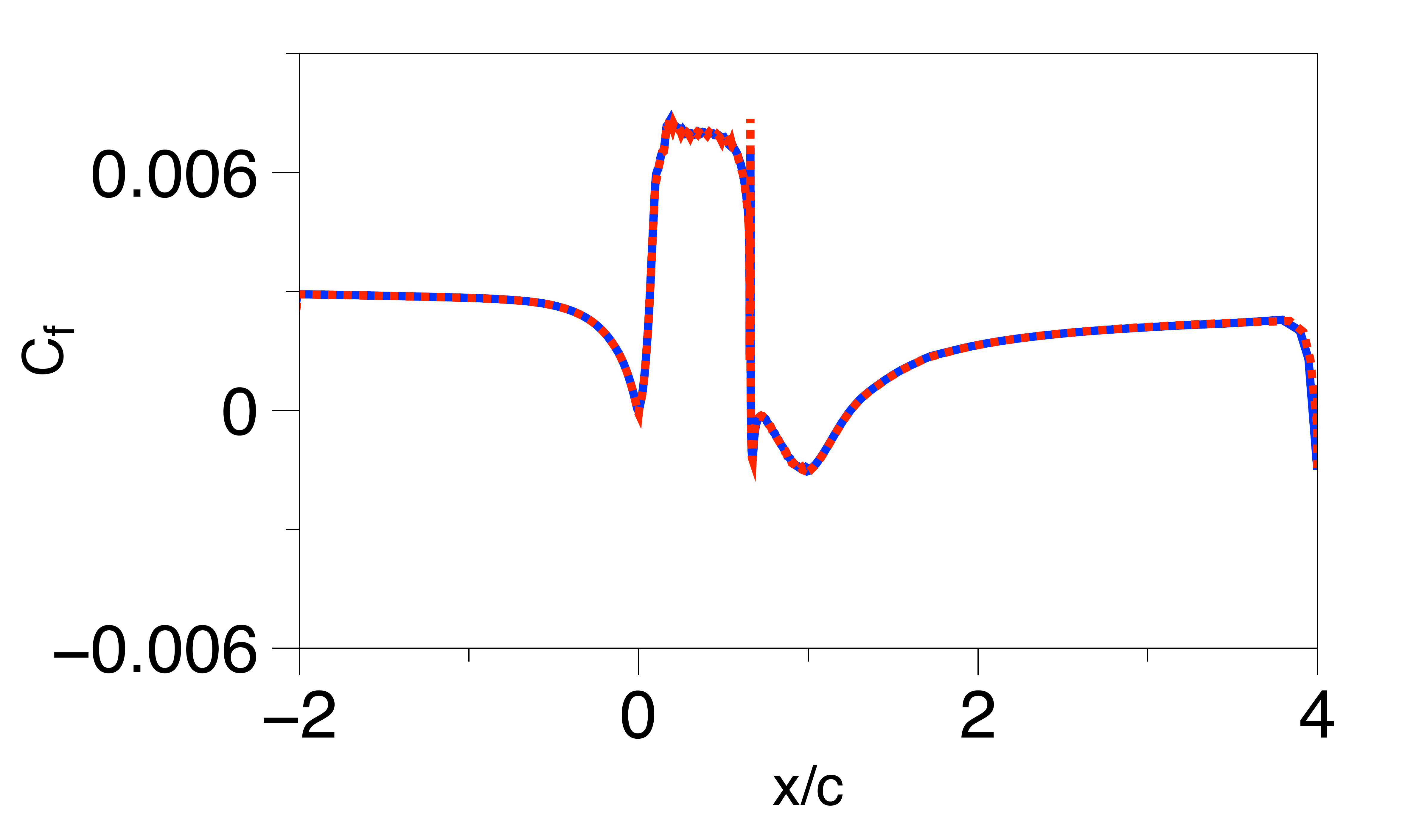}} 
 \label{fig:grid_convergence_hill}
\caption{Effect of grid on SA model predictions for the uncontrolled case, 
for (a) the wall-static pressure coefficient and (b) the skin-friction coefficient. Two grids 
are shown: coarse grid ($\solid$)\,; fine grid ($\dashed$). }
\end{figure*}

\begin{figure}[h!]
 \centering
\includegraphics[scale=0.5]{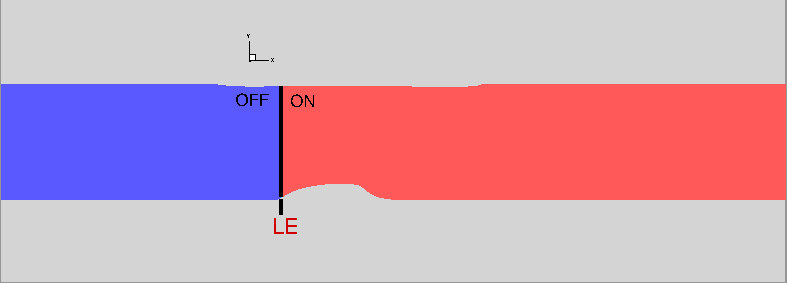}
\caption{  Separation zones showing  where the  filtering scheme is active (on) or not  (off).}
 \label{fig:ZONE_FILTERING}
\end{figure}

\begin{figure*}[h!]
\flushleft
\subfloat[]
{\label{fig:cp_zf}
\includegraphics[scale=0.13]{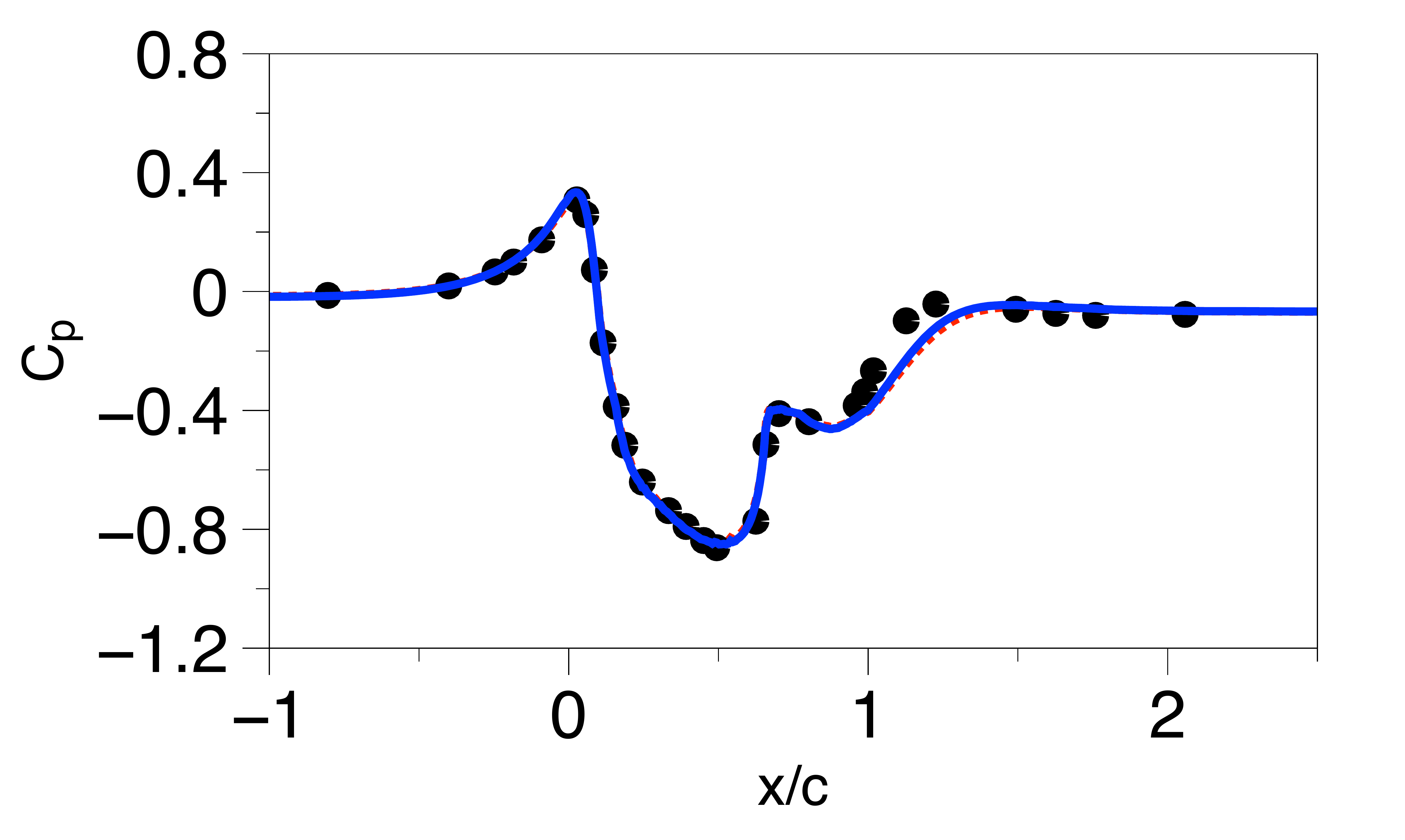}} 
\subfloat[]
{\label{fig:cf_zf}
\includegraphics[scale=0.13]{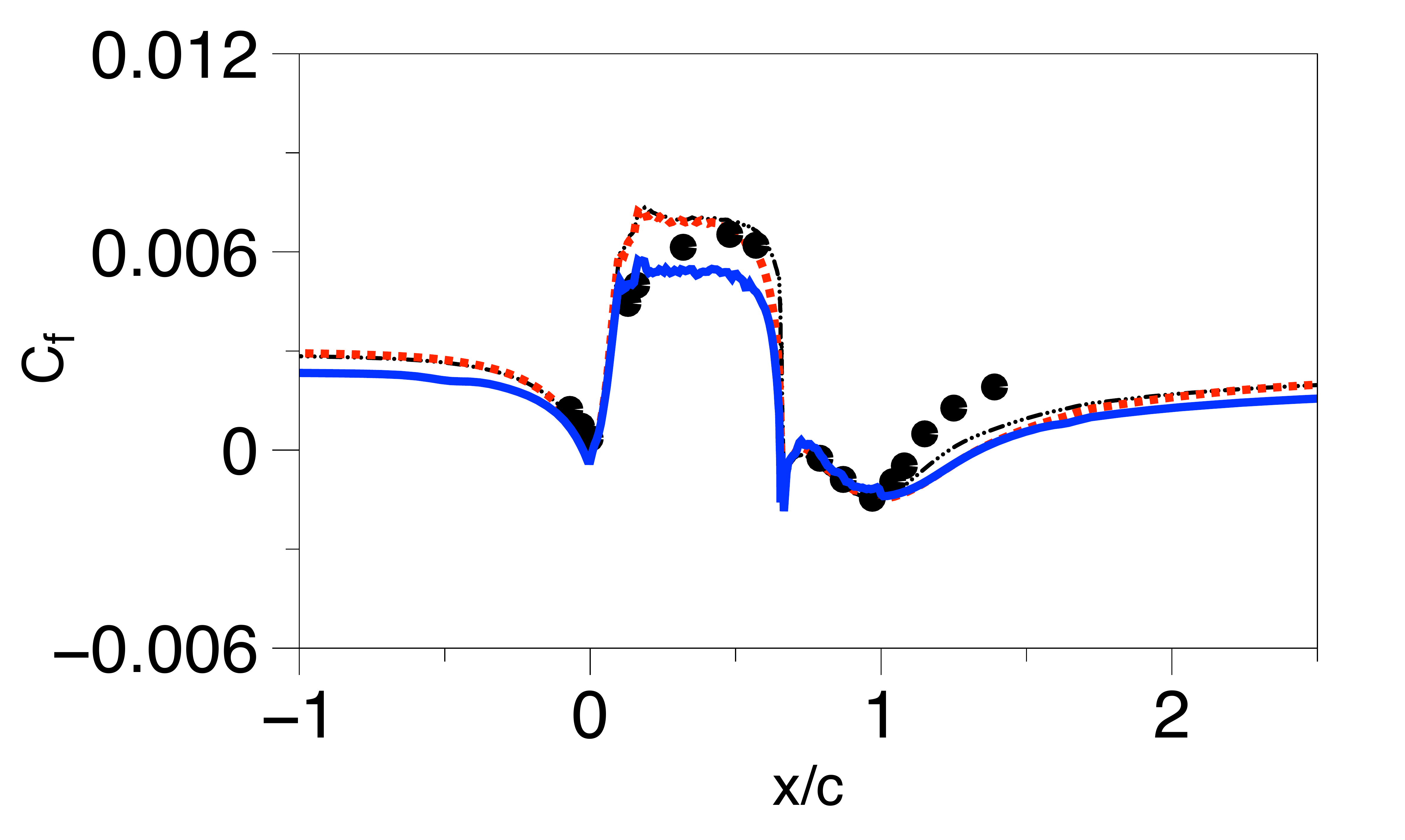}} 
\caption{ Effect of zonal filtering on  (a) the wall static-pressure coefficient and (b) 
the wall skin-friction coefficient for  the no-flow control case. SA ($\dashed$)  and 
ASBM-SA ($\chndotdotdot$) are compared  to the ASBM-SA predictions   when filtering is active 
in the entire domain ($\solid$) and the experimenal values (symbols) of Greenblatt et al. 
\cite{Greenblatt2004}. } 
 \label{fig:cfcp_zf}
\end{figure*}

\begin{figure}[h!]
 \centering
\includegraphics[scale=0.2]{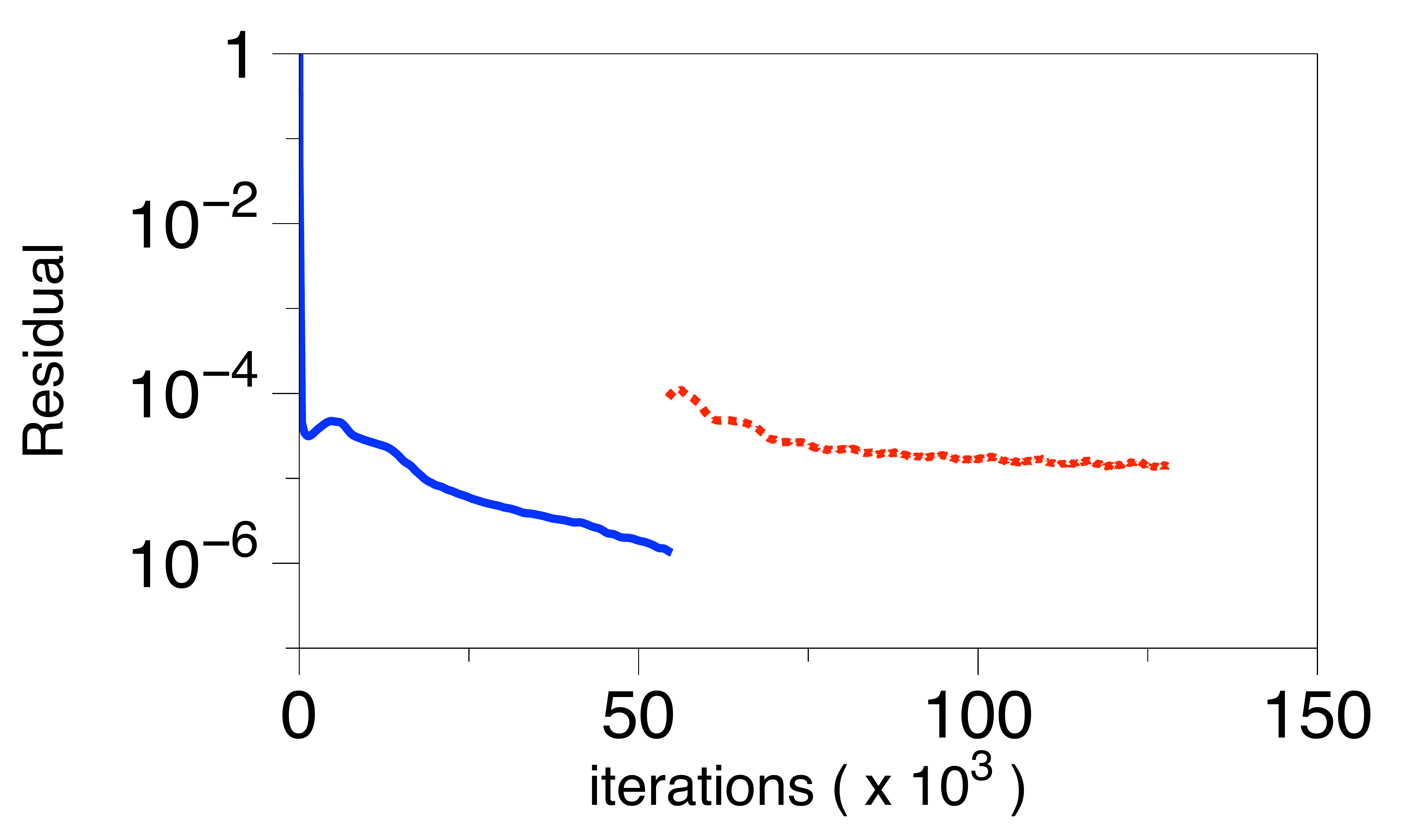}
\caption{Time history of the streamwise mean velocity residual for the uncontrolled case. 
The sudden jump in the residual levels indicates the point where the ASBM coupling is switched on.}
 \label{fig:time_convergence}
\end{figure}

\begin{figure*}[h!]
\centering
\subfloat[]
{\label{fig:str_sa_cavity_off}
\includegraphics[scale=0.25]{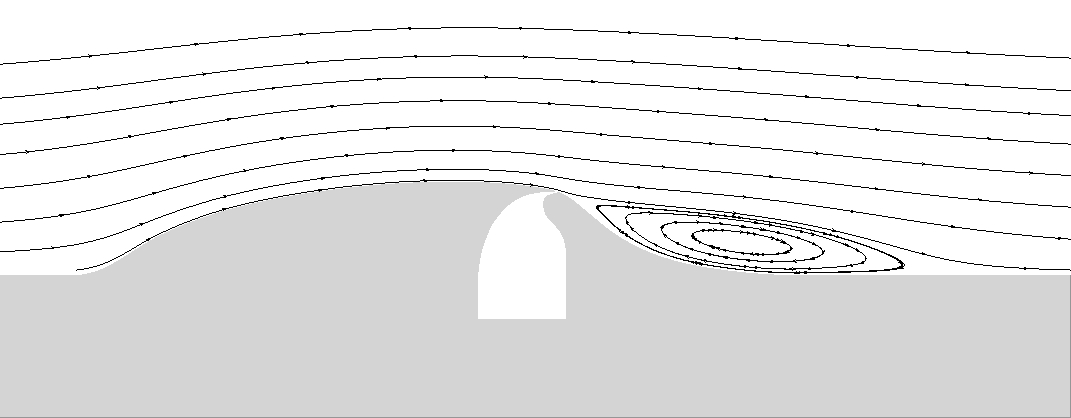}} \\
\subfloat[]
{\label{fig:str_sa_nocavity_off}
\includegraphics[scale=0.25]{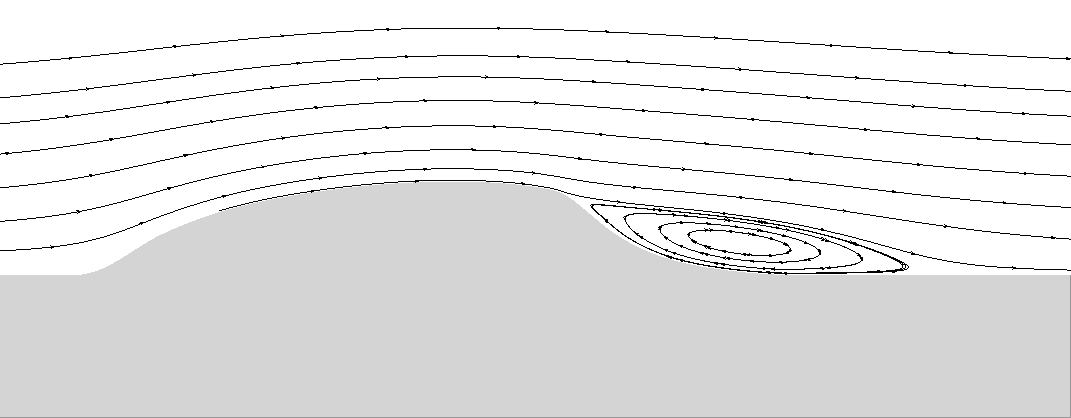}} 
\caption{SA model predictions for the streamlines of the mean flow  approaching  the hump 
(a) in the presence of the cavity and (b) in the absence of the cavity.} 
 \label{fig:streamlines_off}
\end{figure*}

\begin{figure}[h!]
 \centering
\includegraphics[scale=0.2]{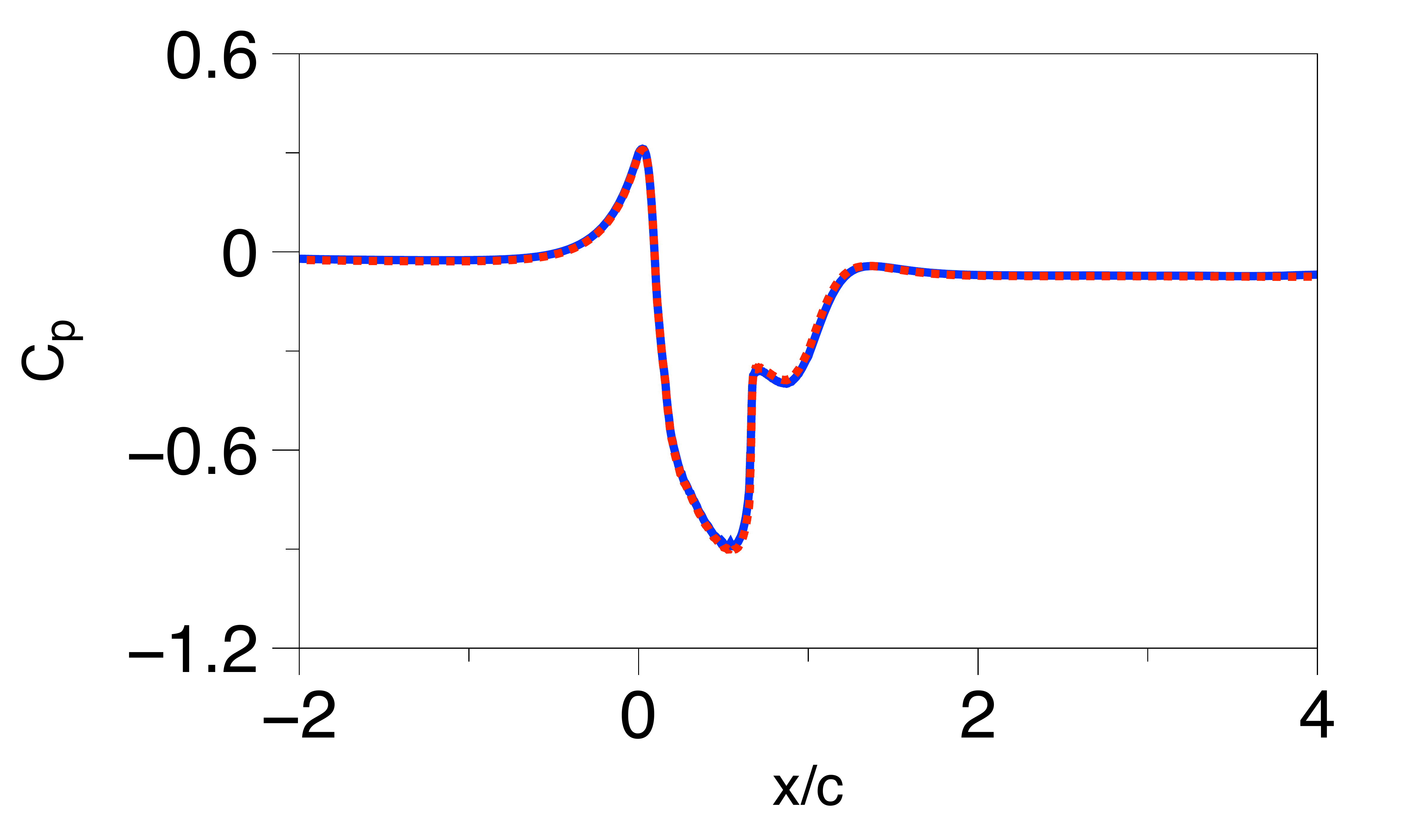}
\caption{ SA model predictions  in the presence ($\solid$) or absence ($\dashed$)  of the 
cavity for the wall-static pressure coefficient.}
 \label{fig:xc_cp_sa_cav_nocav}
\end{figure}

The case of flow over a hump having the shape of ``Modified Glauert" hill is considered 
next. This case was originally conceived for testing the ability of active control to  
reduce the size of the existing recirculation bubble. However, from a turbulence modeling 
perspective, even the uncontrolled case is interesting  due to the presence of strong  
separation, which proves to be challenging to turbulence engineering models. Thus, in our 
numerical experiments, we considered first the uncontrolled case of flow. 
Simulations have been performed  using SA and ASBM-SA models, which are compared to the 
experimental work of Greenblatt et al. \cite{Greenblatt2004}  At the inlet surface, profiles
for the variables are obtained from the SA solution for the turbulent boundary layer 
corresponding to $Re_{\delta} \approx 68200$ as described in the previous subsection. At 
the floor surface, as well as at the wall surfaces inside the cavity,  solid wall (no-slip) 
boundary conditions were applied. A penalty condition is imposed at the outlet surface to 
ensure that mass flow exits the domain properly, while slip conditions are used at the top 
surface and periodic conditions for the spanwise direction. Two grids were considered to 
conduct a grid-sensitivity analysis. The coarser grid contains approximately 
103,000 grid points, whereas the finer grid posses approximately 160,000 grid points. Mesh 
details are shown in Figure~\ref{fig:hill_meshing}, together with a zoomed view of the cavity region.
\FloatBarrier
Figures~\ref{fig:xc_cp_mesh_comp}-b  show SA model predictions for the wall-static pressure 
coefficient $c_{p}=(p-p_{0})/\frac{1}{2}\rho U^{2}_{\infty}$ and skin-friction coefficient 
$c_{f}=\tau_{w}/\frac{1}{2}\rho U^{2}_{\infty}$  respectively.  Based on these results, the 
coarser grid is shown to be  fine enough for the current case. 
\FloatBarrier
Due to the algebraic nature of the ASBM closure, numerical  difficulties are encountered for 
the cases where strongly separated flows are considered, as the present ones. During previous 
works, a filtering scheme was applied in order to smoothen the profiles, improving that 
way the numerical stability of the solution. However, use of this scheme in the current case 
led to a mismatch between SA and ASBM-SA predictions for the skin-friction coefficients in 
regions where a good agreement was expected, such as upstream of the leading edge of the hill 
and downstream of the recirculation region. As a result, we separated the domain 
into two zones, one prior the leading edge where the filtering scheme is not active, and the 
region downstream the leading edge where the scheme is switched on (Figure~\ref{fig:ZONE_FILTERING}).
Figure~\ref{fig:cfcp_zf}  shows a comparison between SA  and ASBM-SA predictions using both 
approaches for the filtering scheme, revealing the significant role that the filtering details 
play on the skin-friction distribution all along the bottom surface. In contrast, the 
pressure coefficient remains unaffected by the choice of filtering scheme.
\FloatBarrier
Figure~\ref{fig:time_convergence} displays the evolution of the maximum mean streamwise 
velocity residual. The residual is divided by its initial value, denoted by subscript 0, 
and is defined by 
\begin{equation}
\text{ Residual }= \max \bigg[  \frac{ V \times \Delta U_{x}/\Delta t }{  (V \times 
\Delta U_{x}\Delta t)_{0}  }  \bigg]\,, \hspace{0.3 cm} \Delta U_{x} = U^{n+1}_{x}-U^{n}_{x}\,,
\end{equation} 
\noindent where $n$ refers to the $n$-th iteration, $\Delta t$ to the time step and 
$V$ to the volume of the corresponding cell. For both SA and ASBM-SA models, a drop of at 
least 5 orders of magnitude for the residuals compared to the initial field is achieved, 
which is believed to be sufficient to provide time-converged solutions.
\FloatBarrier
In order to both accelerate our simulations and overcome some stability issues related to  
the ASBM-SA computations inside the cavity in the case of no-suction, additional computations 
using similar meshes but without the presence of the cavity were conducted. For these cases, we 
considered a solid-wall condition along the slot exit. Figures~\ref{fig:str_sa_cavity_off}-b 
show SA predictions for the mean velocity streamlines at the vicinity of the hump in the 
presence and absence of the cavity respectively, demonstrating the trivial discrepancies 
between the two approaches. Figure~\ref{fig:xc_cp_sa_cav_nocav} shows the corresponding 
comparison for the wall-static pressure coefficient, which again reveals the negligible 
effect on the results of the cavity absence when the flow control is inactive (no suction).

\FloatBarrier

\begin{figure}[h!]
 \centering
\includegraphics[scale=0.20]{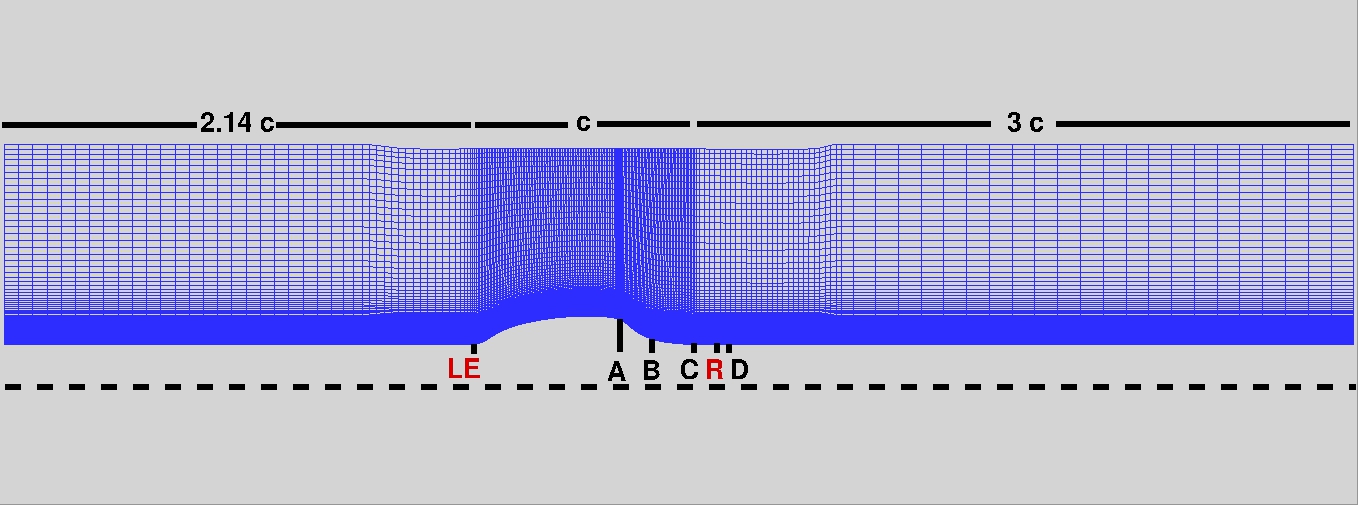}
\caption{ Geometrical and mesh details  in the absence of cavity.  Data is extracted at four 
stations, denoted as A,\,B,\,C and D. The leading-edge point (LE) and the re-attachment 
point (R) are also shown.} 
 \label{fig:stations_view}
\end{figure}
\begin{figure*}[h!]
\centering
\subfloat[]
{\label{fig:cp_nf}
\includegraphics[scale=0.13]{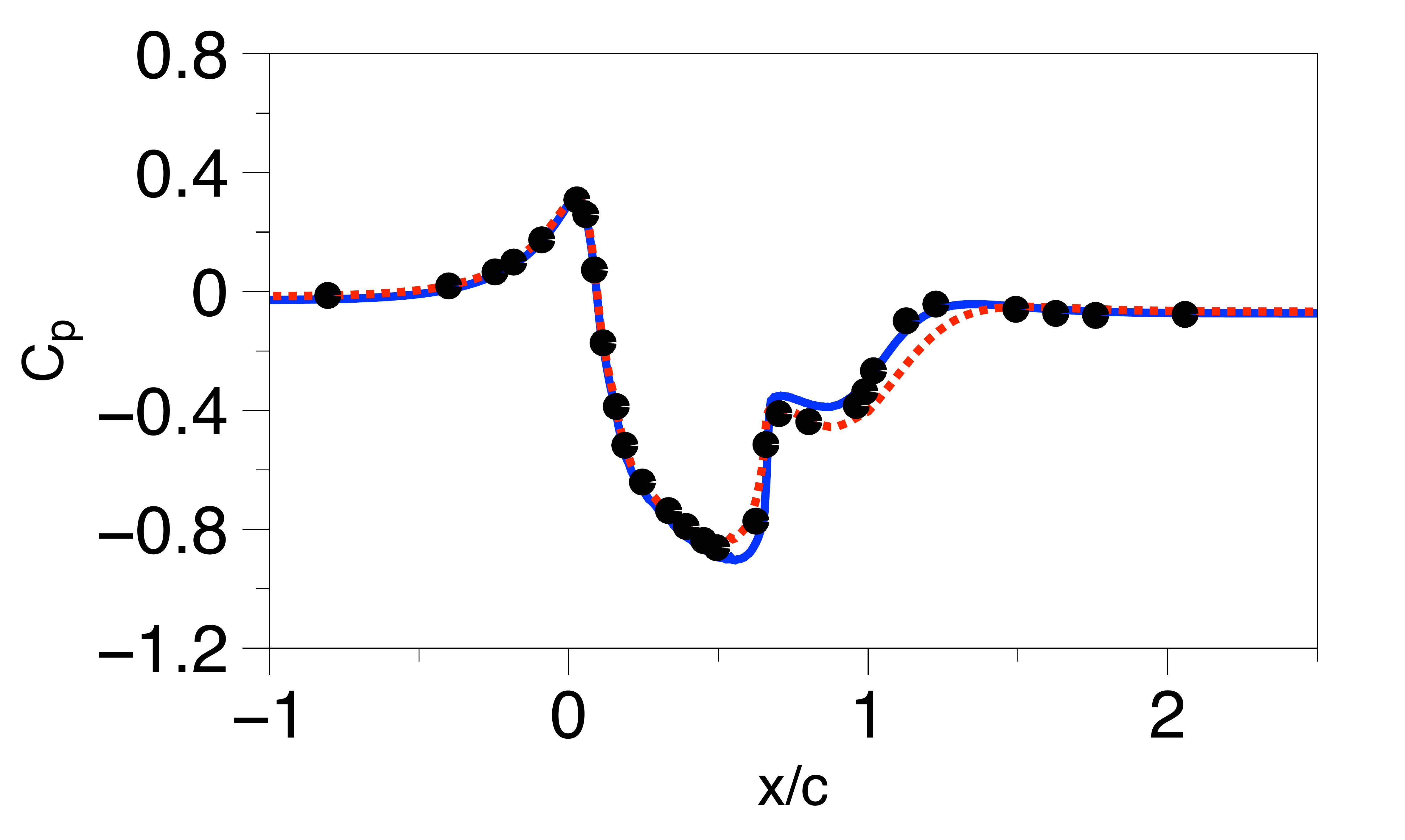}} 
\subfloat[]
{\label{fig:cf_nf}
\includegraphics[scale=0.13]{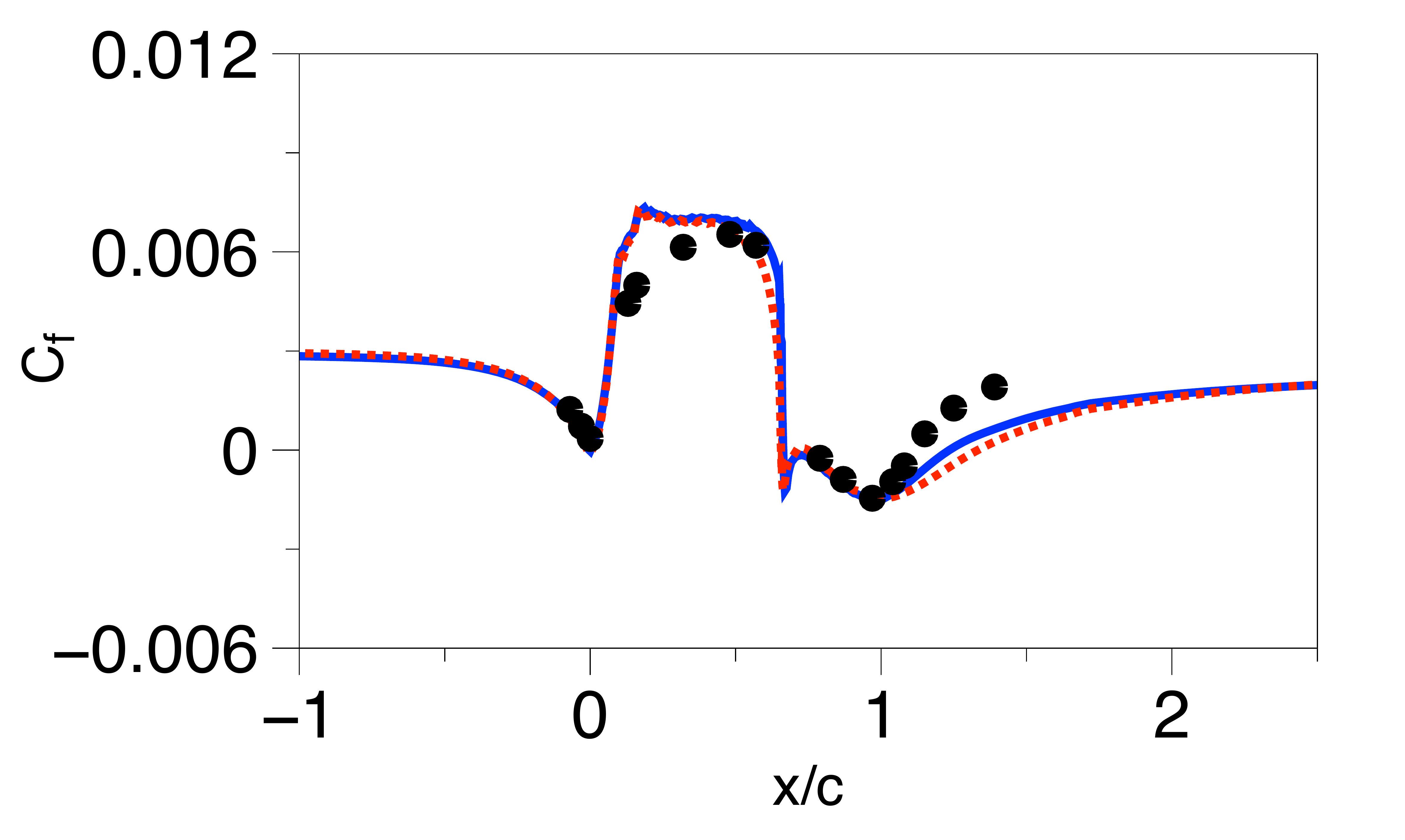}} 
\caption{SA ($\solid$)  and ASBM-SA ($\dashed$) model predictions for  the no-flow control 
case for (a) the wall static-pressure coefficient and (b) the wall skin-friction coefficient. 
Comparison is made to experimental values (symbols) of Greenblatt et al. \cite{Greenblatt2004}.} 
 \label{fig:coefs_hill}
\end{figure*}
\begin{figure*}[h!]
\centering
\subfloat[station A, $x/c=0.66$]
{\label{fig:umean_nf_066}
\includegraphics[scale=0.13]{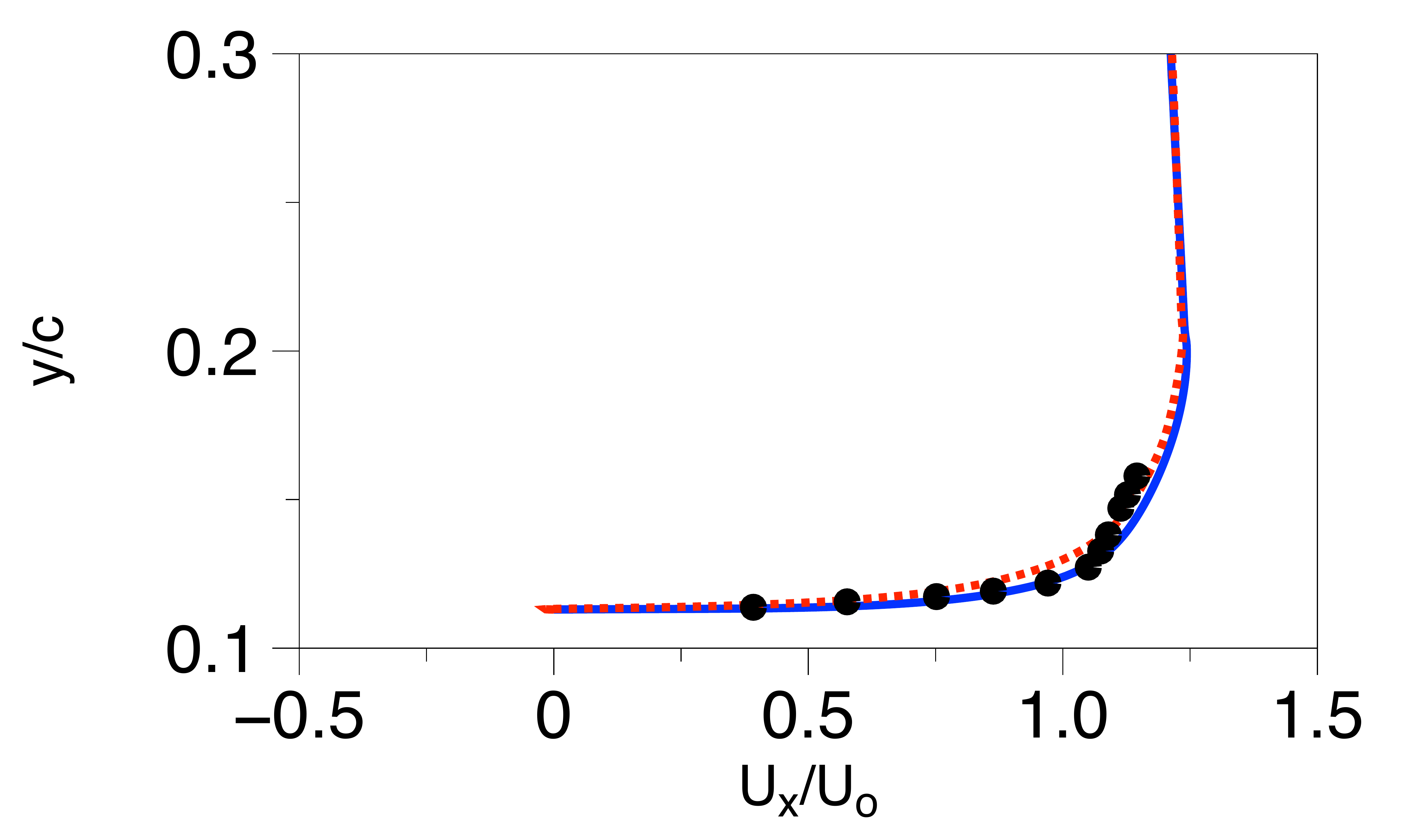}} 
\subfloat[station B, $x/c=0.8$]
{\label{fig:umean_nf_080}
\includegraphics[scale=0.13]{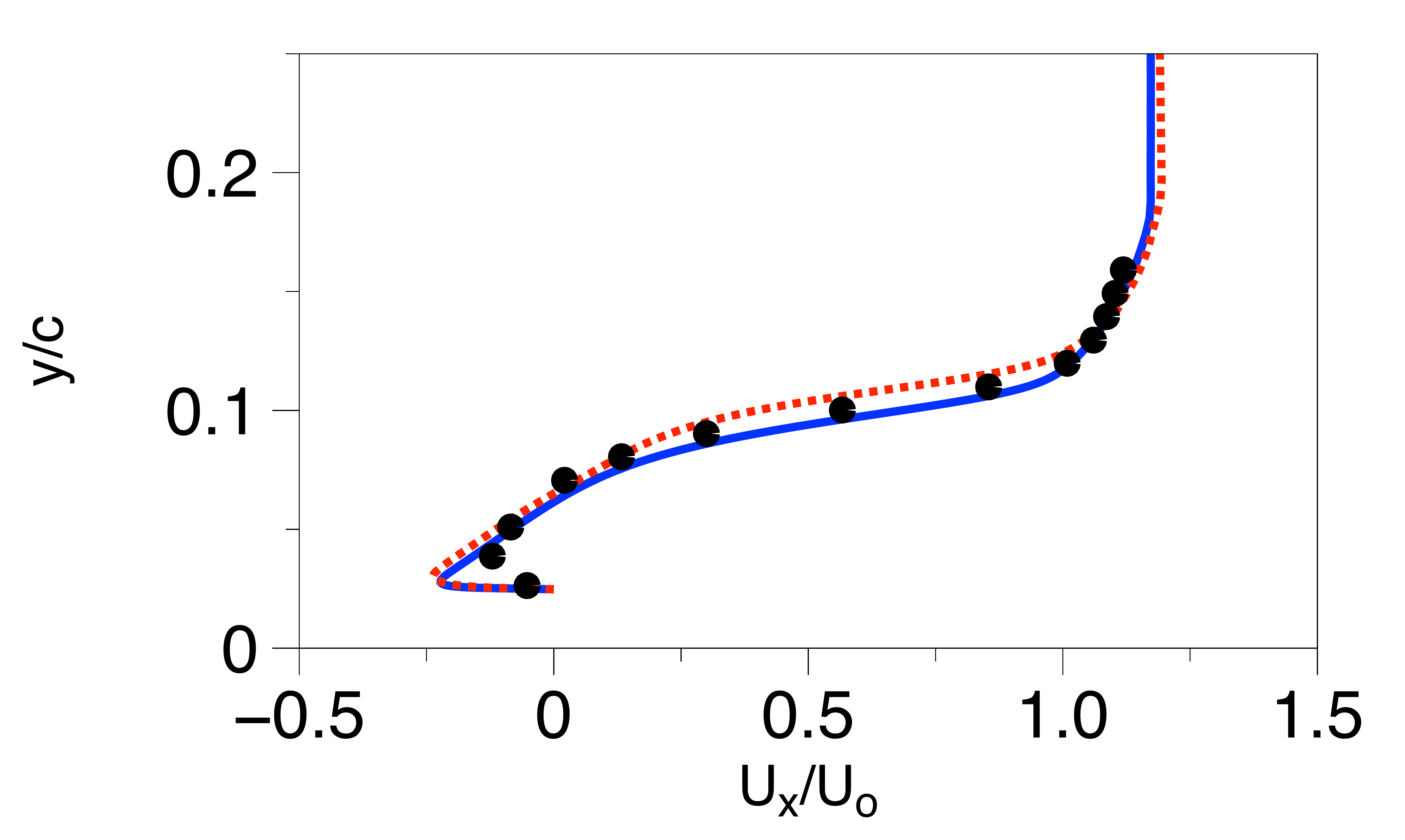}} \\
\subfloat[station C, $x/c=1.0$]
{\label{fig:umean_nf_10} 
\includegraphics[scale=0.13]{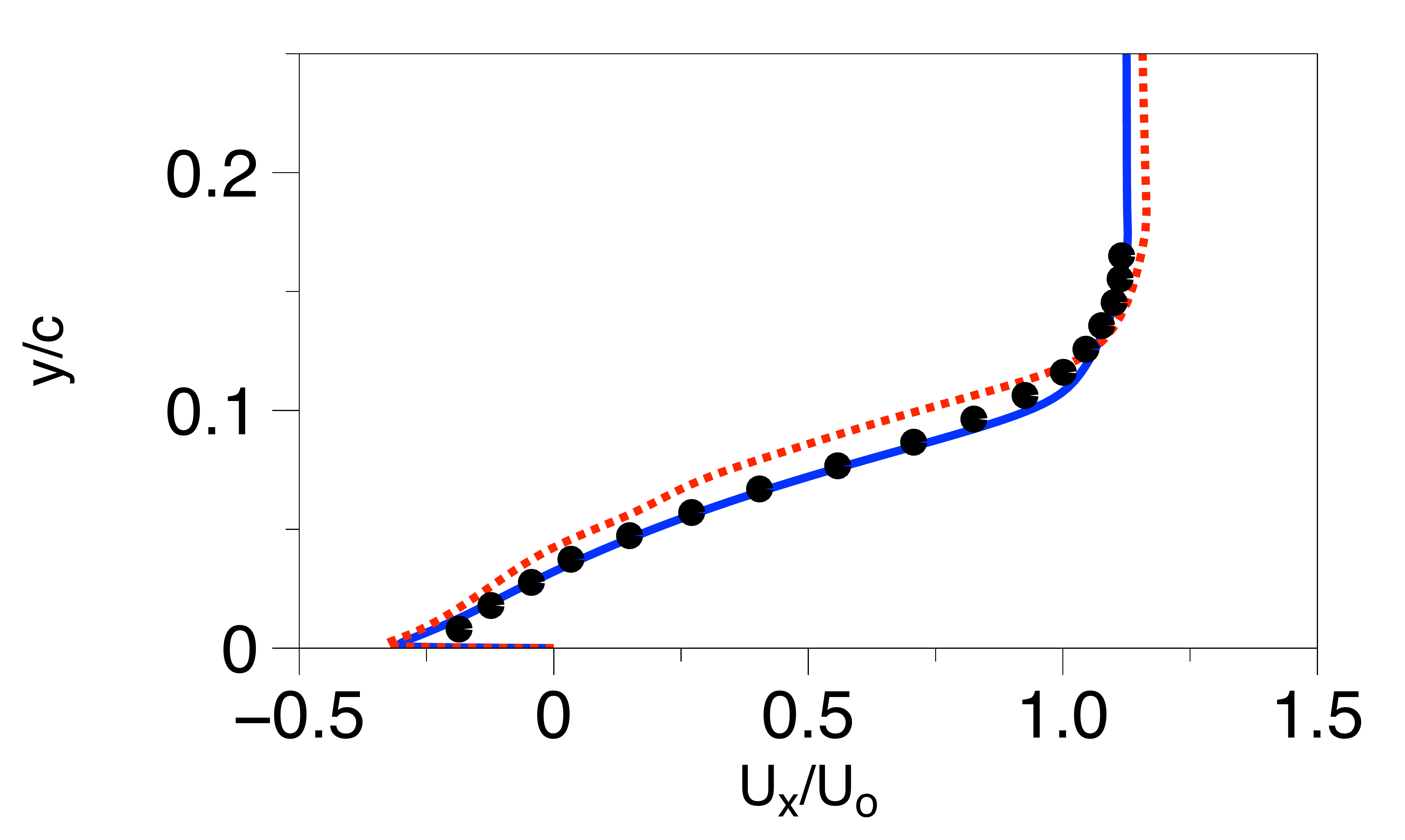}}
\subfloat[station D, $x/c=1.2$]
{\label{fig:umean_nf_12}
\includegraphics[scale=0.13]{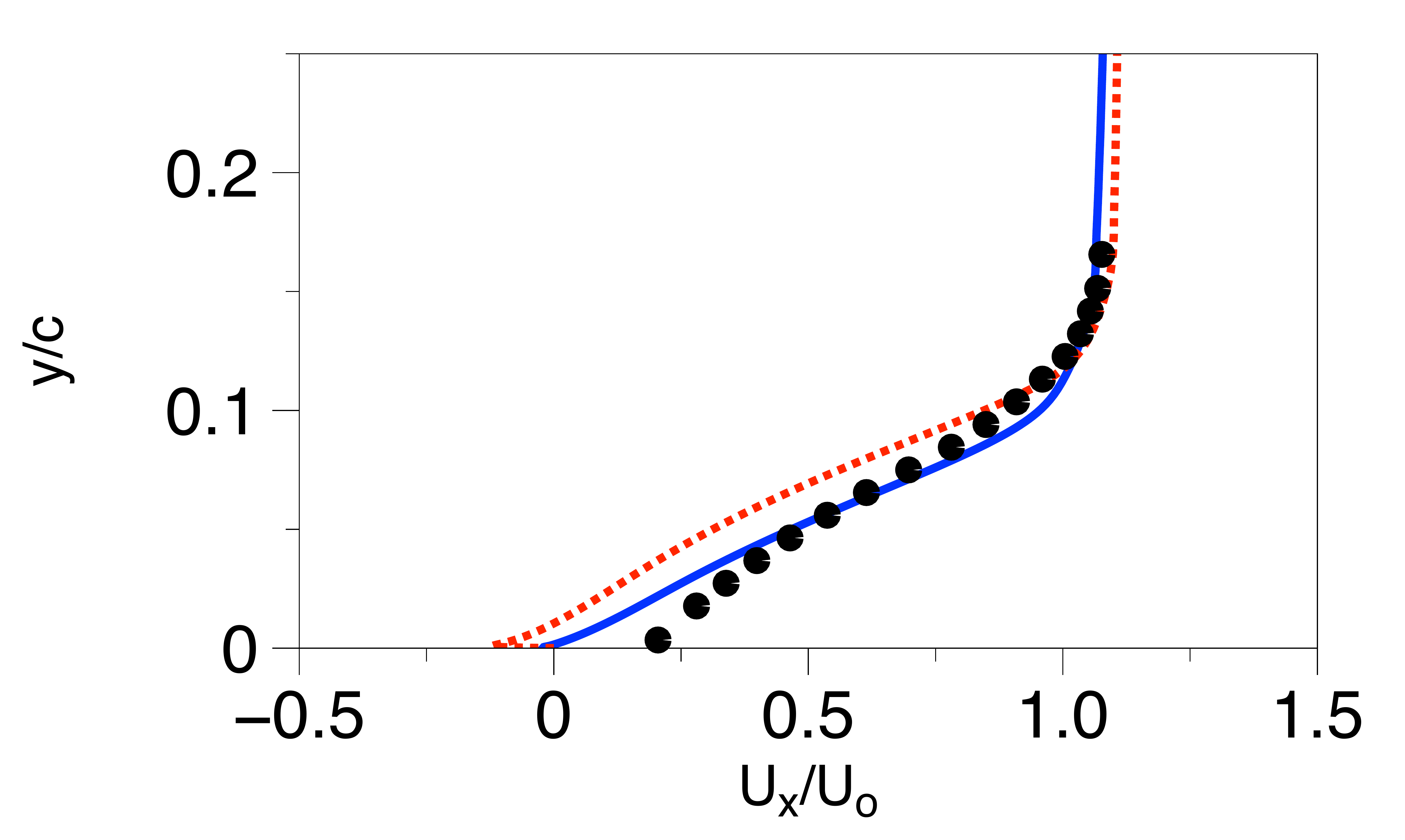}} 
\caption{Turbulent flow over the  ``Glauert-Goldschmied" 2D hill for the no-flow control case. 
Model predictions for the streamwise mean velocity $U_{x}$ at various $x$-stations for SA 
($\solid$) and ASBM-SA ($\dashed$)  closures. Comparison is made to experimental values of 
Greenblatt et al. \cite{Greenblatt2004}.} 
 \label{fig:Ux_hill}
\end{figure*}

In the following figures, all turbulent and mean quantities (except $c_{f}$, $c_{p}$) are 
normalized based on the chord length $c$ and the reference inlet freestream velocity $U_{\infty}$. 
Setting the leading edge of the hump as the origin of the streamwise distance ($x/c=0$), 
profiles are extracted  at three different stations inside the recirculation bubble as 
measured by the experiments  ($x/c=0.66, \ 0.8, \ 1.0$) and one station at the recovery 
region ($x/c=1.2$), denoted as stations $A,\ B,\ C,\ D$ respectively 
(Figure~\ref{fig:stations_view}).   
\FloatBarrier
Figure~\ref{fig:coefs_hill} shows the predictions of the ASBM-SA and SA  closures for the 
variation of the pressure ($c_p$) and skin friction ($c_f$) coefficients along the wall 
surface. Comparison is made to experimental measurements.  The ASBM-SA model captures accurately 
the peak magnitude of  the pressure coefficient around $x/c \approx 0.57$. 
For $x/c$ ranging from -1 to 1.1, that is from the inlet up to about the re-attachment point, the  ASBM-SA 
provides slightly improved $c_p$ predictions when compared to SA. Right after re-attachment though, 
the ASBM-SA predicts a slightly delayed recovery of $c_p$ as compared to the experiments 
(and the SA closure).  The two models produce comparable agreement with the experiments for 
the skin-friction coefficient.  ASBM-SA provides an improvement right after the upstream 
edge of the hill ($x/c \approx 0.2$), while  SA  predicts correctly the magnitude near the 
sharp geometry change that occurs around $x/c \approx 0.65$.  According to Table~\ref{Table:Cases}, 
both models overpredict the recirculation bubble, with ASBM having the tendency to delay the 
re-attachment of the flow further downstream, an observation which is in agreement  with 
previous cases  in which separated flows over 2D hills were considered, such as the ``Witch 
of Agnesi" hump, as described in detail in \cite{Panagiotou2015}.
Figure~\ref{fig:Ux_hill} shows results for the streamwise  mean velocity $U_{x}$ at the four 
stations. As shown in Figures~\ref{fig:umean_nf_066}-b, at the first two stations ASBM-SA 
provides slightly improved predictions relative to the SA model, while SA is in better 
agreement with the experimental data  at the next two stations, mostly due to the  greater 
delay of re-attachment in the ASBM-SA predictions.

\begin{figure*}[h!]
\centering
\subfloat[station A, $x/c=0.66$]
{\label{fig:uu_nf_066}
\includegraphics[scale=0.13]{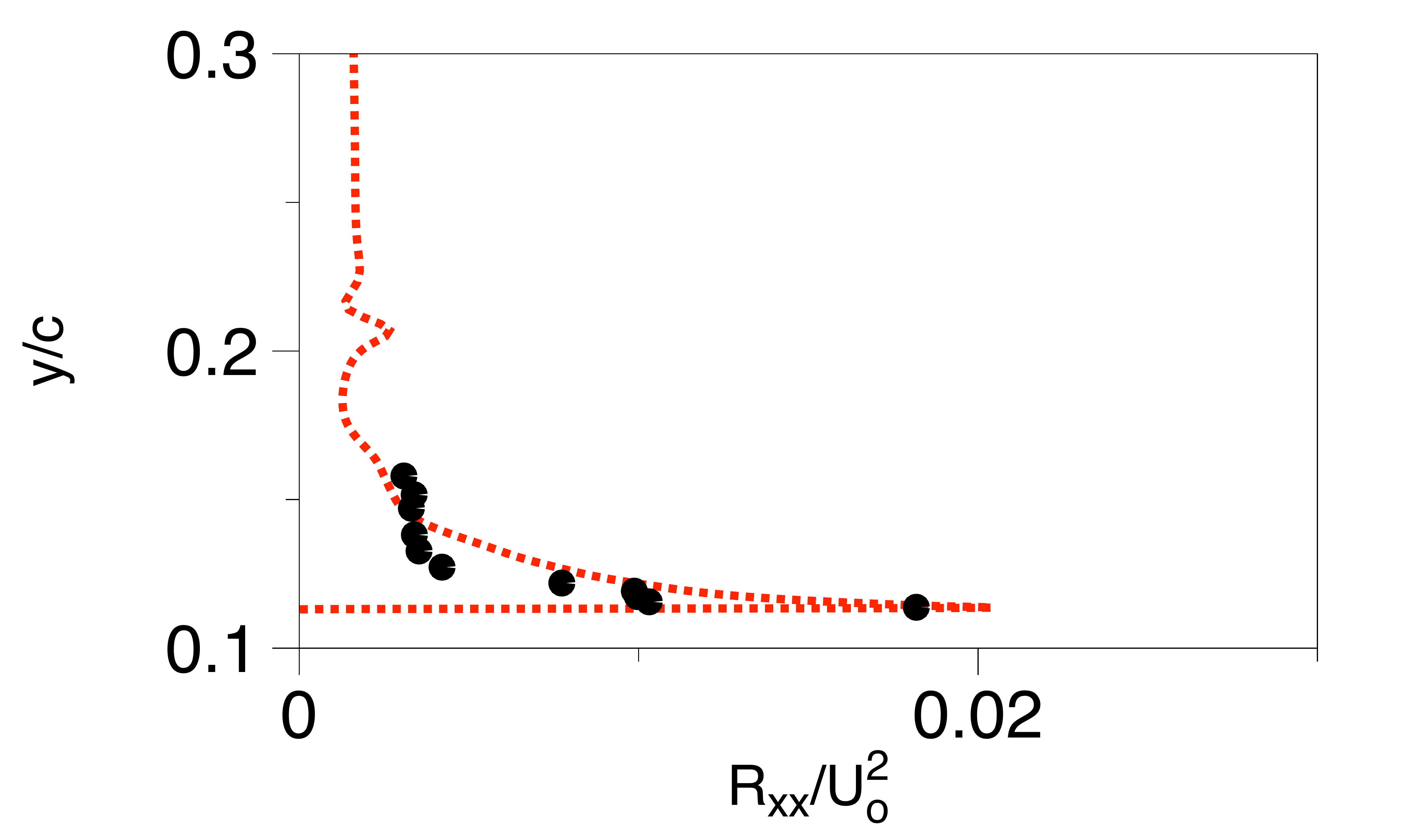}} 
\subfloat[station B, $x/c=0.8$]
{\label{fig:uu_nf_080}
\includegraphics[scale=0.13]{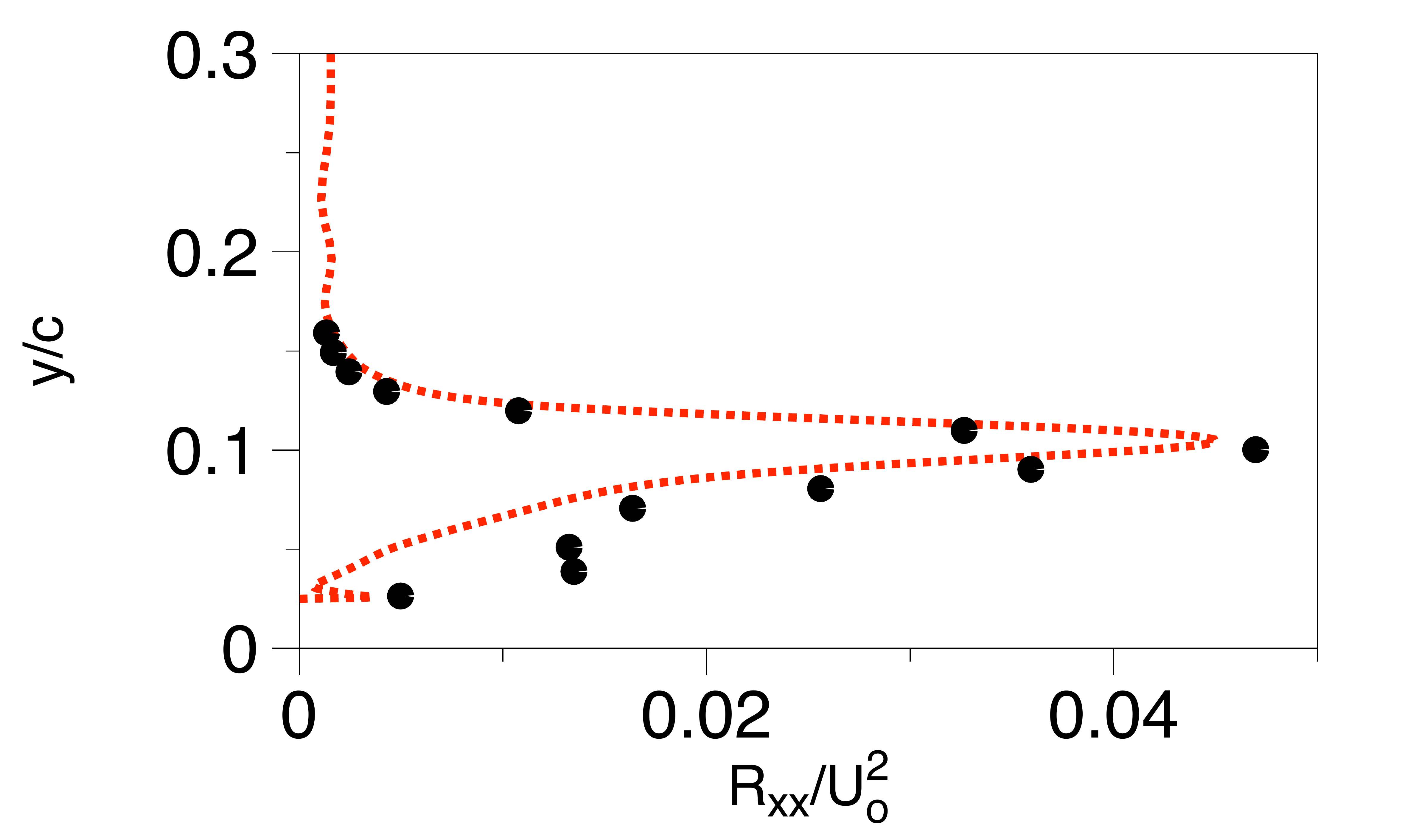}} \\, 
\subfloat[station C, $x/c=1.0$]
{\label{fig:uu_nf_10} 
\includegraphics[scale=0.13]{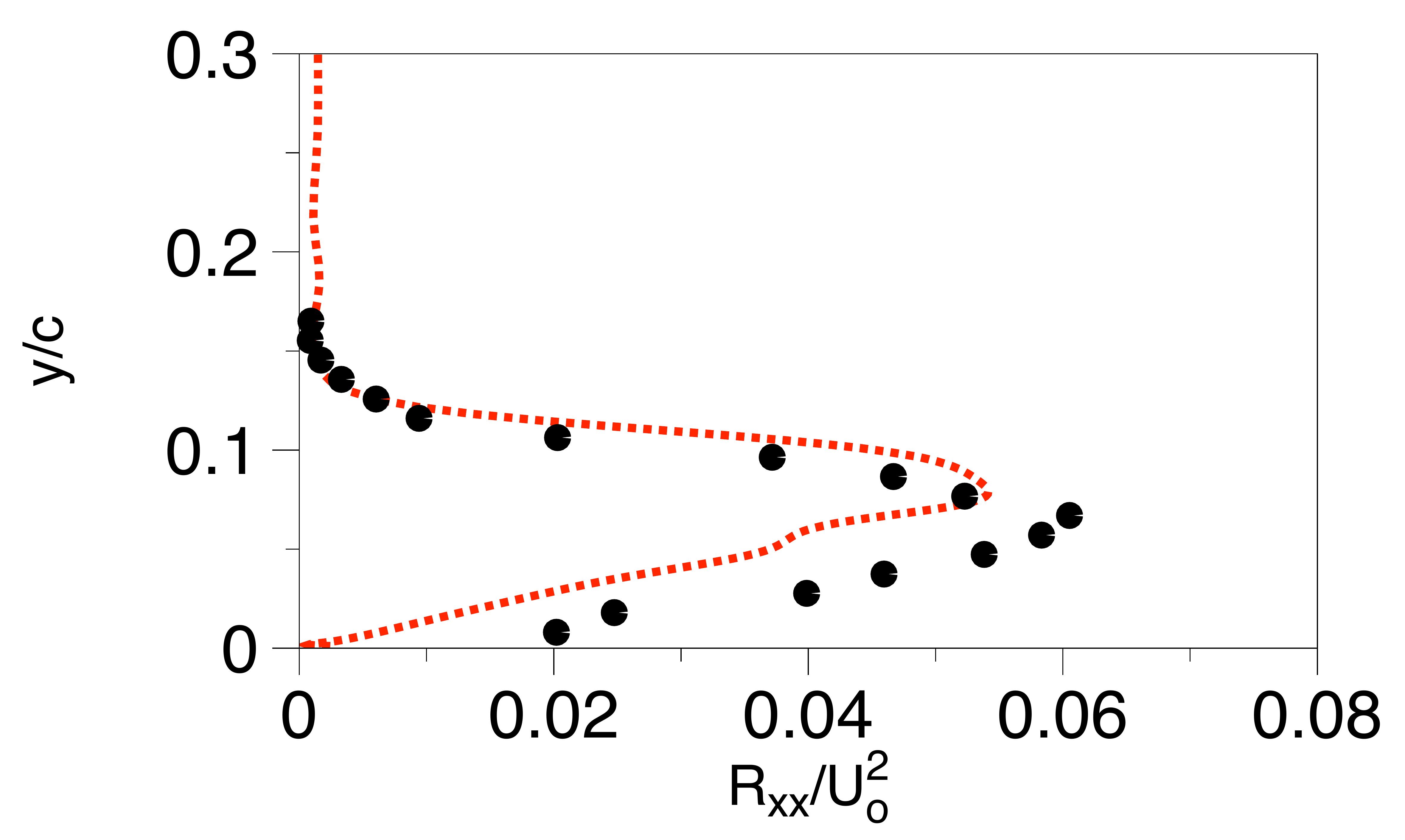}}
\subfloat[station D, $x/c=1.2$]
{\label{fig:uu_nf_12}
\includegraphics[scale=0.13]{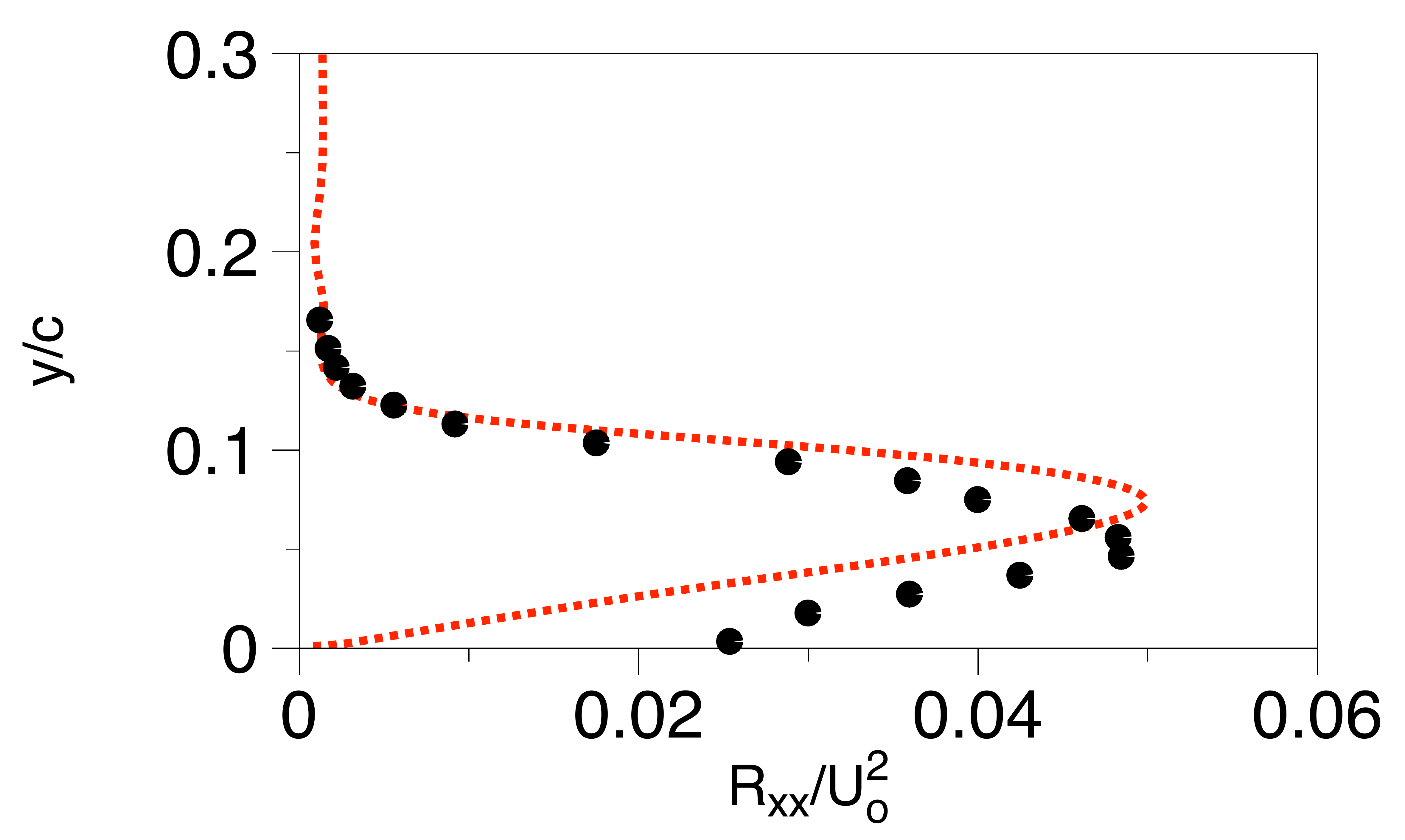}} 
\caption{Turbulent flow over the ``Glauert-Goldschmied" 2D hill for the no-flow control case. 
ASBM-SA model predictions (lines) for the streamwise Reynolds stress component $R_{xx}$ at 
various $x$-stations are shown. Comparison is made to experimental values (symbols) of
Greenblatt et al. \cite{Greenblatt2004}. } 
 \label{fig:rxx_hill}
\end{figure*}

\begin{figure*}[h!]
\flushleft
\subfloat[station A, $x/c=0.66$]
{\label{fig:vv_nf_066}
\includegraphics[scale=0.13]{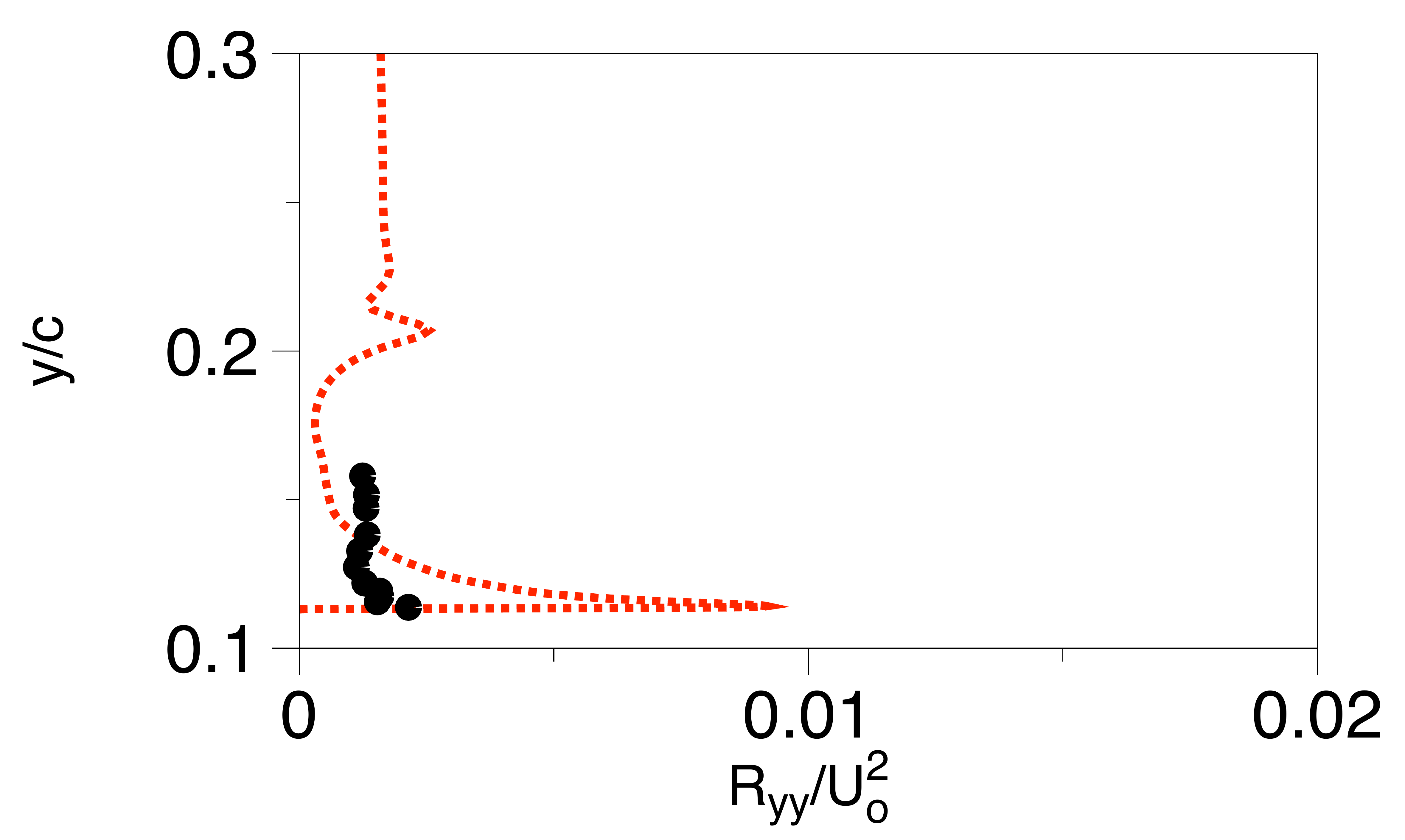}} 
\subfloat[station B, $x/c=0.8$]
{\label{fig:vv_nf_080}
\includegraphics[scale=0.13]{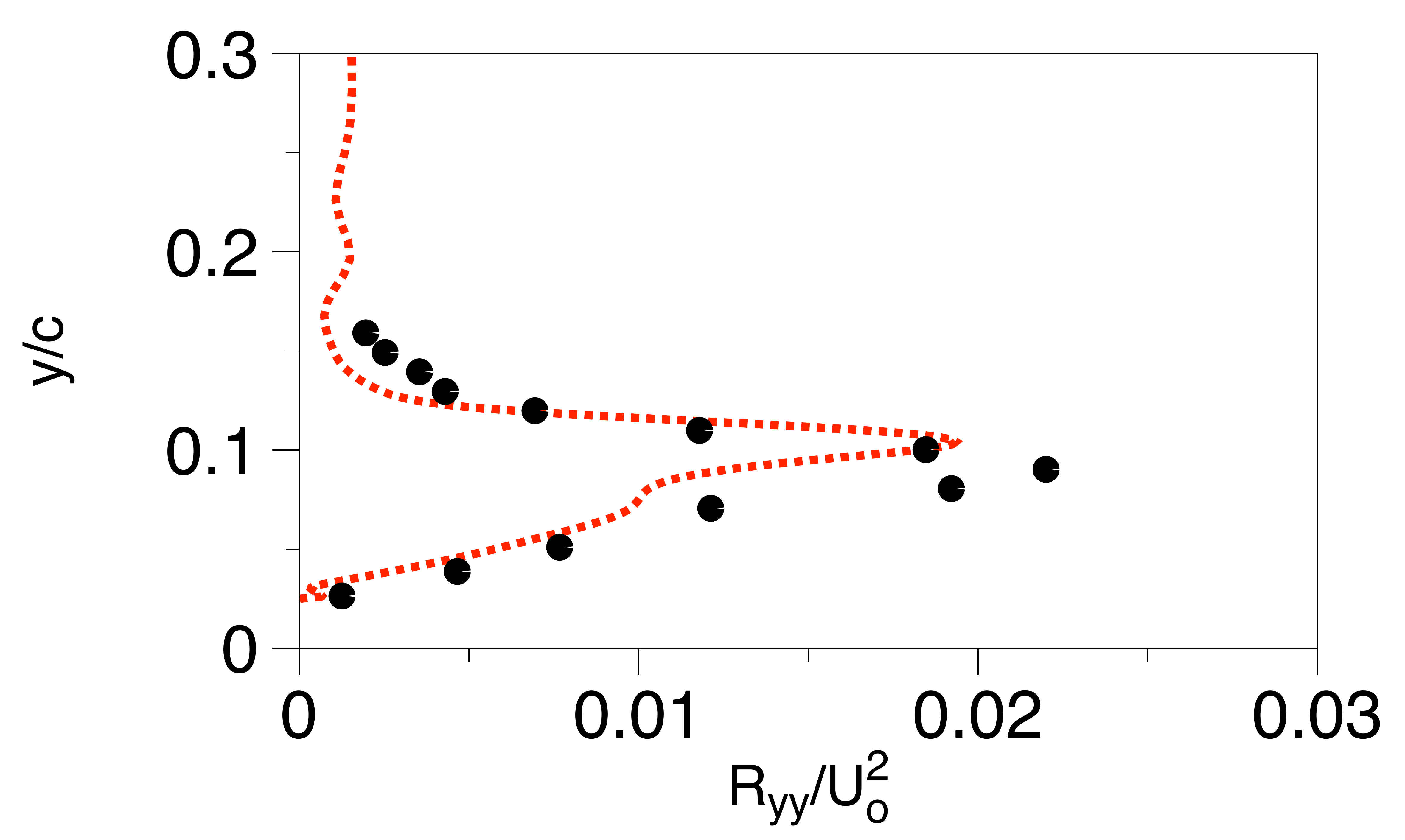}} \\
\subfloat[station C, $x/c=1.0$]
{\label{fig:vv_nf_10} 
\includegraphics[scale=0.13]{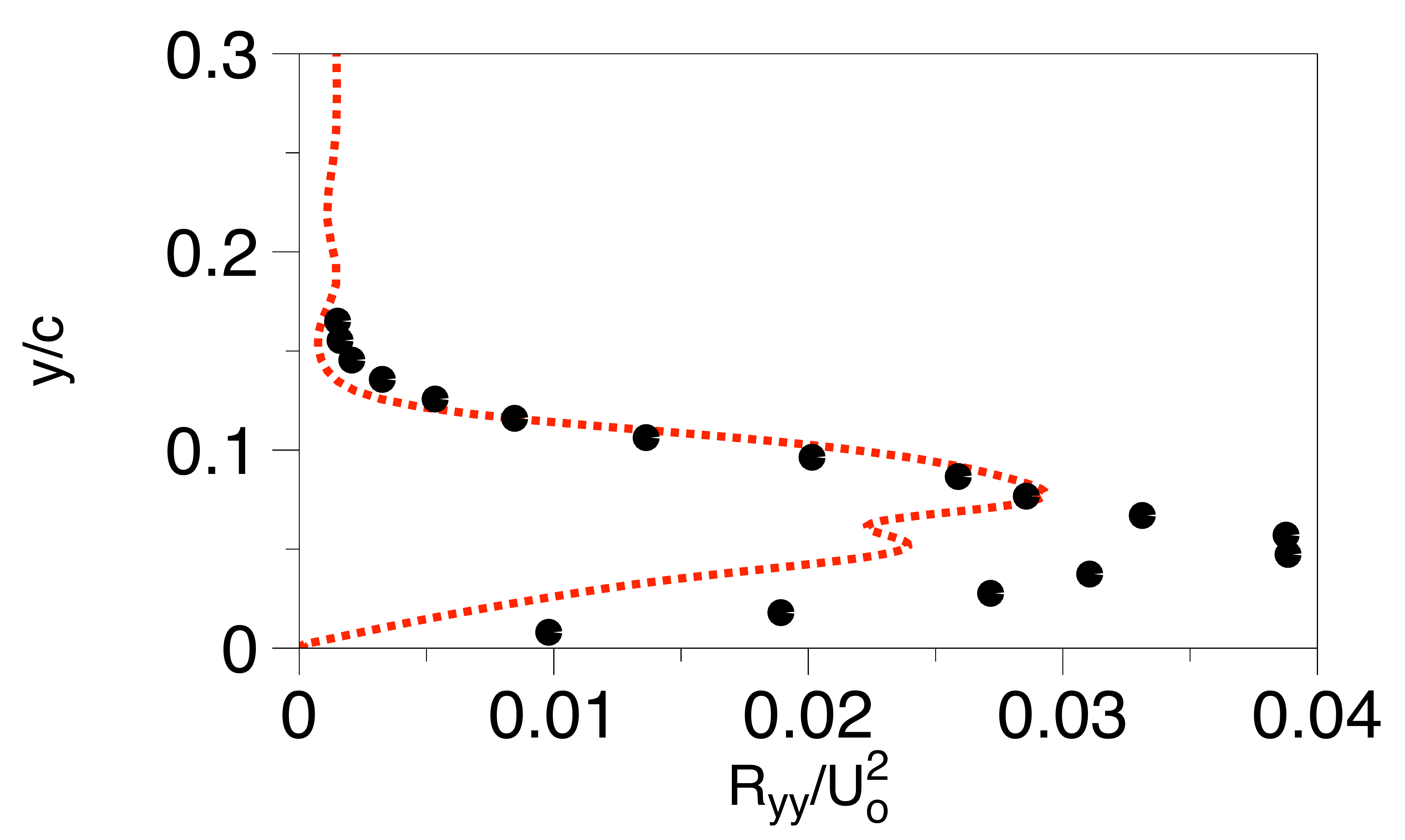}}
\subfloat[station D, $x/c=1.2$]
{\label{fig:vv_nf_12}
\includegraphics[scale=0.13]{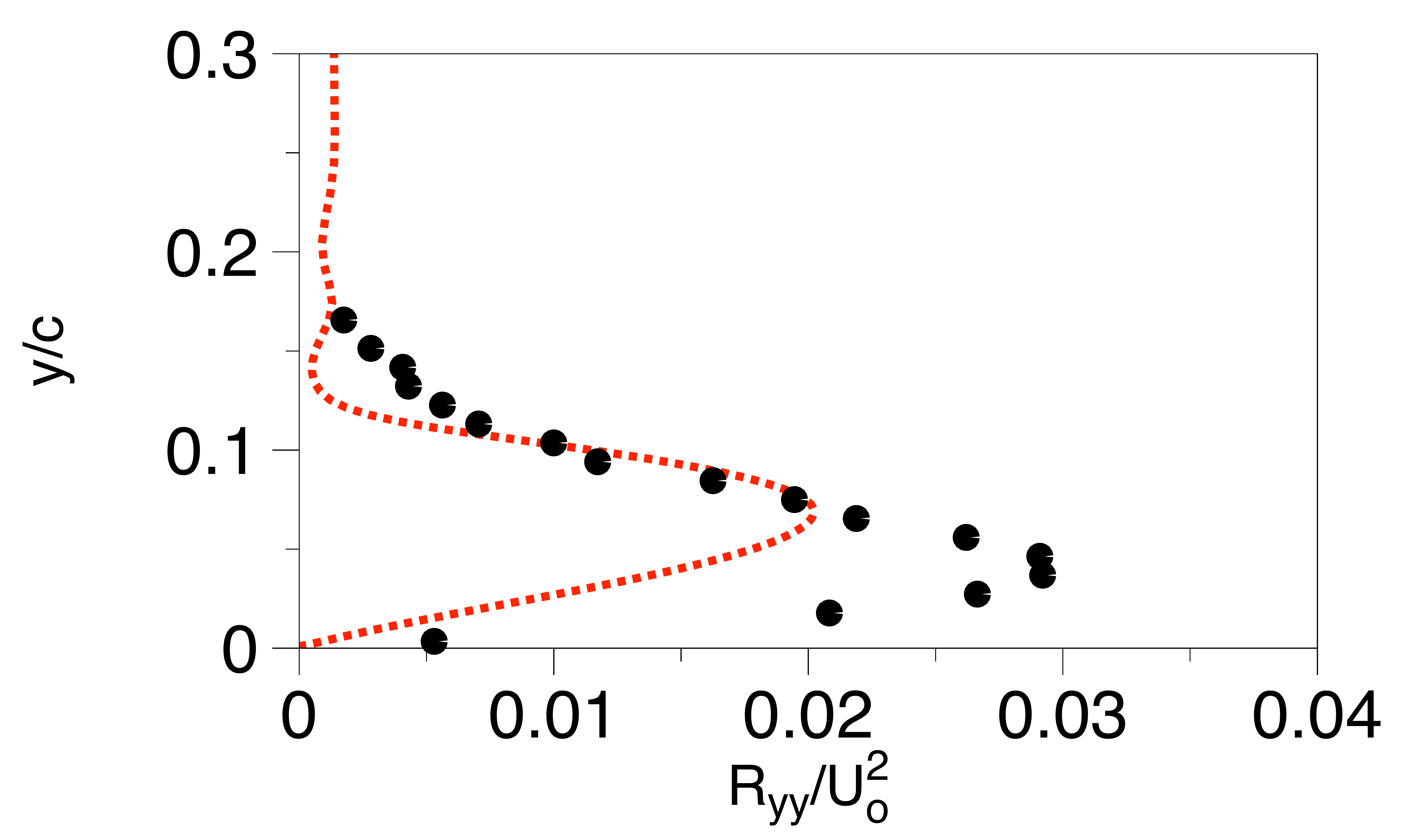}} 
\caption{Turbulent flow over the  ``Glauert-Goldschmied" 2D hill for the no-flow control 
case. ASBM-SA model predictions (lines) for the transverse Reynolds stress component  $R_{yy}$ 
at various $x$-stations are shown. Comparison is made to experimental values (symbols) of
Greenblatt et al. \cite{Greenblatt2004}.} 
 \label{fig:ryy_hill}
\end{figure*}

\begin{figure*}[h!]
\flushleft
\subfloat[station A, $x/c=0.66$]
{\label{fig:uv_nf_066}
\includegraphics[scale=0.13]{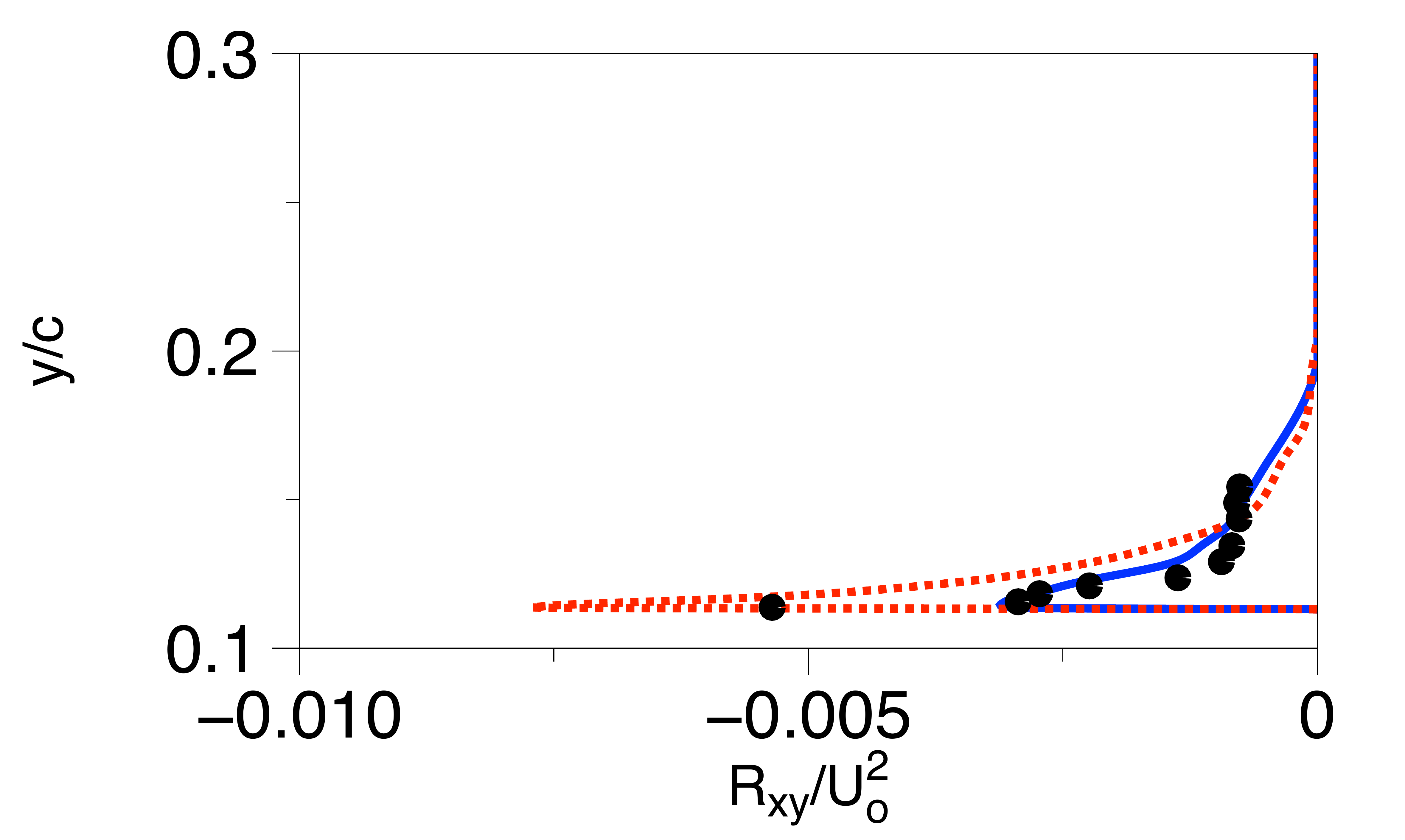}} 
\subfloat[station B, $x/c=0.8$]
{\label{fig:uv_nf_080}
\includegraphics[scale=0.13]{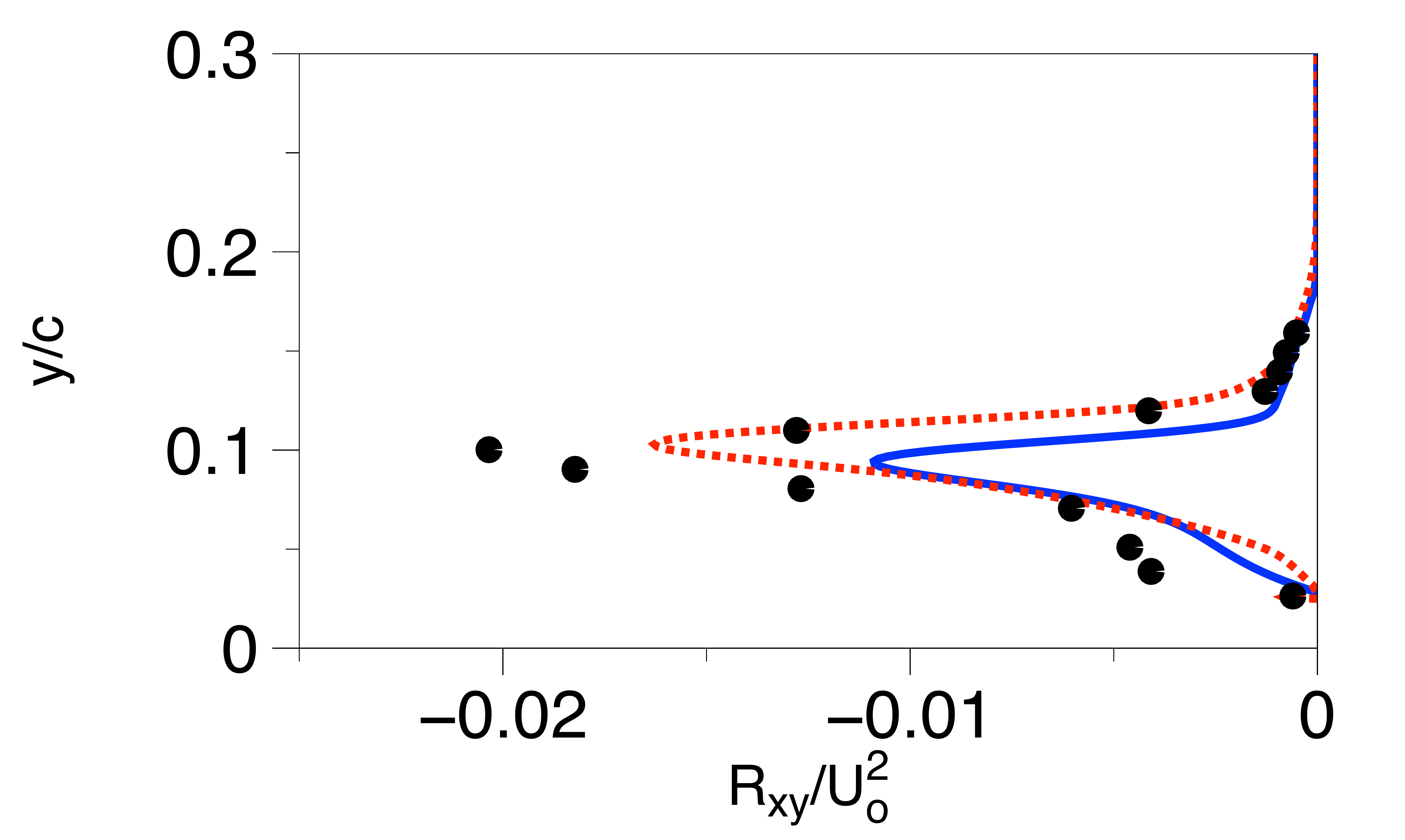}} \\
\subfloat[station C, $x/c=1.0$]
{\label{fig:uv_nf_10} 
\includegraphics[scale=0.13]{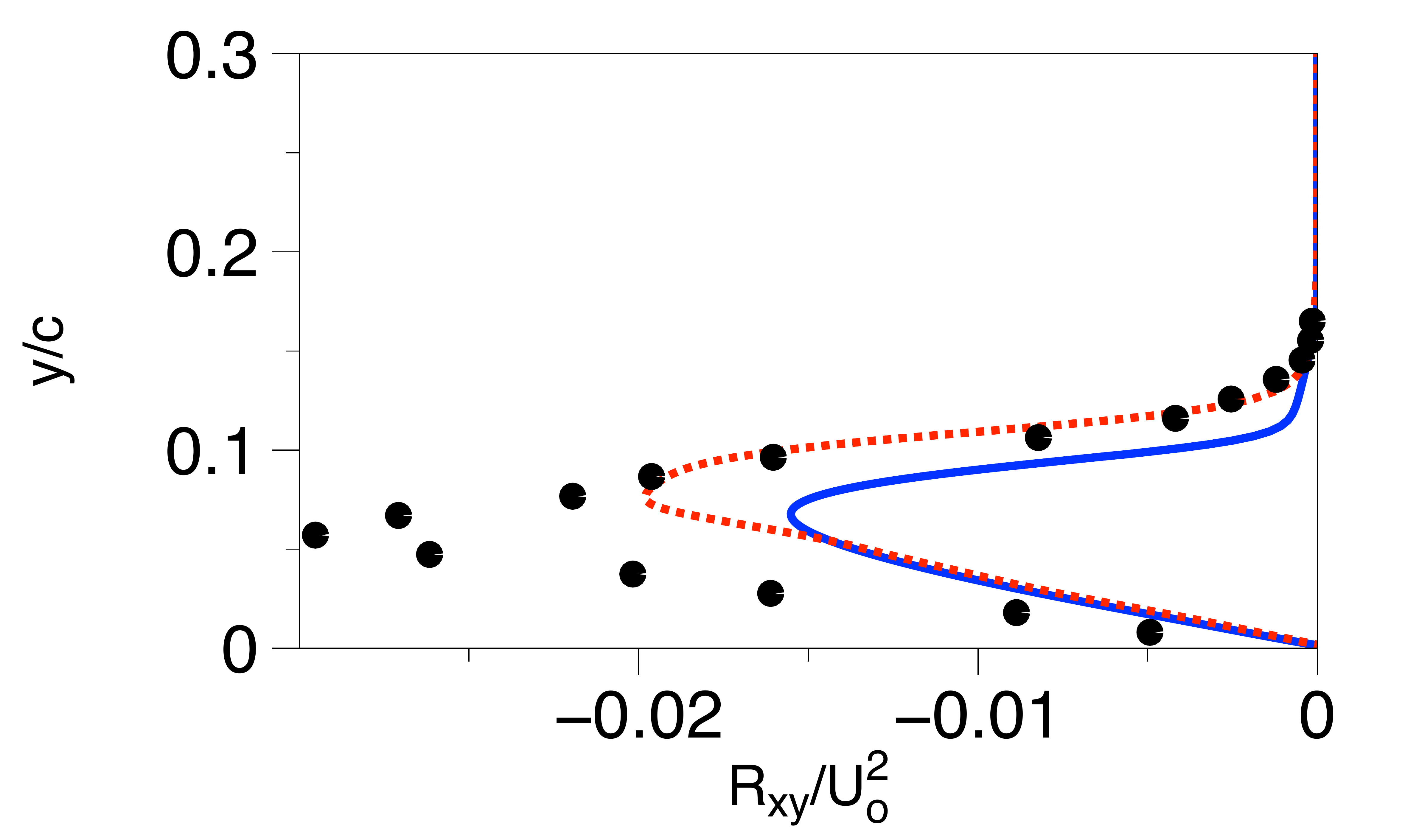}}
\subfloat[station D, $x/c=1.2$]
{\label{fig:uv_nf_12}
\includegraphics[scale=0.13]{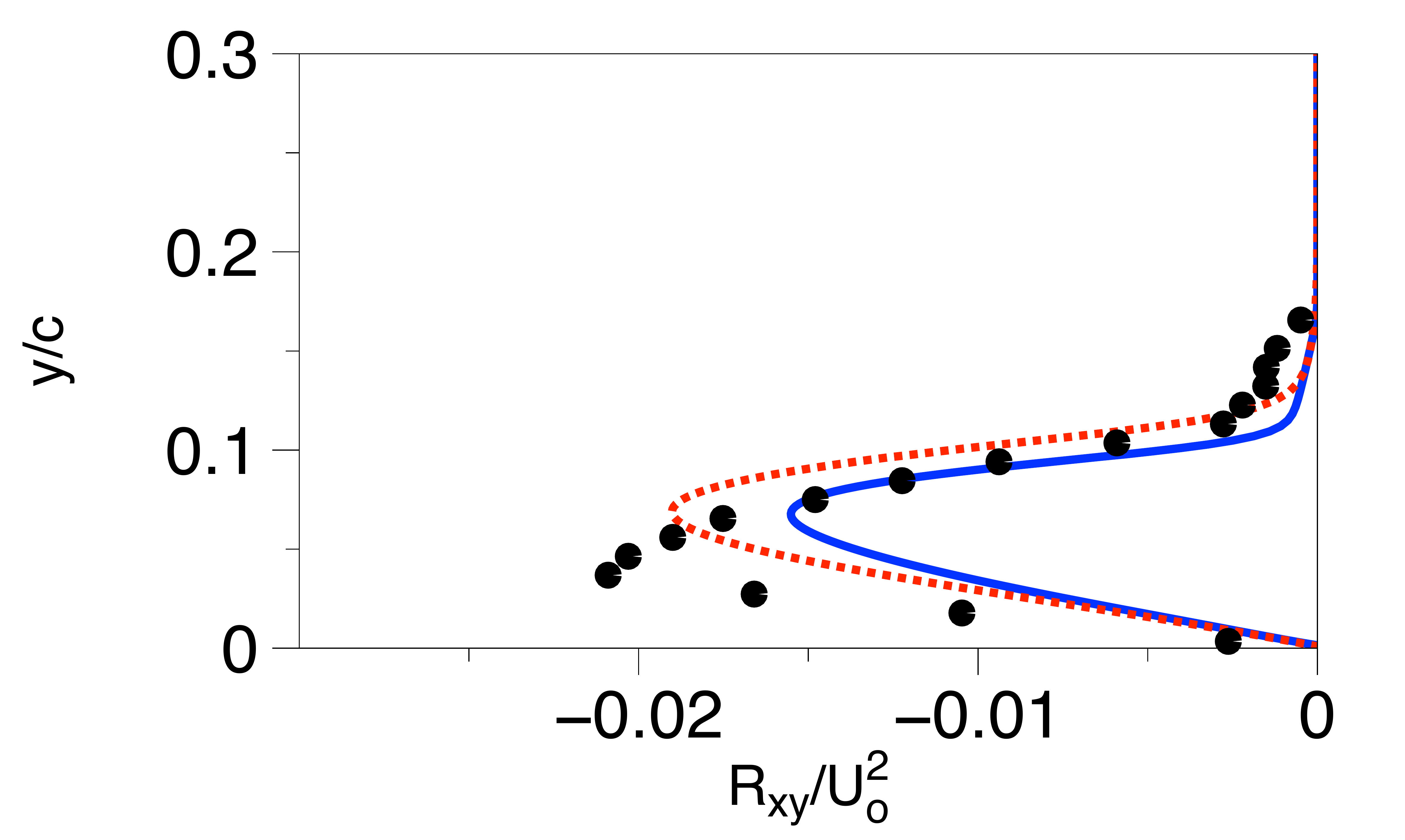}} 
\caption{Turbulent flow over the  ``Glauert-Goldschmied" hill for the no-flow control case. 
Model predictions for the shear  stress component  $R_{xy}$ at various $x$-stations for SA 
and ASBM-SA closures. Comparison is made to experimental values of 
Greenblatt et al.  \cite{Greenblatt2004}.} 
 \label{fig:rxy_hill}
\end{figure*}

Next, we consider the performance of the ASBM-SA closure for the turbulent intensities and 
the fluctuating shear stress with respect to  experimental results. The SA model is included 
only in the comparison for the fluctuating shear stress component, since it cannot provide 
predictions for the turbulent intensities.
Figure~\ref{fig:rxx_hill} displays the streamwise Reynolds stress component $R_{xx}$. At 
station A, ASBM-SA yields reasonable  predictions, being able to capture the near-wall peak magnitude. 
At the remaining three stations ($x/c=0.8, 1.0, 1.2$), ASBM-SA is able to capture 
satisfactorily  the peak magnitude. We note that the  wiggles near $y/c=0.2$ at station A 
(Figure~\ref{fig:uu_nf_066}) originate from the algebraic expressions for the estimation of 
the turbulent kinetic energy (not shown here). Following term by term the algebraic procedure 
for the calculation of kinetic energy, we found that this issue is most likely related to 
the local mean velocity gradients. These wiggles are also present in previous works 
\cite{Panagiotou2015}, for which similar findings were deduced. We also point out that the 
location $y/c=0.2$  at which the wiggles appear is close to the interface between two grid 
blocks. Overall, we believe that this is a localized effect that does not affect the quality 
of the solution.
\FloatBarrier
Figure~\ref{fig:ryy_hill} depicts the corresponding predictions for the transverse Reynolds 
stress component $R_{yy}$. ASBM-SA closure strongly overpredicts the near-wall magnitude 
at station A, while a fair agreement with the experiments is achieved at the remaining stations.   
\FloatBarrier
Figure~\ref{fig:rxy_hill} shows SA and ASBM-SA predictions for the fluctuating shear stress 
component. At the first station, ASBM-SA exhibits a similar behavior as for the transverse 
Reynolds stress component. At the remaining three stations, ASBM-SA provides noticable 
improvement relative to the SA model, in both the near-wall and freestream regions. This 
improvement is evident in the whole range of the recirculation bubble, suggesting a satisfactory 
response of the hybrid model to the strong anisotropic effects that characterize this region. 
\FloatBarrier

\subsubsection{ Control via steady suction  }
\begin{figure}[h!]
 \centering
\includegraphics[scale=0.13]{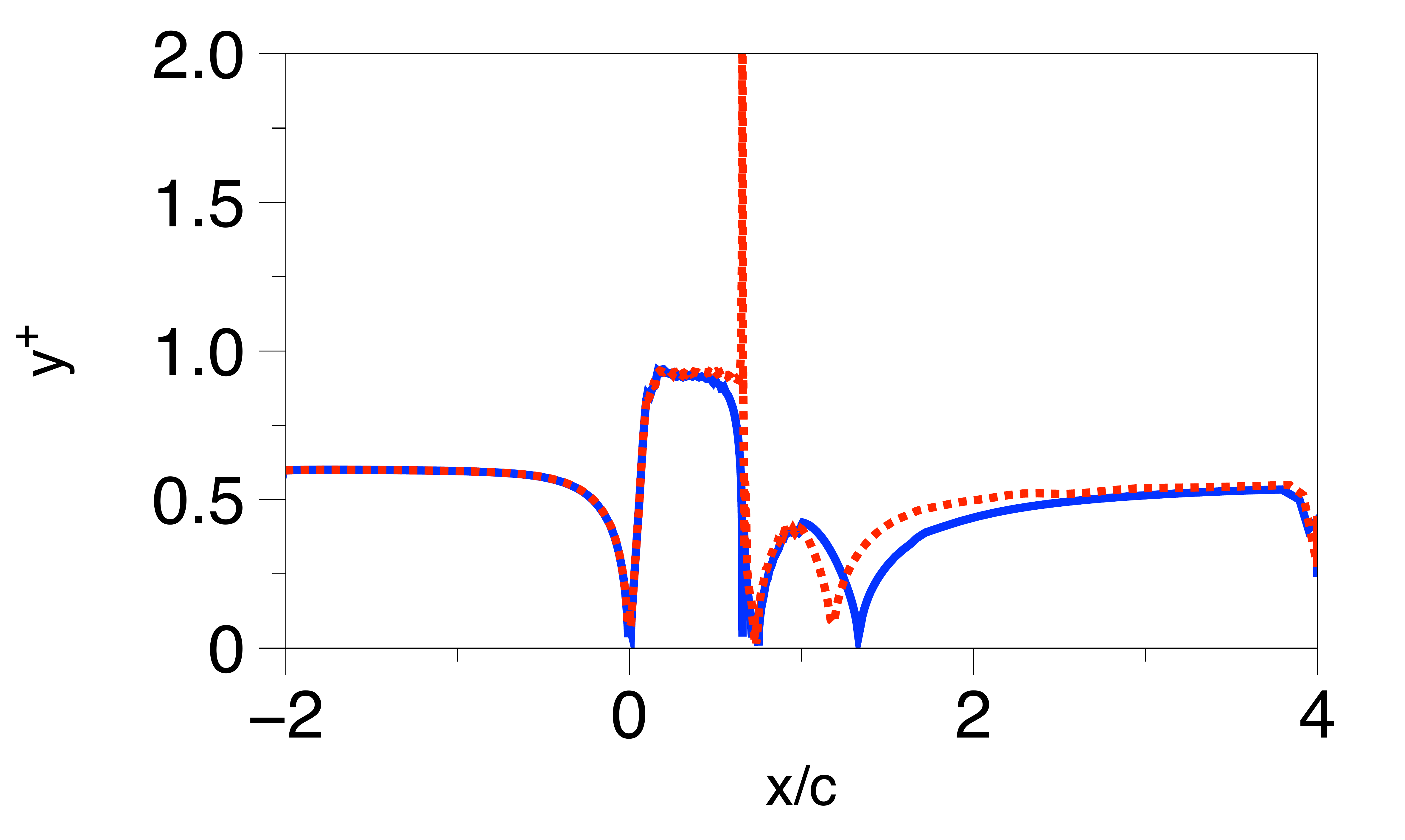}
\caption{ ASBM-SA model predictions for the streamwise variation at the bottom surface for 
the normal spacing at the wall normalized in wall-units. Results are shown  for both the 
no-flow control ($\solid$) and the  steady-suction $(\dashed)$ cases.}  
 \label{fig:yplus_comparison}
\end{figure}

As a first step, we define the steady mass transfer momentum coefficient  
\begin{equation}
c_{\mu}= \frac{ \rho h U^{2}_{jet}  }{ 1/2 c \rho U^{2}_{o}   }\,,
\end{equation}
\noindent where $U_{jet}$ denotes the jet velocity. For the current case, $c_{\mu}$ is set 
equal to $0.241\, \%$, corresponding to a constant mass flow rate of $\dot{m}=0.01518 \ kg/s$ 
being sucked through the slot, in order to match the experimental conditions of Greenblatt 
et al. \cite{Greenblatt2004}. Figure~\ref{fig:yplus_comparison} shows ASBM-SA model predictions  
for the wall-normal spacing  along the floor surface for the uncontrolled and controlled cases, 
needed to ensure that our solutions are obtained in sufficiently resolved grids. As expected, 
a sharp increase in $y^{+}$ levels occurs at the location of the slot exit where no-wall is 
present.
\FloatBarrier

\begin{figure*}[h!]
\centering
\subfloat[]
{\label{fig:cp_ss}
\includegraphics[scale=0.2]{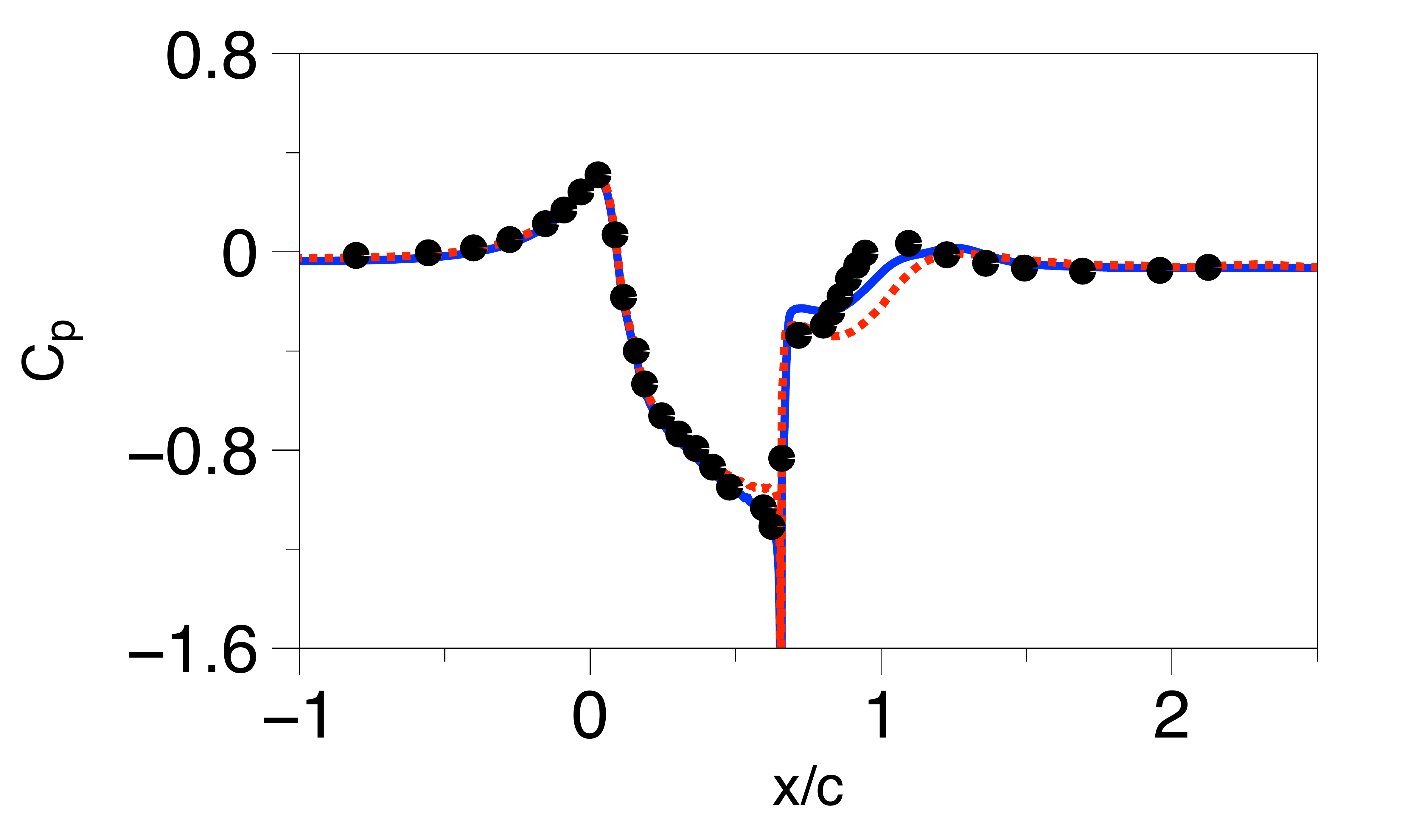}} 
\caption{SA ($\solid$)  and ASBM-SA ($\dashed$) model predictions for the steady-suction 
case for the wall static-pressure coefficient. Comparison is made to experimental values 
of Greenblatt et al. \cite{Greenblatt2004}.} 
 \label{fig:coefs_on_hill}
\end{figure*}
\begin{figure*}[h!]
\centering
\subfloat[station A, $x/c=0.66$]
{\label{fig:umean_ss_066}
\includegraphics[scale=0.13]{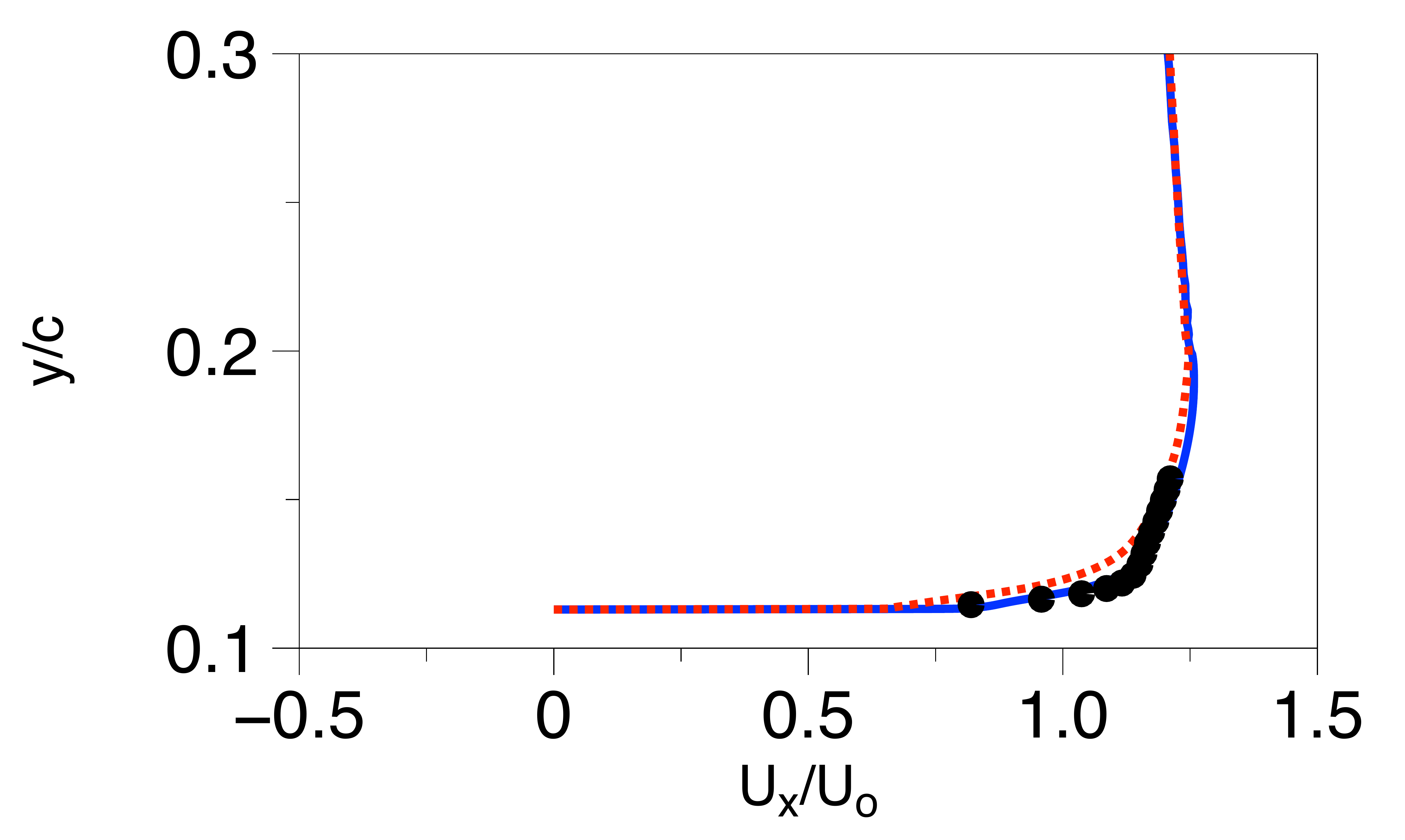}} 
\subfloat[station B, $x/c=0.8$]
{\label{fig:umean_ss_080}
\includegraphics[scale=0.13]{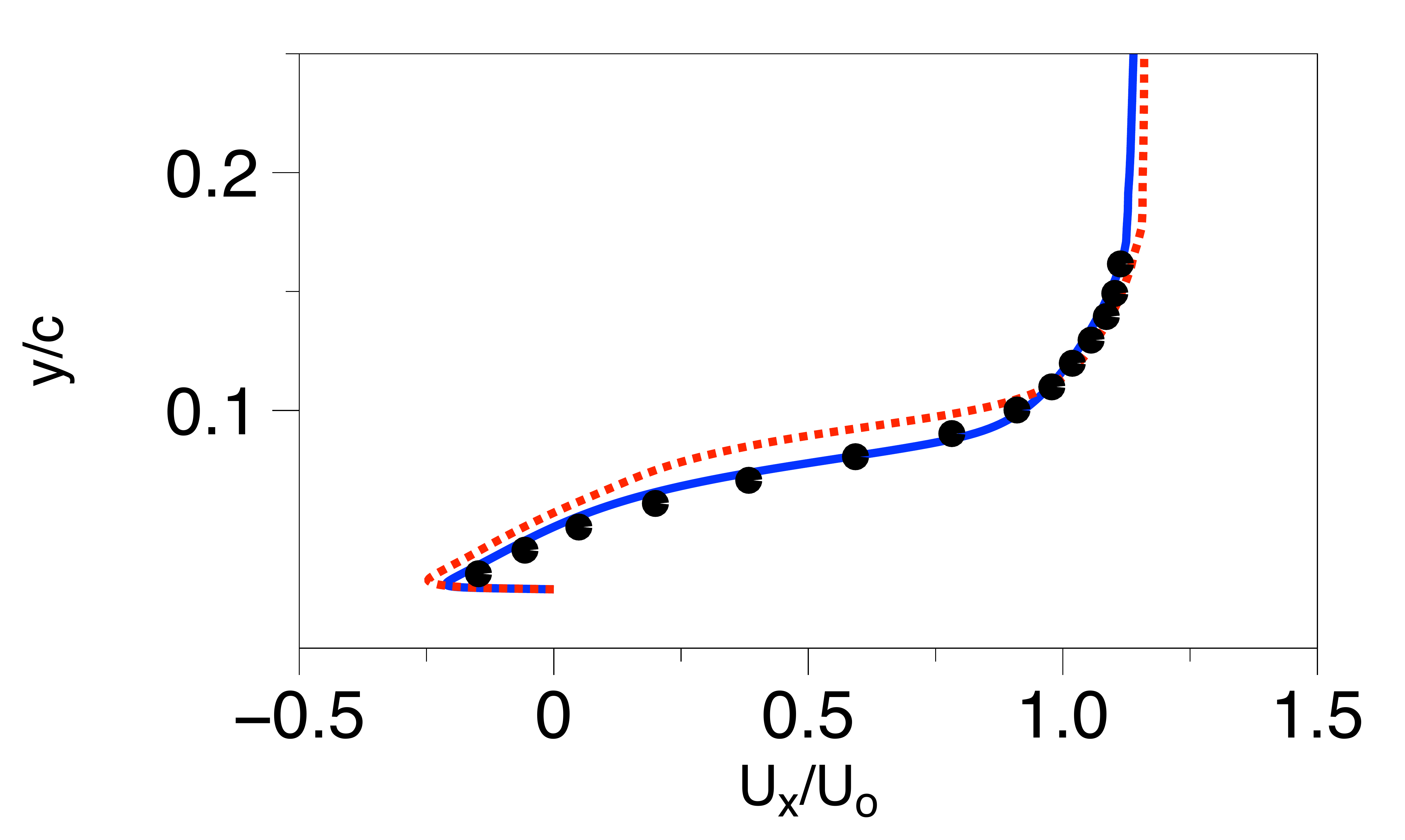}} \\
\subfloat[station C, $x/c=1.0$]
{\label{fig:umean_ss_10} 
\includegraphics[scale=0.13]{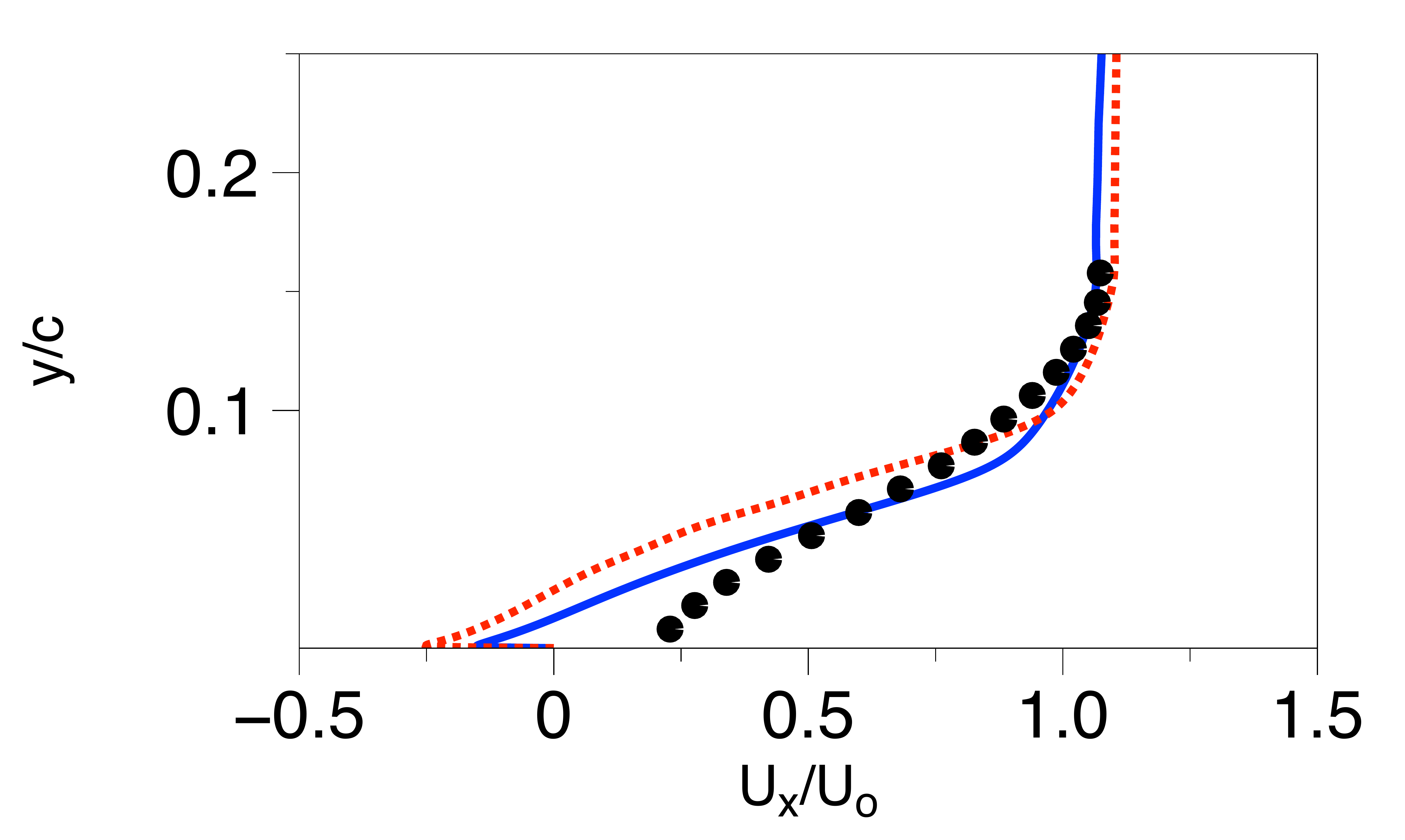}}
\subfloat[station D, $x/c=1.2$]
{\label{fig:umean_ss_12}
\includegraphics[scale=0.13]{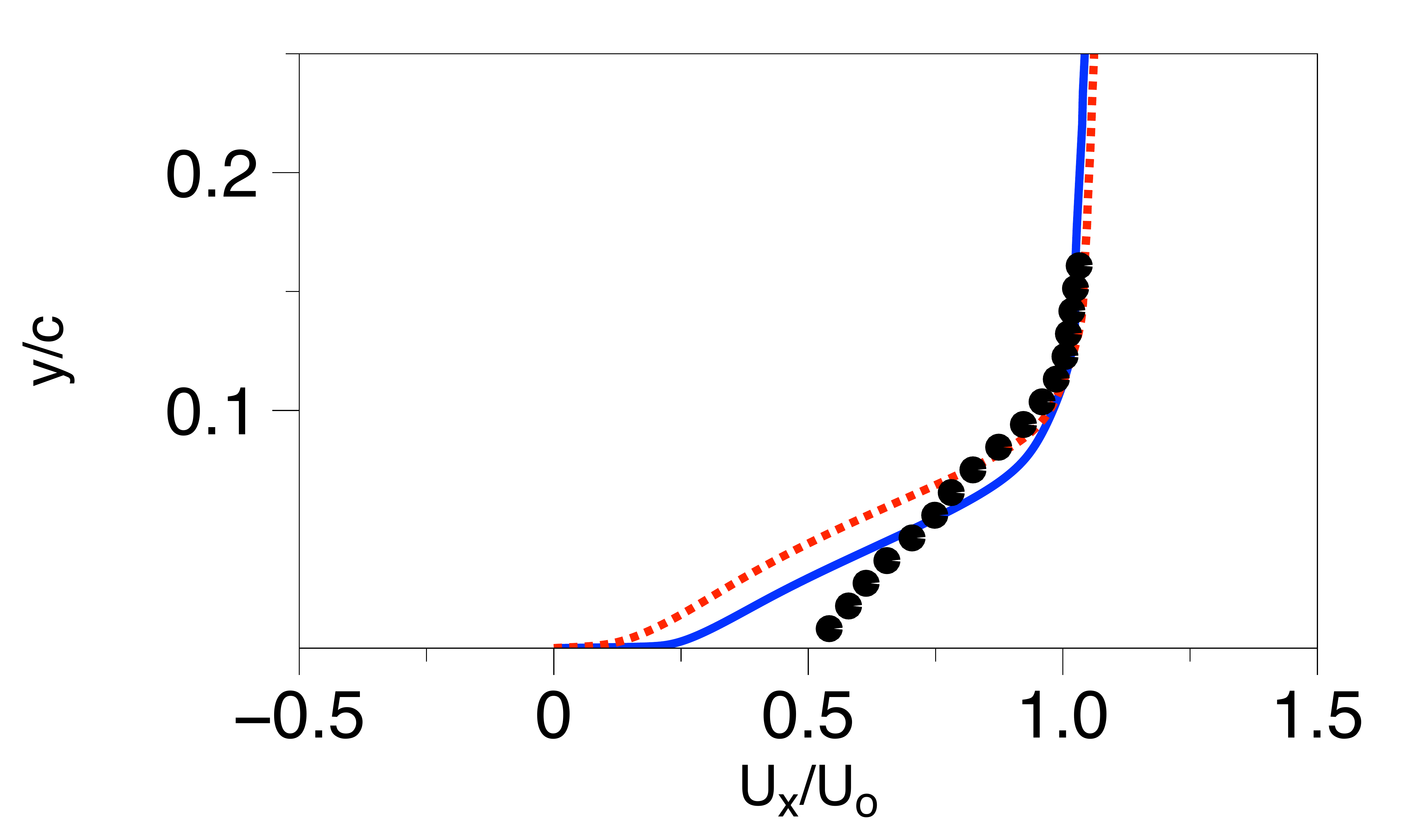}} 
\caption{Turbulent flow over the  ``Glauert-Goldschmied" hill for the steady-suction case. 
Model predictions for the streamwise mean velocity $U_{x}$ at various $x$-stations for SA 
and ASBM-SA closures. Comparison is made to experimental values of 
Greenblatt et al.  \cite{Greenblatt2004}.} 
\label{fig:Ux_on_hill}
\end{figure*}
\begin{figure*}[h!]
\centering
\subfloat[station A, $x/c=0.66$]
{\label{fig:vmean_ss_066}
\includegraphics[scale=0.13]{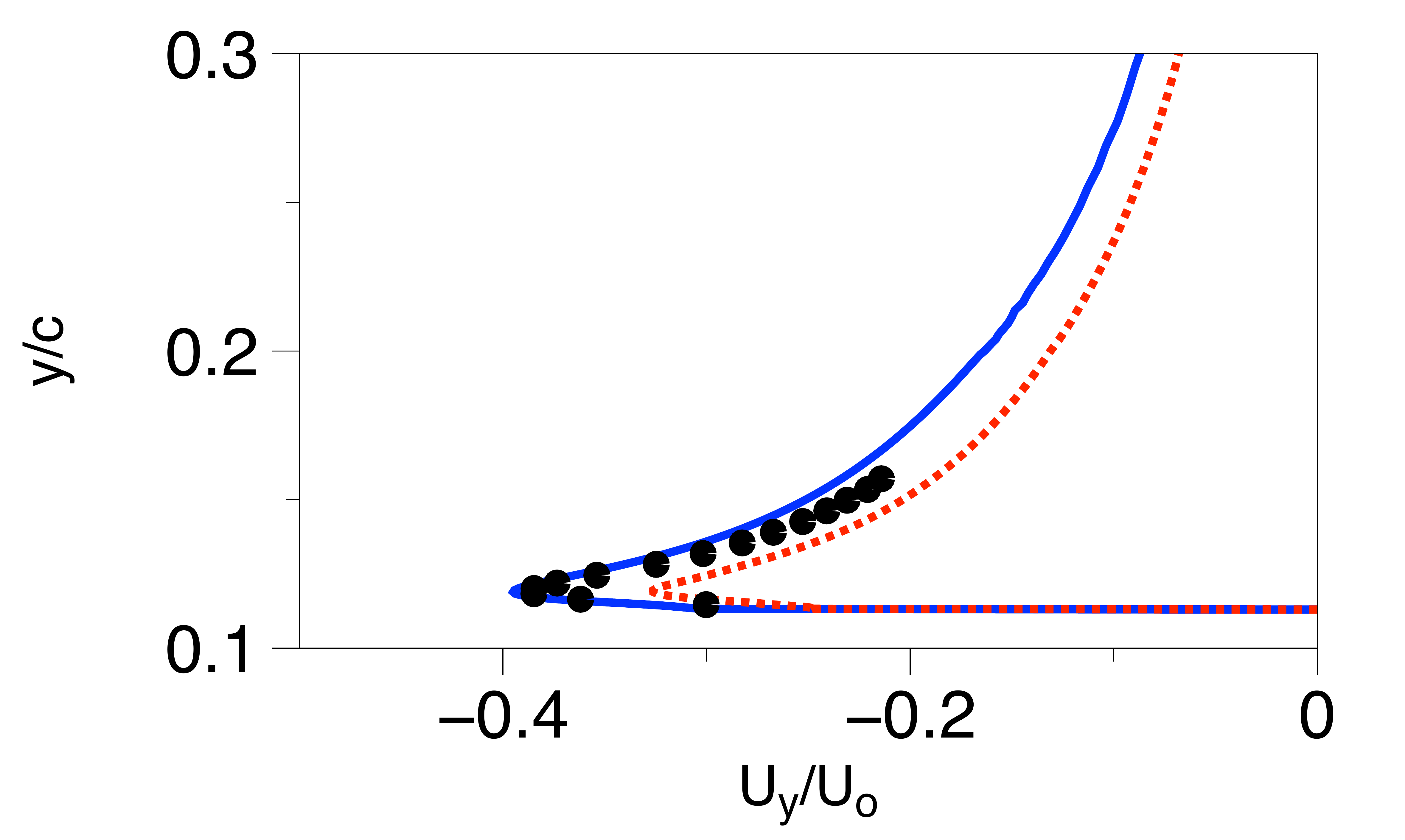}} 
\subfloat[station B, $x/c=0.8$]
{\label{fig:vmean_ss_080}
\includegraphics[scale=0.13]{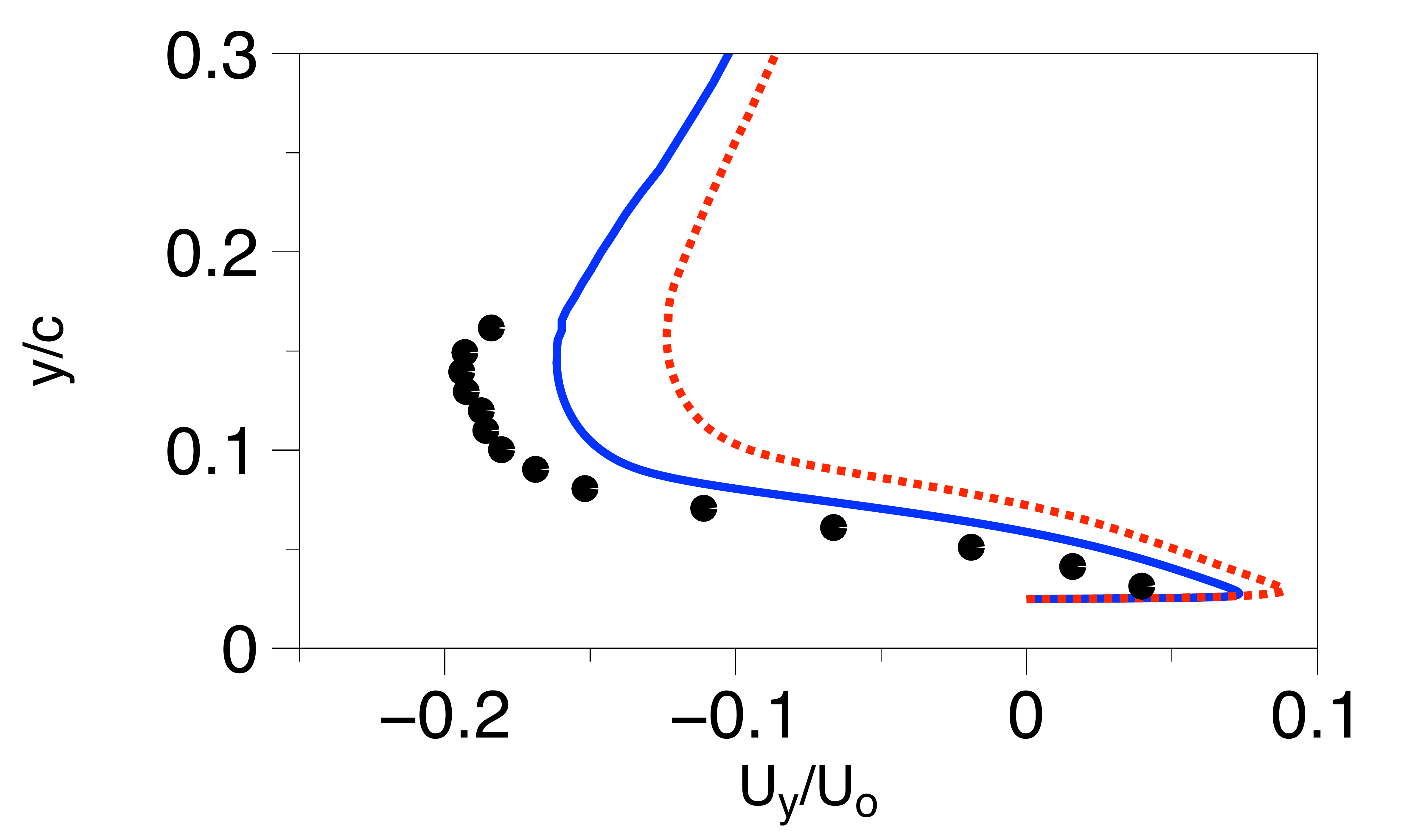}} \\
\subfloat[station C, $x/c=1.0$]
{\label{fig:vmean_ss_10} 
\includegraphics[scale=0.13]{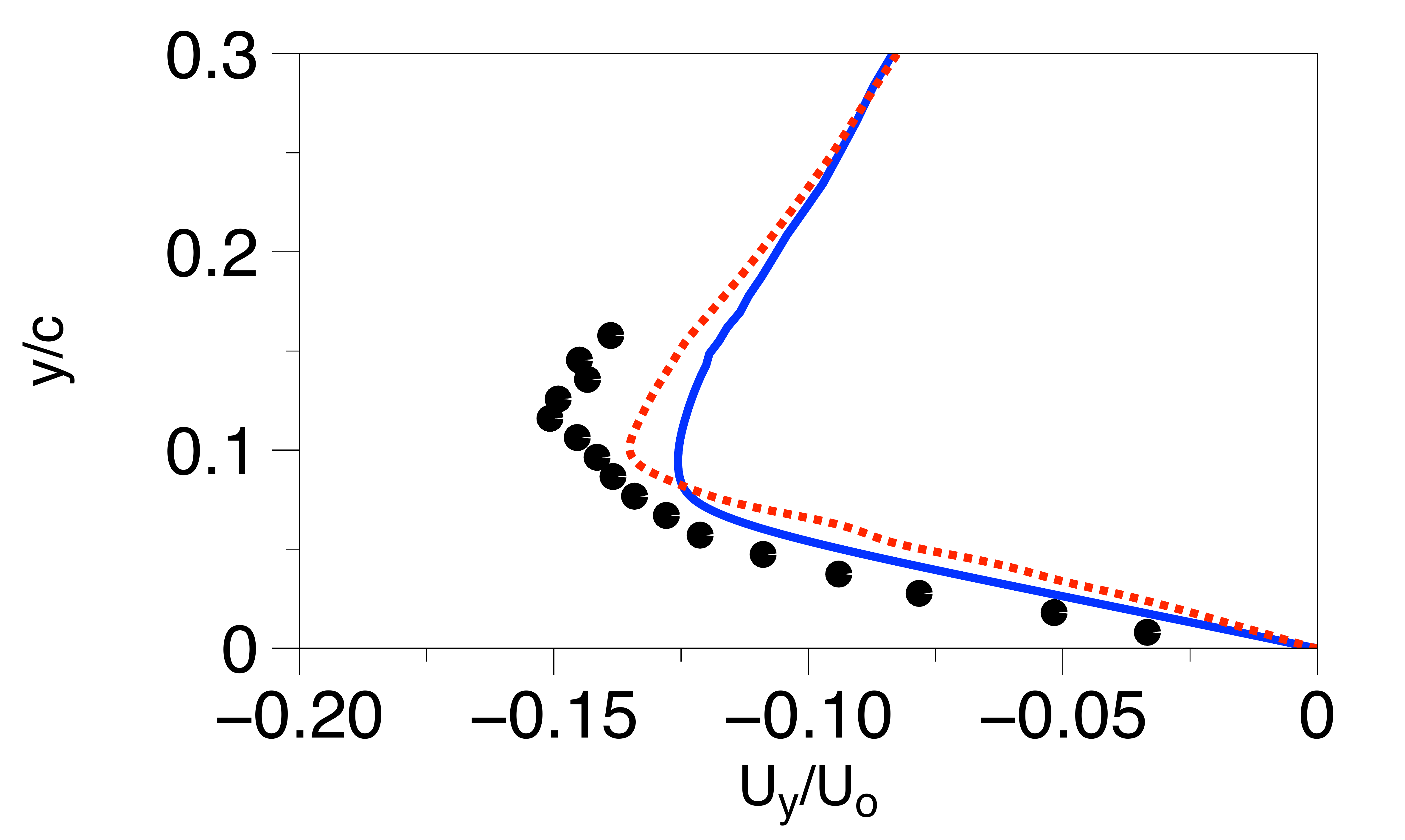}}
\subfloat[station D, $x/c=1.2$]
{\label{fig:vmean_ss_12}
\includegraphics[scale=0.13]{NHUY_10_on}} 
\caption{Turbulent flow over the ``Glauert-Goldschmied" hill for the steady-suction case. 
Model predictions for the transverse mean velocity $U_{y}$ at various $x$-stations for SA 
and ASBM-SA closures. Comparison is made to experimental values of 
Greenblatt et al. \cite{Greenblatt2004}.} 
 \label{fig:Uy_on_hill}
\end{figure*}

Results for the wall-static pressure coefficient are shown in Figure~\ref{fig:coefs_on_hill}, 
where the predictions of the SA and ASBM-SA models are compared to the experimental 
data of Greenblatt et al. \cite{Greenblatt2004}. The ASBM-SA closure manages to capture 
accurately the magnitude and the location of the first sharp change right after the slot, 
located around $x/c \approx 0.67$, followed by a small recovery delay, as compared to the 
SA model, till the trailing edge of the hill ($x/c = 1$)  where the two models coincide again.
\FloatBarrier
Figure~\ref{fig:Ux_on_hill} displays results for the streamwise mean velocity $U_{x}$ at 
the four stations. As shown, SA provides slightly better agreement with the experiments than 
the ASBM-SA model. Figure~\ref{fig:Uy_on_hill} shows the corresponding comparison for 
the transverse mean velocity $U_{y}$. In general, SA predictions are in better agreement 
with the experimental data compared to ASBM-SA model.  
\FloatBarrier
\FloatBarrier
Table~\ref{Table:Cases} shows details regarding the recirculation region. SA predicts more 
accurately the re-attachment point for both cases, providing an indication why SA closure 
obtains better results than the hybrid model for the mean statistics. 
\begin{table}[h!]
\begin{center}
\small
\begin{tabular}{l*{6}{c}r}\hline
Case              &    Model      &    sep.loc.        &    sep.loc.   & Error    &  reatt.loc.            &    reatt.loc.   &   Error   \\
                  &               &    experiment      &    CFD        &  ($\%$)  &  experiment            &    CFD          &   ($\%$)  \\
\hline
no-flow control   &    SA         & $\approx  0.67 $   & 0.663         &  1.0     &  $1.11 \pm 0.003   $   &    1.235        &   11.3    \\
no-flow control   &    ASBM-SA    & $\approx  0.67 $   & 0.656         &  2.1     &  $ 1.11 \pm 0.003  $   &    1.330        &   19.8    \\
steady suction    &    SA         & $\approx  0.68 $   & 0.676         &  0.6     &  $ 0.94 \pm 0.005  $   &    1.113        &   18.4    \\
steady suction    &    ASBM-SA    & $\approx  0.68 $   & 0.665         &  2.2     &  $ 0.94 \pm 0.005  $   &    1.180        &   25.5    \\ \hline
\end{tabular}
 \end{center}
 \caption{ Details of SA and ASBM-SA model predictions regarding the recirculation bubble 
for each case. Comparison is made to the experimental work of Greenblatt et 
al \cite{Greenblatt2004}. }
\label{Table:Cases}
\end{table}
\FloatBarrier

\begin{figure*}[h!]
\centering
\subfloat[station A, $x/c=0.66$]
{\label{fig:uu_ss_066}
\includegraphics[scale=0.13]{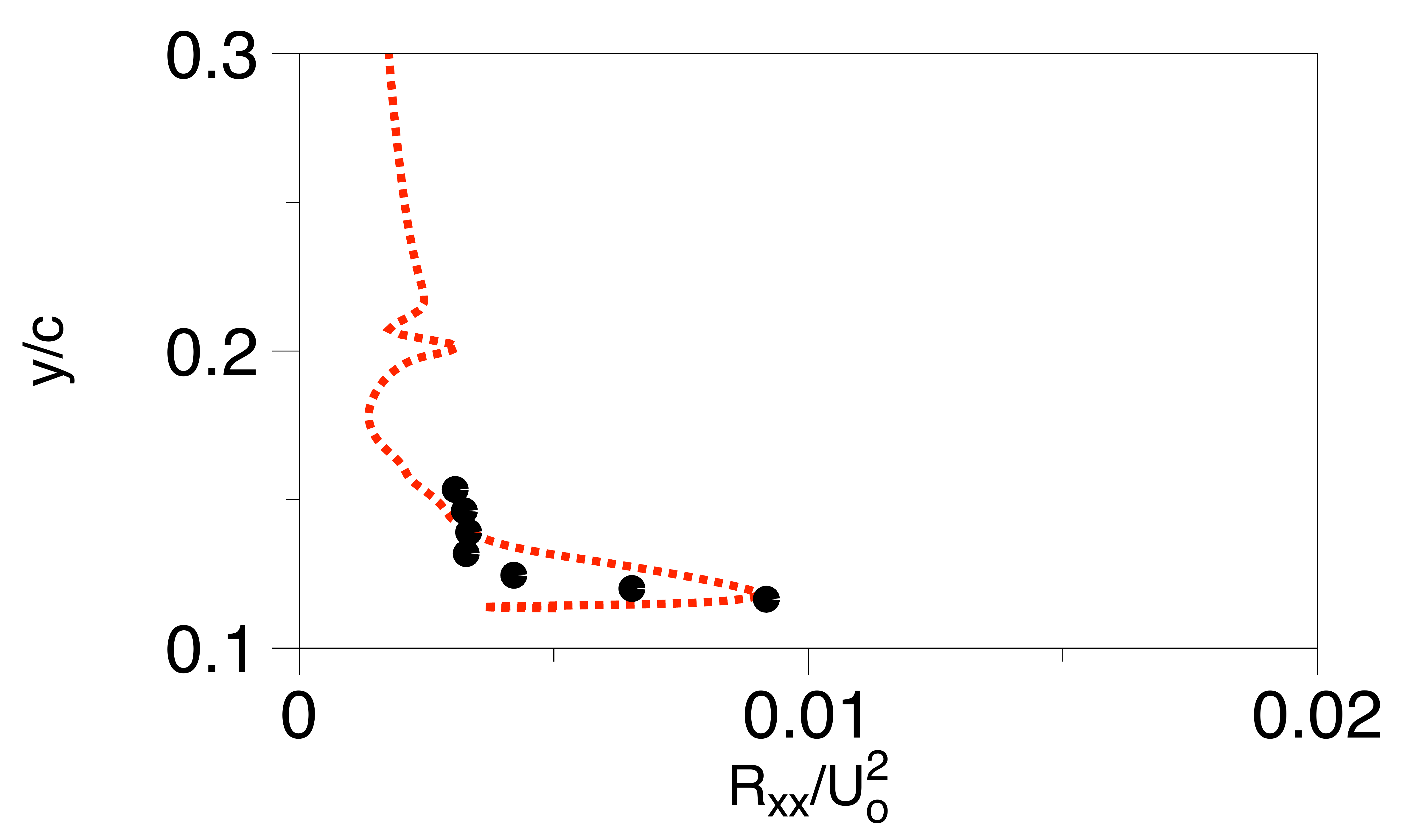}} 
\subfloat[station B, $x/c=0.8$]
{\label{fig:uu_ss_080}
\includegraphics[scale=0.13]{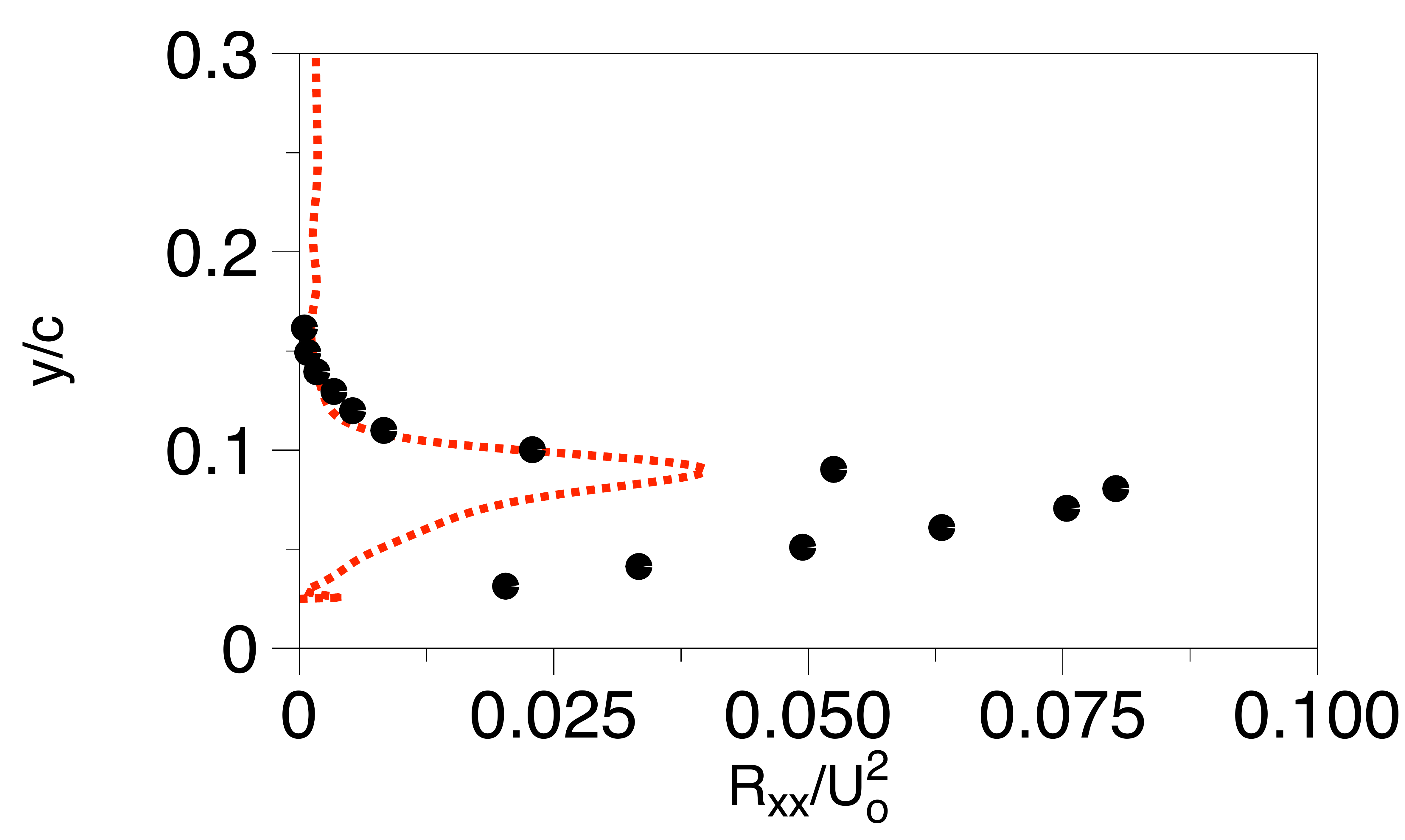}} \\
\subfloat[station C, $x/c=1.0$]
{\label{fig:uu_ss_10} 
\includegraphics[scale=0.13]{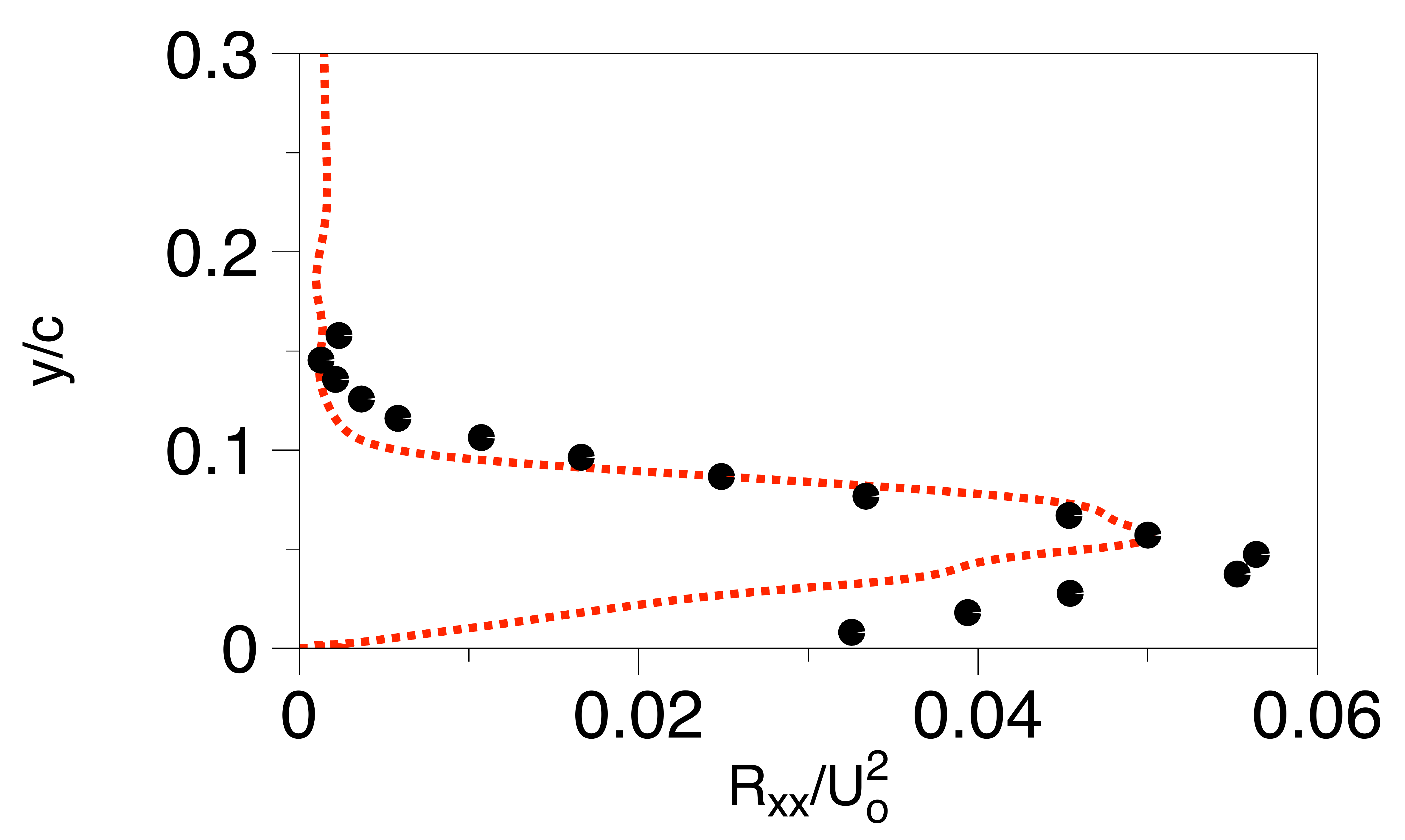}}
\subfloat[station D, $x/c=1.2$]
{\label{fig:uu_ss_12}
\includegraphics[scale=0.13]{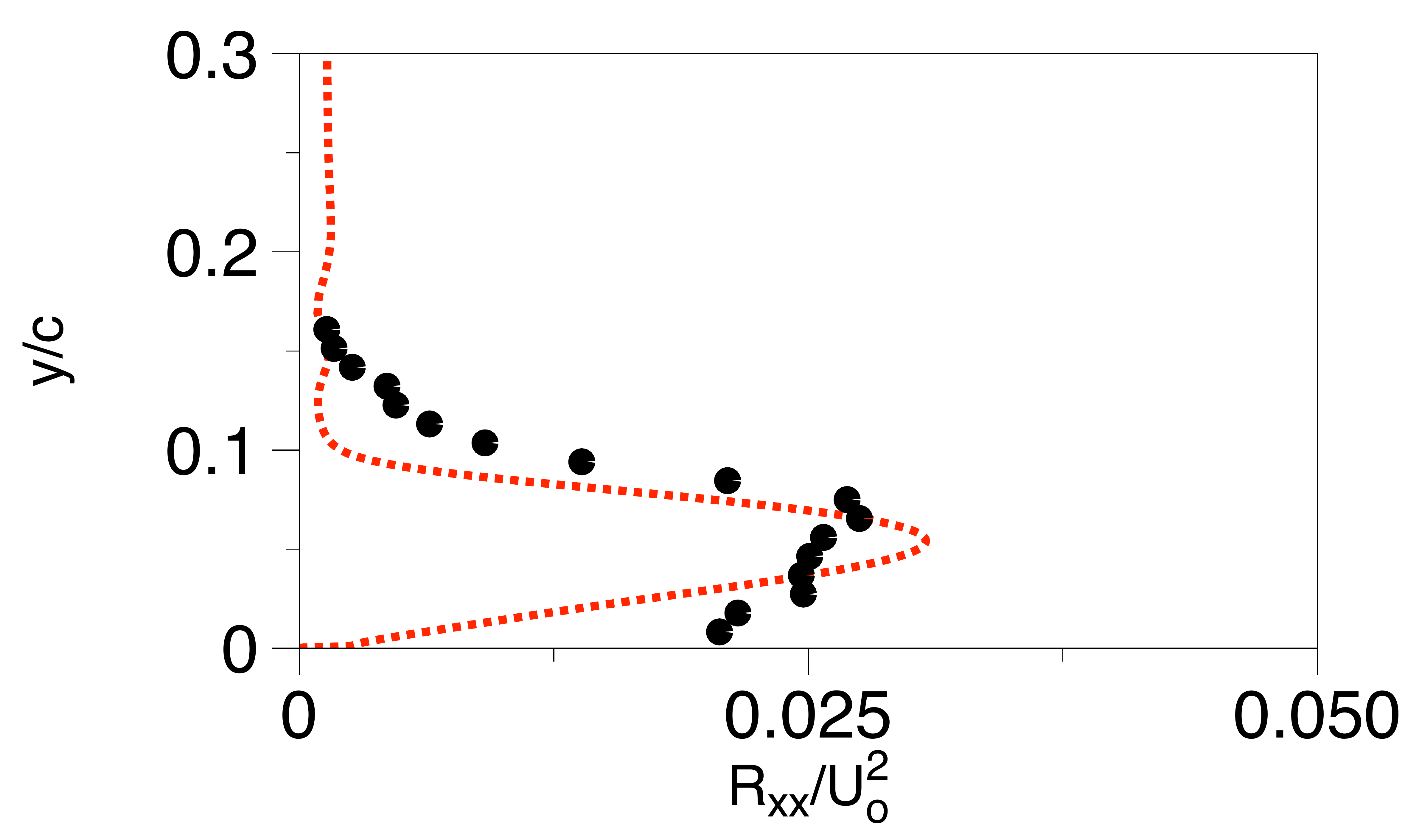}} 
\caption{Turbulent flow over the  ``Glauert-Goldschmied" hill for the steady-suction case. 
Model predictions for the streamwise Reynolds stress component $R_{xx}$ at various $x$-stations
for SA and ASBM-SA closures. Comparison is made to experimental values of 
Greenblatt et al. \cite{Greenblatt2004}.} 
\label{fig:rxx_on_hill}
\end{figure*}
\begin{figure*}[h!]
\flushleft
\subfloat[station A, $x/c=0.66$]
{\label{fig:vv_ss_066}
\includegraphics[scale=0.13]{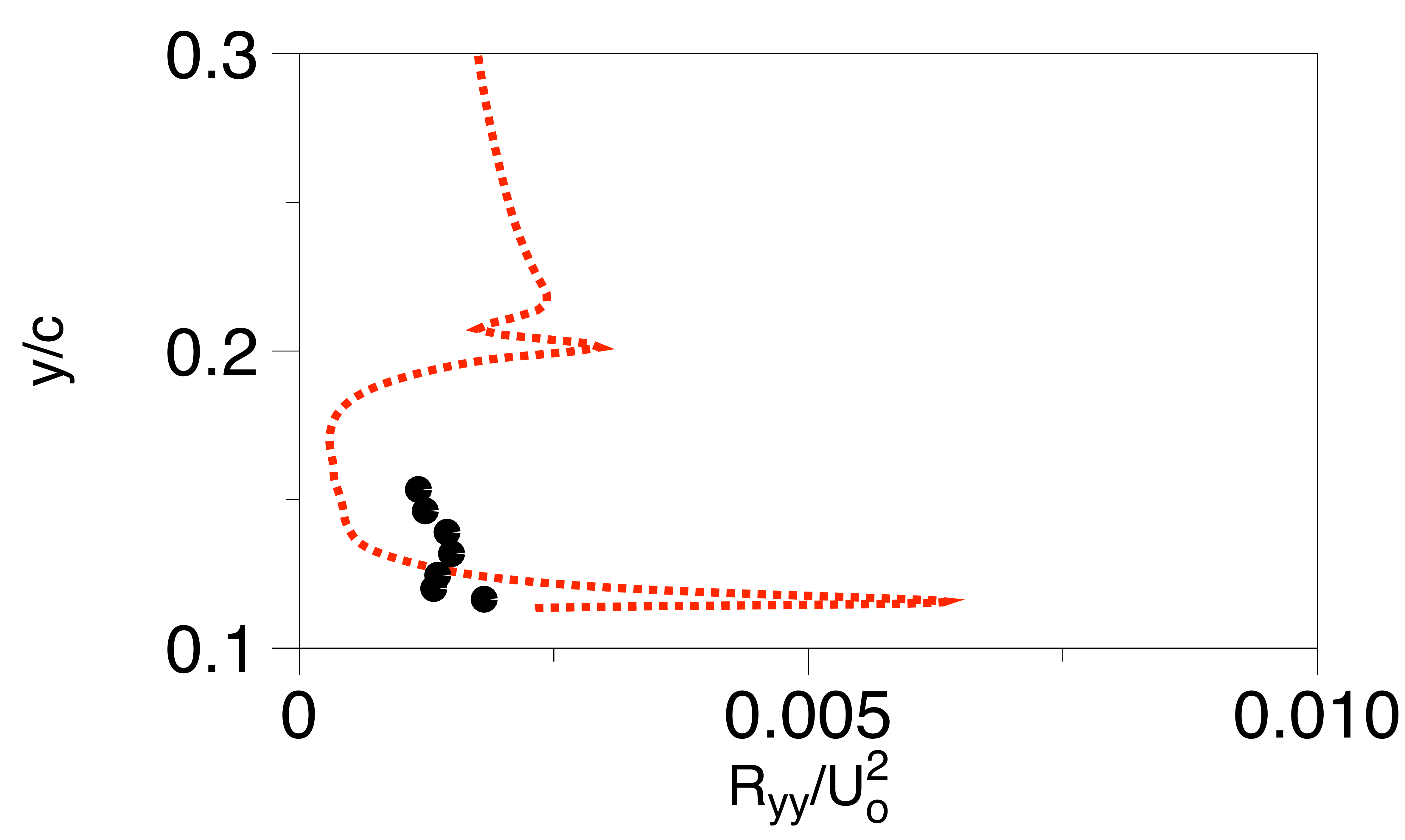}} 
\subfloat[station B, $x/c=0.8$]
{\label{fig:vv_ss_080}
\includegraphics[scale=0.13]{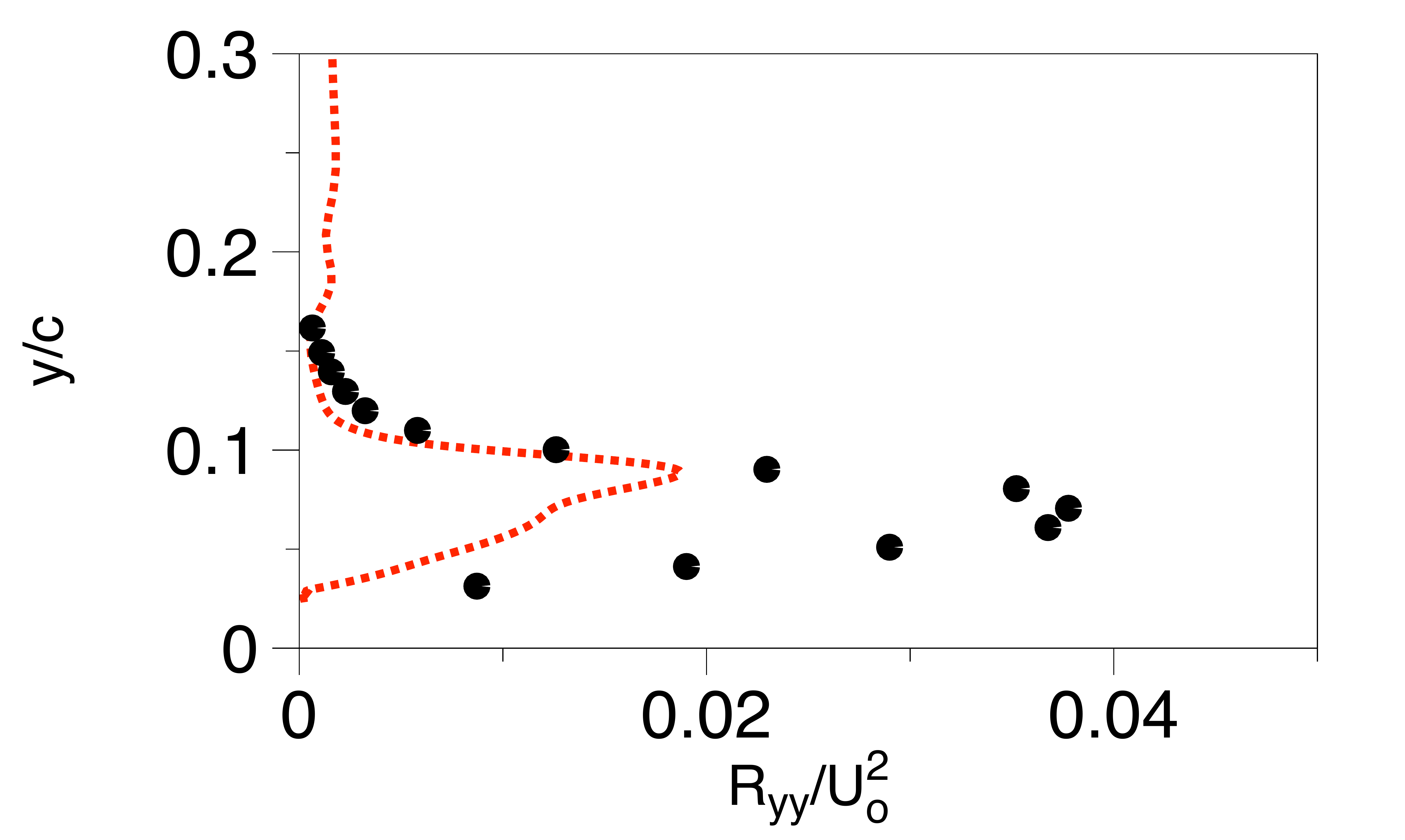}} \\
\subfloat[station C, $x/c=1.0$]
{\label{fig:vv_ss_10} 
\includegraphics[scale=0.13]{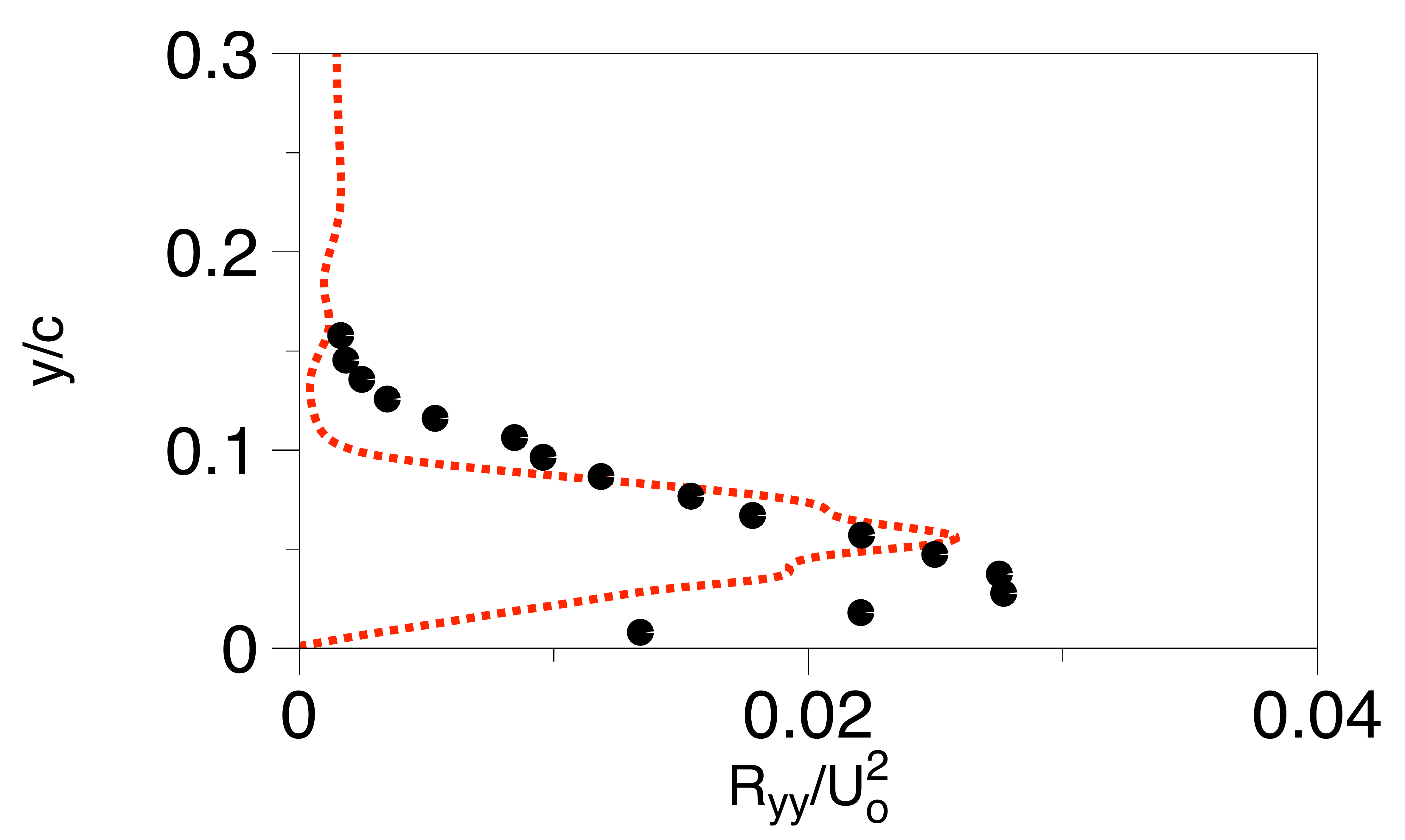}}
\subfloat[station D, $x/c=1.2$]
{\label{fig:vv_ss_12}
\includegraphics[scale=0.13]{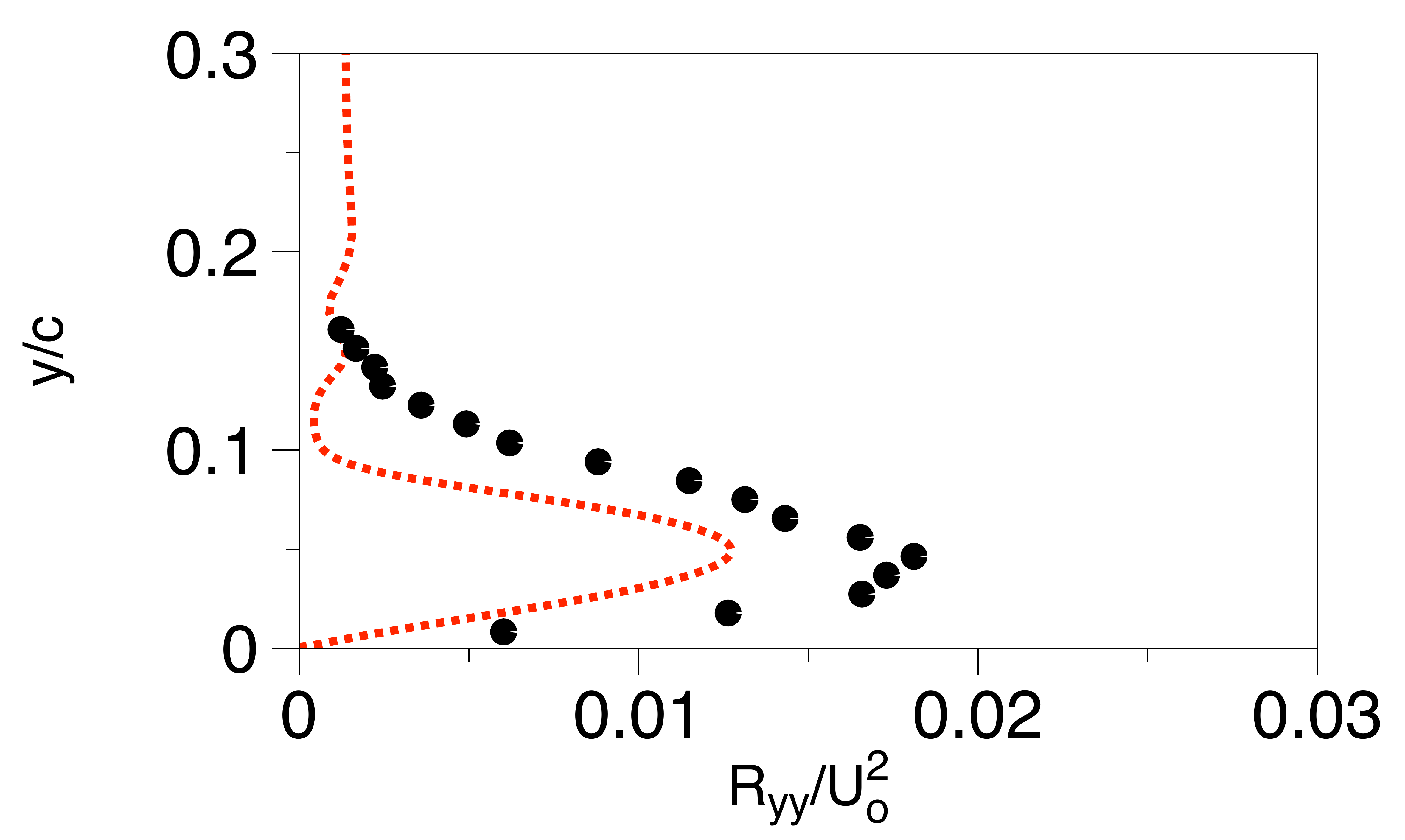}} 
\caption{Turbulent flow over the ``Glauert-Goldschmied" hill for the steady-suction case. 
Model predictions for the transverse Reynolds stress component  $R_{yy}$ at various 
$x$-stations for SA and ASBM-SA closures. Comparison is made to experimental values 
\cite{Greenblatt2004}.} 
 \label{fig:ryy_on_hill}
\end{figure*}
\begin{figure*}[h!]
\flushleft
\subfloat[station A, $x/c=0.66$]
{\label{fig:uv_ss_066}
\includegraphics[scale=0.13]{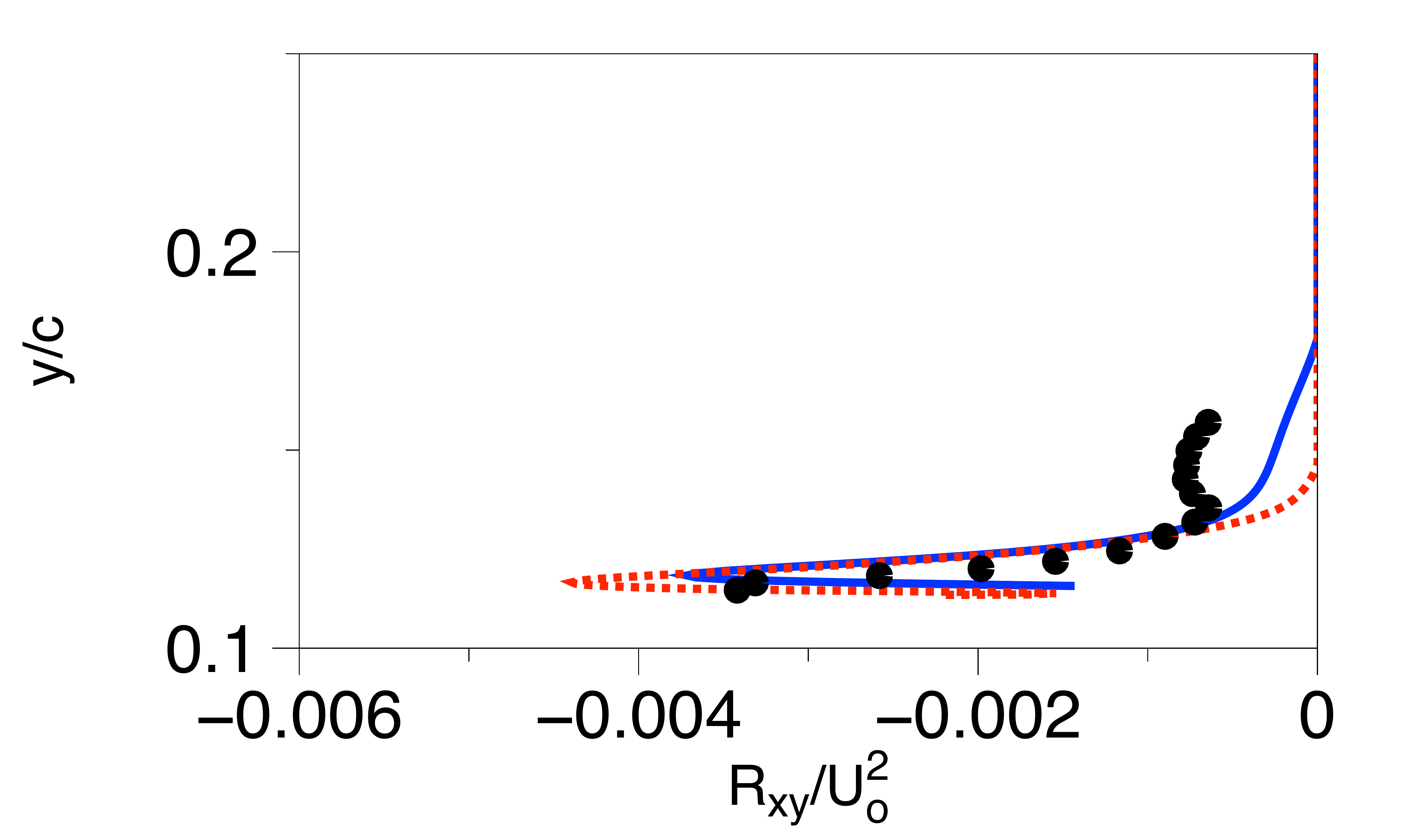}} 
\subfloat[station B, $x/c=0.8$]
{\label{fig:uv_ss_080}
\includegraphics[scale=0.13]{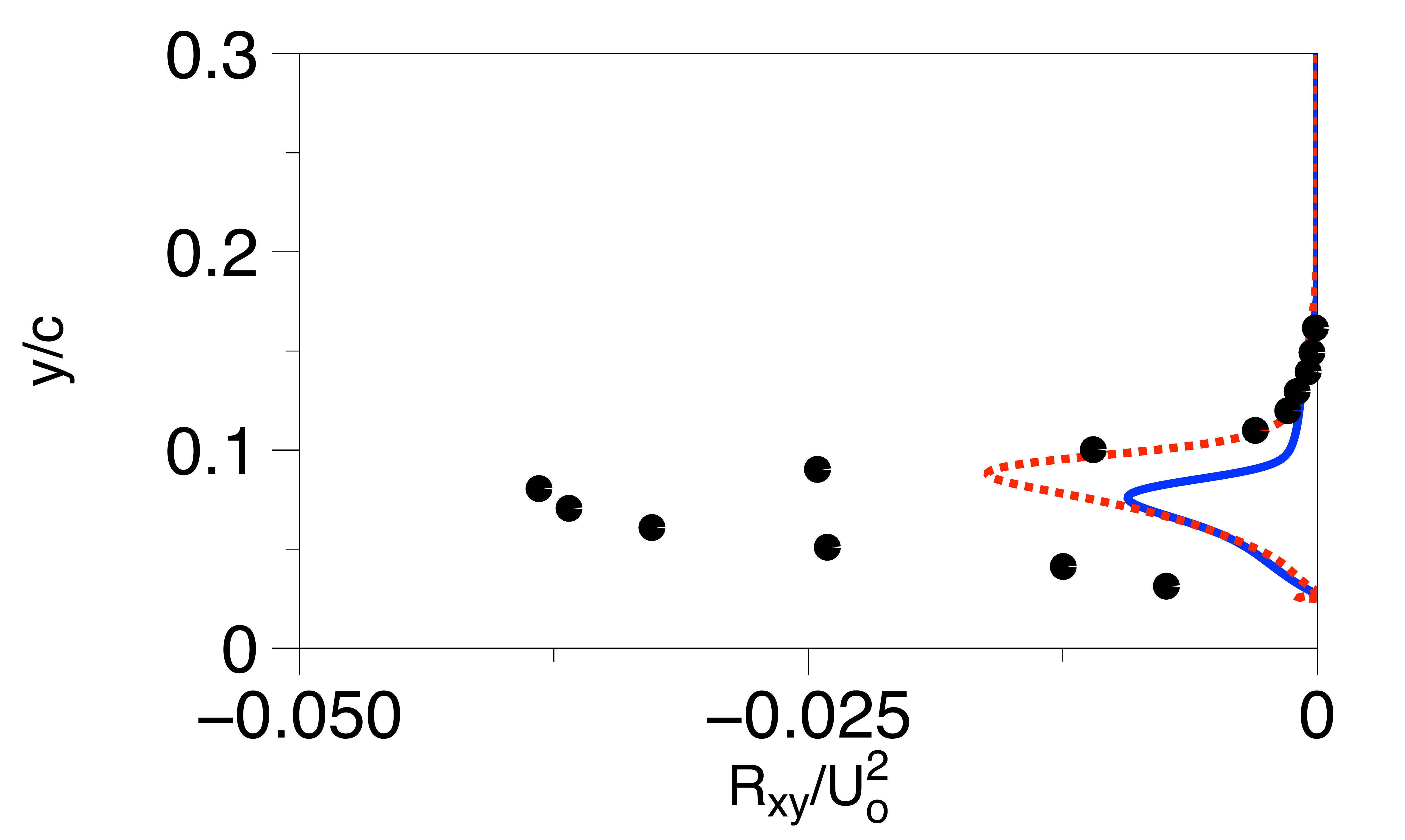}} \\
\subfloat[station C, $x/c=1.0$]
{\label{fig:uv_ss_10} 
\includegraphics[scale=0.13]{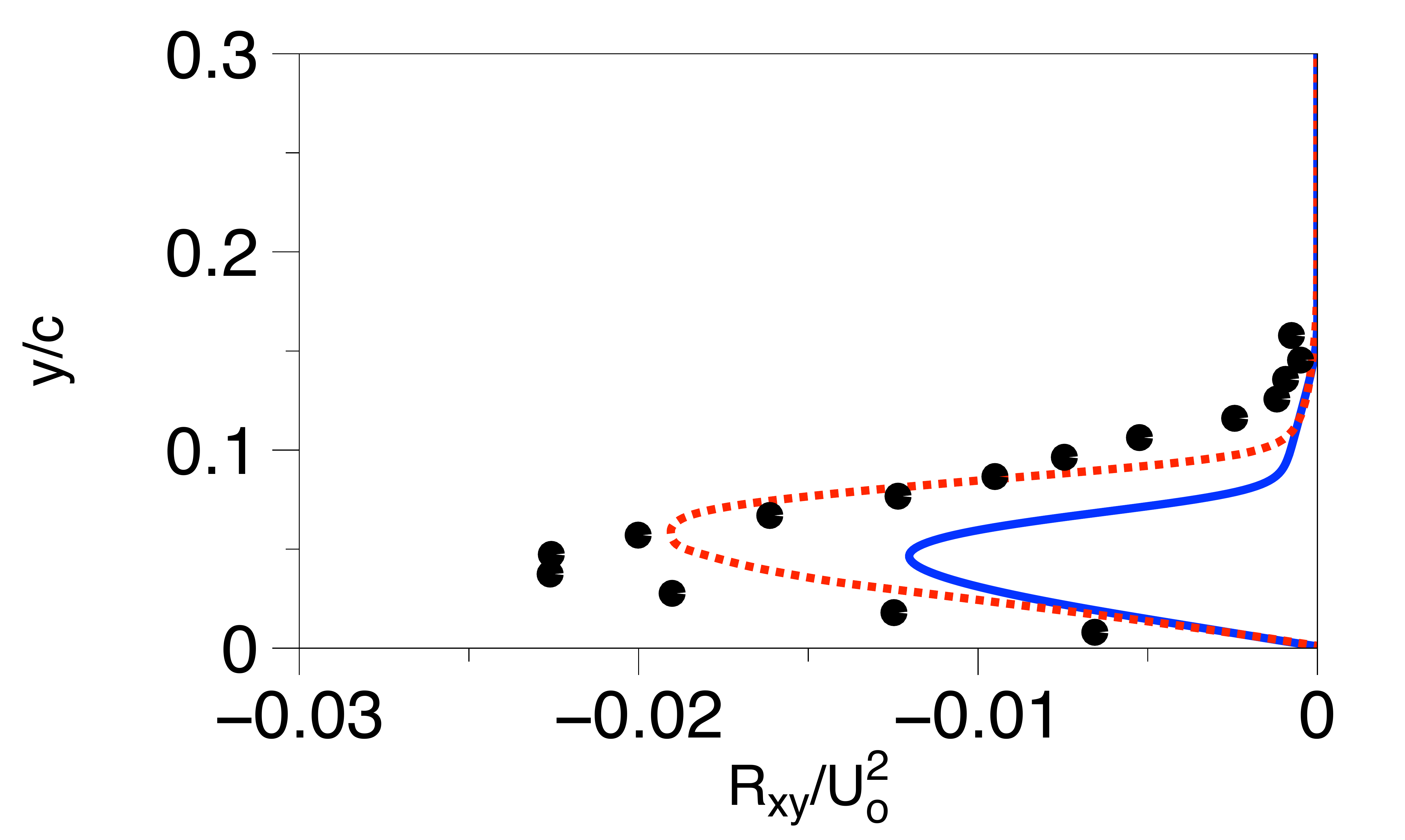}}
\subfloat[station D, $x/c=1.2$]
{\label{fig:uv_ss_12}
\includegraphics[scale=0.13]{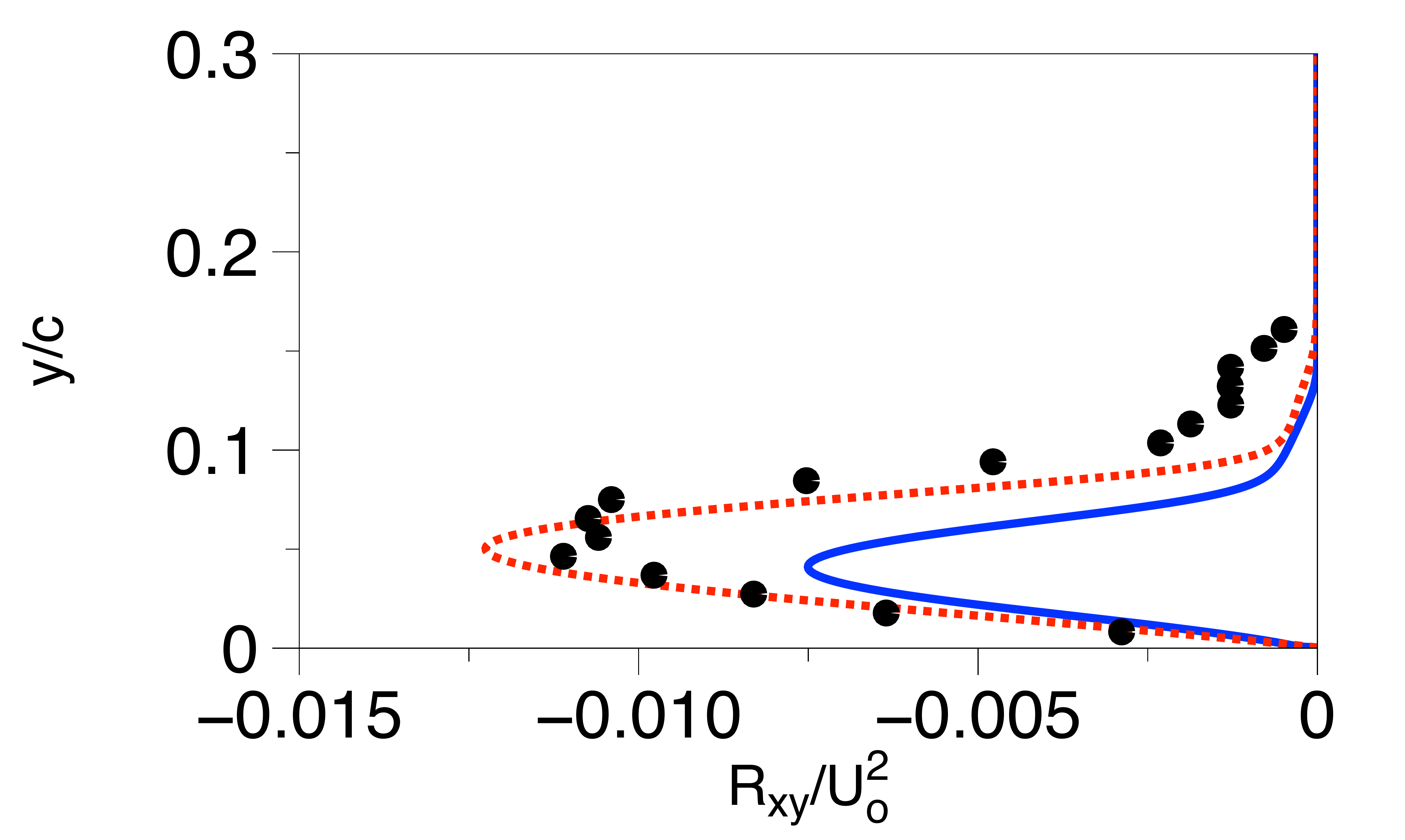}} 
\caption{Turbulent flow over the ``Glauert-Goldschmied" hill for the steady-suction case. 
Model predictions for the fluctuating shear stress component  $R_{xy}$ at various $x$-stations 
for SA and ASBM-SA closures. Comparison is made to experimental values \cite{Greenblatt2004}.} 
 \label{fig:rxy_on_hill}
\end{figure*}
\begin{figure*}[h!]
\centering
\subfloat[]
{\label{fig:str_asbmsa_cavity_off}
\includegraphics[scale=0.3]{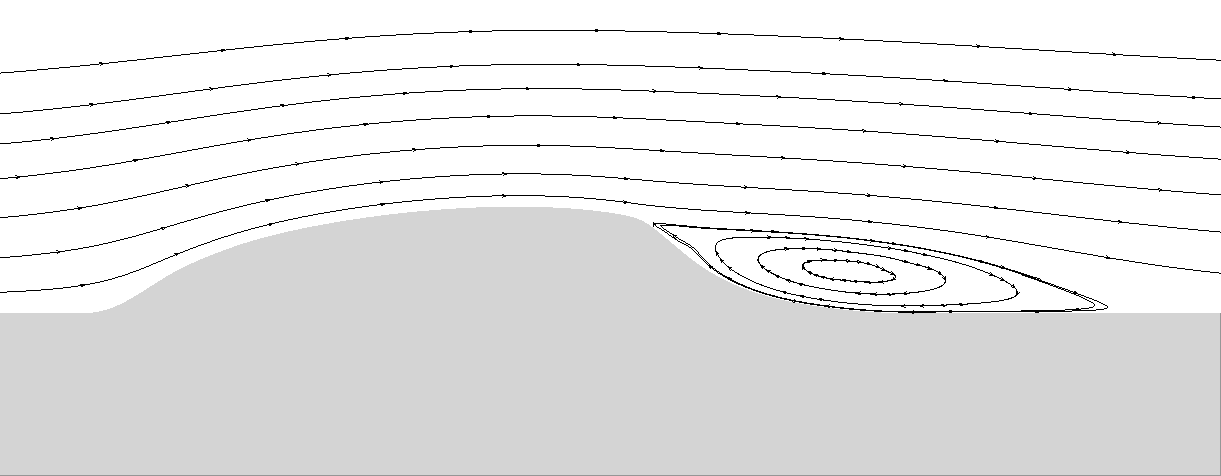}} \\
\subfloat[]
{\label{fig:str_asbmsa_nocavity_on}
\includegraphics[scale=0.30]{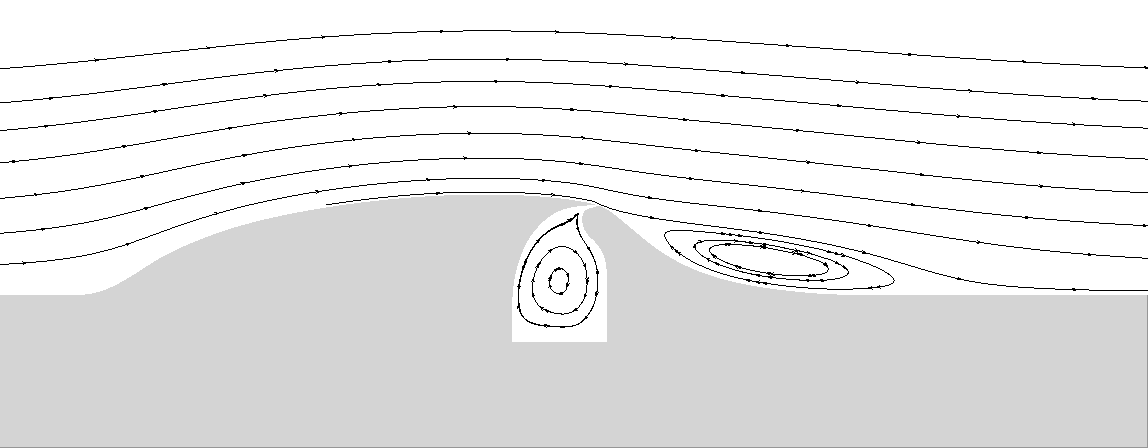}} 
\caption{ASBM-SA model predictions for the streamlines of the mean velocity for (a) the 
uncontrolled case and (b) the controlled case .  } 
 \label{fig:streamlines_on_off}
\end{figure*}

Figure~\ref{fig:rxx_on_hill}  shows the corresponding comparison for the streamwise Reynolds 
stress component $R_{xx}$. At three of the four stations, ASBM-SA correctly predicts the 
near-wall peak magnitude and the freestream values,  yielding a fair agreement with the 
experiments.   
\FloatBarrier
In Figure~\ref{fig:ryy_on_hill},  the agreement is qualitative  between the ASBM-SA predictions
and the  experimental measurements for the transverse Reynolds stress component $R_{yy}$. 
Combining these results with the analogous ones for $R_{xx}$ reveals the sensitivity of the 
algebraic model to the anisotropic nature of the flow.
\FloatBarrier
Results for the fluctuating shear stress component $R_{xy}$ are shown in 
Figure~\ref{fig:rxy_on_hill}. As shown,   the hybrid  ASBM-SA model is able to provide  
significantly improved predictions compared to the SA closure in the whole range of the 
recirculation region.  Overall, ASBM-SA provides a satisfactory agreement with experiments. 
\FloatBarrier
The active control effect on the recirculation bubble is visualized in 
Figure~\ref{fig:streamlines_on_off}. ASBM-SA model predictions for the streamlines of the 
mean velocity for both cases are shown, revealing  a noticeable reduction of the bubble size. 
\FloatBarrier

\section{Summary and Conclusions }\label{summary_conclusions}

The ASBM-SA closure has been tested for the case of a flow over a two-dimensional smooth hill 
in the shape of a ``Modified Glauert-Goldschmied" hump both in the presence and absence of 
separation control. For both cases considered, the ASBM-SA model produced satisfactory 
predictions for the streamwise Reynolds stress component $R_{xx}$, while a qualitative 
agreement with the experiments was achieved for the transverse component $R_{yy}$. ASBM-SA 
closure provided improved predictions compared to SA for the shear stress at all stations. 
Regarding the mean quantities, the predictions of both closures are comparable, providing fair 
agreement with the experiments. Overall, the hybrid model managed to capture satisfactory the 
traits of these highly anisotropic flows, while maintaining the high robustness of the SA model, 
at a good convergence rate. 
Use of  separated zones for the activation 
of the filtering scheme resulted to smoother mean velocity profiles in the recirculation region, 
and more meaningful skin-friction profiles upstream and downstream the separated region.\\
As part of future work we intend to  ascertain the performance of the hybrid model over 
challenging two-dimensional flows, such as turbulent flows around a wall-mounted cube and 
plane wall jets, while an extension to three-dimensional smooth hills will be attempted. 
Regarding the refinement issues encountered in the current and previous works, further 
consideration is needed to understand why ASBM-SA delays further the re-attachment compared 
to SA for all validation cases considered until now, even though it gives better predictions 
for the  shear stress over the entire region of the recirculation.  We have already started 
developing more advanced filtering schemes, suitable for highly deformed meshes, since we 
believe that the choice of filtering scheme plays a role on the delay of flow's re-attachment, 
and contributes to  yielding larger recirculation bubbles.

\newpage 
\appendix

\newpage
\bibliography{references.bib}
\bibliographystyle{unsrt}
\end{document}